\documentclass[prd,aps,reprint,onecolumn,groupedaddress,tightenlines,nofootinbib,eqsecnum,preprintnumbers,longbibliography,12pt]{revtex4-2}
\usepackage{epsfig,amsfonts,mathrsfs,amsmath,amssymb,graphicx,color,slashed,multirow}
\usepackage{amsmath,latexsym,amssymb,graphicx,slashed,hyperref,color,enumerate,url,etoolbox,cancel,gensymb,physics}
\hypersetup{colorlinks,citecolor= nicegreen,linkcolor= nicered}
\usepackage[usenames,dvipsnames]{xcolor}
\usepackage{comment}

\definecolor{nicered}{rgb}{0.7,0.1,0.1}
\definecolor{nicegreen}{rgb}{0.1,0.5,0.1}
\definecolor{niceblue}{rgb}{0.1,0.2,0.6}

\usepackage{cancel}
\usepackage{bbm}
\usepackage{physics}
\usepackage{braket}
\usepackage{dsfont}
\usepackage{cancel}
\usepackage{cleveref}
\definecolor{DarkerRed}{rgb}{0.666,0,0}
\definecolor{DarkerGreen}{rgb}{0,0.4,0 }
\definecolor{DarkerBlue}{rgb}{0,0,0.4 }
\definecolor{DarkerPurple}{rgb}{0.45,0.1,0.45 }
\definecolor{DarkerOrange}{rgb}{0.66,0.35,0.3 }

\usepackage[normalem]{ulem} 

\begin{document}

\def\Carleton{Ottawa-Carleton Institute for Physics, Carleton University, Ottawa, ON K1S 5B6, Canada}

\title{Anomaly mediation in Seiberg-Witten theories
}

\author{Cyrus Tearlach Robertson Orkish}
\email{cyrusrobertsonorkish@cmail.carleton.ca}
\author{Daniel Stolarski}
\email{stolar@physics.carleton.ca}

\affiliation{\Carleton}


\begin{abstract}
We study the coupling of anomaly mediated supersymmetry breaking (AMSB) to $\mathcal{N}=2$ supersymmetric theories with $SU(2)$ gauge group.  Perturbatively, $\mathcal{N}=1$ supersymmetry (SUSY) is preserved in the UV description with the AMSB terms being equivalent to a supersymmetric mass term for the adjoint chiral multiplet, as considered in the original Seiberg-Witten papers. We show, however, that nonperturbative  instanton contributions break supersymmetry completely. In the IR description, the theory with AMSB is qualitatively similar to its $\mathcal{N}=1$ analog, exhibiting monopole condensation and confinement. However, the effective Lagrangian breaks supersymmetry completely, and the values of the condensates are different from their $\mathcal{N}=1$ analogs. Under physically reasonable assumptions, it can be shown that the vacua must change to those of the perturbatively identical $\mathcal{N}=1$-preserving deformation as the SUSY-breaking scale crosses the strong coupling scale. This change must be caused by effects that cannot in principle be described within the effective weakly coupled IR theory. It is not a phase transition, and the vacua appear to lie in the same universality class at any finite SUSY-breaking scale. Nonetheless, this work highlights the subtle manner in which AMSB coupling to nonperturbative effects can change the vacuum field configuration and the global symmetries of a theory.

\end{abstract}

\maketitle

\tableofcontents

\section{Introduction}
Since the advent of Yang-Mills~\cite{Yang:1954ek} theory more than 70 years ago, the quest for an analytic understanding of strongly coupled gauge theory has become the one of the most notorious theoretical problems in particle physics. Quantum ChromoDynamics (QCD) is the prototypical example of such a theory. It is extraordinarily well studied experimentally, and many decades of theoretical efforts have yielded a rich, empirically validated picture of QCD at low energies~\cite{Gross_2023}. Despite this empirical success, our understanding of the theory remains incomplete, and it is not known how to derive the low-energy effective dynamics starting from the fundamental Lagrangian.

Supersymmetric gauge theories are under much better control. SUperSYmmetry (SUSY) is a powerful constraint, and it can be leveraged to derive exact results that remain out of reach for these theories' nonsupersymmetric cousins, see for example~\cite{Affleck:1984xz, Seiberg_1994, Seiberg_1994_II, Seiberg:1994bp, Konishi_2003}. It is then natural to wonder what one can learn about non-SUSY theories by starting with a SUSY theory and introducing soft SUSY breaking. To some extent, this is a viable approach. However, the weakly coupled effective InfraRed (IR) descriptions of asymptotically free gauge theories are generically quite different from their UltraViolet (UV) descriptions, and there is no general prescription for mapping soft breaking terms from the UV to the IR or vice versa. The SUSY-breaking mechanism known as anomaly mediated SUSY breaking (AMSB)~\cite{Giudice_1998, Randall_1999, Pomarol_1999, Jung_2009} offers a partial solution to this problem. AMSB introduces SUSY breaking by promoting the conformal anomaly of a theory to a superconformal anomaly. Up to a single dimensionful parameter that sets the scale, the SUSY-breaking terms are completely determined by the Lagrangian, $\beta$ functions, and anomalous dimensions of the underlying SUSY theory at the scale of SUSY breaking. SUSY breaking can be parameterized in terms of the gravitino mass, $m_{3/2}$. This suggests the following strategy: \begin{itemize}
\item Introduce AMSB to your UV SUSY Lagrangian. Typically, the large $m_{3/2}$ limit will decouple objects like the gaugino and the squarks, leaving behind a more familiar non-SUSY gauge theory. 
\item Introduce AMSB to your effective IR SUSY Lagrangian. At small $m_{3/2}$, one can calculate how the SUSY breaking alters the low-energy phase of the theory. It typically introduces a scalar potential that eliminates all but a finite number of vacua. As $m_{3/2}$ approaches the EFT cutoff scale, this description of the vacuum will break down, but one hopes that the gross features of the small SUSY-breaking limit will reflect those of the large SUSY-breaking limit, in which we recover the nonsupersymmetric theory of interest.
\end{itemize}
This strategy was employed by Murayama in his seminal work~\cite{Murayama_2021}. Since then, many more papers have explored the IR behaviour of various SUSY theories coupled to AMSB (e.g.~\cite{Cs_ki_2021, Bai_2022,  delima2024sconfiningsusyqcdanomalymediation, Kondo:2025njf, csáki2025phasetransitionsunusualvalues}.) The original Murayama paper found that introducing AMSB to SQCD with 3 flavours and 3 colours gives a positive mass to the squarks and the gaugino. In the limit of large-$m_{3/2}$, they decouple and the resulting Lagrangian is that of QCD with three massless quarks. At small $m_{3/2}$, AMSB corrections to the supersymmetric low-energy theory are calculable, and the resulting theory breaks chiral symmetry and confines. 

However, this does not rule out the possibility that a phase transition occurs as $m_{3/2}$ crosses the strong coupling scale.  There are known examples of theories where a phase transition does indeed occur. For example, in~\cite{Bai_2022}, Bai and one of the authors of this paper found that a phase transition is guaranteed in a particular $SU(5)$ Georgi-Glashow type theory. The predictive power of the AMSB technique for non-supersymmetric theories hinges on this issue. While small $m_{3/2}$ analyses are interesting in their own right, any attempt to extrapolate their results to the large $m_{3/2}$ limit is innately conjectural as long as a phase transition cannot be ruled out. See~\cite{dine2022challengesobtainingresultsreal} for a detailed critique of the technique from this lens.

In this paper, we add to this discussion by introducing anomaly mediation to a class of theories that are under better theoretical control than even $\mathcal{N}=1$ SUSY gauge theories. We introduce AMSB to the asymptotically free $\mathcal{N}=2$ SUSY gauge theries with gauge group $SU(2)$, colloquially known as Seiberg-Witten theories~\cite{Seiberg_1994, Seiberg_1994_II}. 

As toy models for real world physics, $\mathcal{N}=2$ theories present an appealing balance. $\mathcal{N}=4$ Super Yang-Mills (SYM) is integrable, but it is conformal and so it is of limited interest to this research programme. $\mathcal{N}=1$ theories are more realistic, but it is not generally possible to exactly determine the IR effective action. $\mathcal{N}=2$ theories can exhibit asymptotic freedom and display rich physical behaviour, yet there exists a procedure to determine the exact IR effective action, though it can become very complicated for high-rank gauge groups. $\mathcal{N}=2$ theories exhibit an exact electromagnetic duality in the IR, and at special points in the space of vacua magnetically charged objects become massless. In the dual description, the theory behaves like a weakly coupled Supersymmetric Quantum ElectroDynamics (SQED) in which the massless magnetic solitons play the role of electron multiplets. Upon the addition of an $\mathcal{N}=1$ preserving mass term, these special points are the only surviving vacua. The massless solitons acquire a VEV, and the effective IR description closely resembles an abelian Higgs model. Since it is the dual photon that acquires a mass, the resulting Meissner effect is a \textit{dual} Meissner effect, implying the confinement of electric charge. As far as we know, the works~\cite{Seiberg_1994, Seiberg_1994_II} constitute the first time confinement was convincingly shown to take place in a 4-dimensional asymptotically free gauge theory. When the $\mathcal{N}=1$ preserving mass $\mu$ responsible for the dual Meissner effect is taken to be very large, $\mathcal{N}=2$ SQCD reduces to its $\mathcal{N}=1$ analog. It is widely believed that the vacua at small deformation are continuously connected to those of the $\mathcal{N}=1$ SQCD limit. As a shorthand, we refer to this deformation as the SW deformation.

Applying AMSB to the $\mathcal{N}=2$ theories, we find a theory which is in many respects very similar to the SW-deformed theory, but it is not identical. For $m_{3/2}\gg \Lambda$ (where $\Lambda$ is the strong coupling scale), we can use the perturbative description of the $\mathcal{N}=2$ theory to calculate the soft masses. We find that AMSB simply reproduces the SW deformation, up to a nonabelian $R$-symmetry transformation that exchanges the two adjoint fermions in the theory. However, in the other limit $m_{3/2}\ll \Lambda$, the anomaly mediated IR Lagrangian differs from the SW-deformed theory in several respects. It is qualitatively similar in that the surviving vacua exhibit the condensation of massless monopoles and dyons. However, the mass spectrum and the values of the condensates are different from the SW-deformed theory, and the interaction terms break SUSY completely. 

In fact, there is a natural explanation for this discrepancy. In $\mathcal{N}=2$ supersymmetric theories, instanton corrections to the $\beta$ function can be defined in a consistent manner~\cite{Seiberg:1988ur}. Incorporating them into the large $m_{3/2}$ calculation of AMSB soft masses, we find that they produce an instanton suppressed mass splitting that breaks the residual SUSY down to $\mathcal{N}=0$ even in the UV. 

There are good reasons to believe that the condensates found at small $m_{3/2}$ must evolve into the condensates of the SW-deformed theory as $m_{3/2}$ is taken to be large. This is not a phase transition because at any nonzero $m_{3/2}$, the massless field content and the global symmetry breaking pattern match those of the SW-deformed theory in the far IR limit where the SUSY-breaking operators (which are associated with massive fields) can be ignored. Regardless, this raises a number of concerns with respect to the overall AMSB programme. Our results for the vacua in the IR are exact as long as $m_{3/2}$ is below the UV cutoff of the IR effective theory, and there is no warning that the condensate will change. In the large $m_{3/2}$ description, the theories look identical in perturbation theory. Even with the instanton corrections incorporated, one can integrate out the heavy fields and at the level of renormalizable operators, the resulting Lagrangian is that of $\mathcal{N}=1$ SQCD. The vacuum structure should then coincide with that of the SW-deformed theory. This suggests that the vacua undergo this evolution when $m_{3/2}\sim\Lambda_{N_F}$ and neither the UV nor IR descriptions are applicable. 

These findings demonstrate that nonperturbative effects that are invisible to both the UV and IR weakly coupled descriptions are capable of changing the ground state of the theory. We also demonstrate, via our UV instanton calculation, that an effect that vanishes to all orders in perturbation theory can nevertheless have an impact on the theory's global symmetry structure at any finite $m_{3/2}$. We were only able to foreshadow the disagreement of the two theories in the small $m_{3/2}$ regime because we included the instantons in our large $m_{3/2}$ calculation. Had we not, it would have appeared to us that the theory is exactly $\mathcal{N}=1$ supersymmetric for $m_{3/2}\gg\Lambda$, breaking $\mathcal{N}=1$ SUSY only after $m_{3/2}$ flows below $\Lambda$. In reality the theory is only $\mathcal{N}=1$ supersymmetric when either $m_{3/2}=0$, or $m_{3/2}\to\infty$ relative to the dynamical scale under consideration. Presumably, this instanton contribution to the UV soft masses in AMSB is completely generic, though whether it impacts the symmetry structure of other theories is another question.

At a technical level, much of our work is closely related to~\cite{csáki2025phasetransitionsunusualvalues}. That work introduces AMSB not to the $\mathcal{N}=2$ theory itself, but to the confining $\mathcal{N}=1$ theory considered in~\cite{Seiberg_1994, Seiberg_1994_II}. The goals and conclusions of~\cite{csáki2025phasetransitionsunusualvalues} are very different from ours, but there is some overlap in the techniques employed. In particular, our anomaly mediated potential in the IR is a special case of theirs. The reader may find other discussions of soft $\mathcal{N}=2$ breaking in~\cite{_LVAREZ_GAUM__1996,  _lvarez_Gaum__1997, N_2_Konishi_1997, N_2_Evans_1997, _LVAREZ_GAUM__1998_II,  C_rdova_2024}, for example. 

The paper proceeds as follows. In Section~\ref{sec:prelim}, we introduce the basic formulae of anomaly mediation and review the salient features of Seiberg-Witten theories. In Section~\ref{sec:AMSB main body}, we couple  $\mathcal{N}=2$ SYM with no matter to AMSB, showing the perturbative preservation of $\mathcal{N}=1$ SUSY, and we identify the IR vacua. In Section~\ref{sec: Nf=0 theory comparison}, we compare the anomaly mediated theory to the SW-deformed theory. This comparison leads us to suggest that the vacua must change to their SW-deformed analogs as $m_{3/2}$ flows to the UV. In Section~\ref{sec: SQCD with AMSB}, we generalize the pure SYM analysis to theories with up to 3 massless fundamental flavours. Section~\ref{sec: discussion NF nonzero} extends the discussion of Section~\ref{sec: Nf=0 theory comparison} to these cases with flavour. Section~\ref{sec:conclusion} summarizes and concludes.

\Cref{sec:convexK,sec:IR rescaling,sec: higgs branch involved derivation} contain several rather technical auxiliary results about AMSB on different pieces of the $\mathcal{N}=2$ moduli space. These results solidify the overall picture of the anomaly mediated theory, but they are somewhat tangential to the main thread of the paper. Appendix~\ref{sec: bare masses for hypermultiplets and AMSB} briefly discusses the generalization to theories with massive matter. The other appendices list various mathematical expressions used throughout the paper.

\section{Preliminaries}
\label{sec:prelim}
In this section, we review salient features of supersymmetric sigma models, anomaly mediation, and $\mathcal{N}=2$ gauge theories. At the beginning of each subsection, we give a short list of references on each topic. Pedagogical reviews of $\mathcal{N}=1$ supersymmetric field theory and superfield notation can be found in~\cite{MARTIN_1998, Wess:1992cp, TongSUSYFT}.

\subsection{Supersymmetric sigma models} \label{sec: SUSY sigma model review}
Supersymmetric gauge theories often admit a continuum of physically inequivalent vacuum field configurations~\cite{Wess:1992cp, gates2001superspacethousandlessonssupersymmetry, TongSUSYFT}. The collection of vacua associated to a given theory is called the moduli space, and the theories of interest in this work all possess nontrivial moduli spaces. For example, the Lagrangian of $\mathcal{N}=2$ SQCD contains a scalar potential $\propto \Tr[\phi^\dagger, \phi]^2$, where $\phi$ is a scalar transforming in the adjoint representation of the gauge group. This potential is minimized as long as $\phi$ lies in the Cartan subalgebra of the gauge group. We will frequently discuss the moduli spaces of the $\mathcal{N}=2$ theories to which we introduce anomaly mediation. This section gives a brief overview of the formalism associated with moduli spaces.

Moduli spaces of vacua are K{\"a}hler manifolds. These are complex manifolds admitting a Hermitian metric whose real and imaginary parts respectively define a Riemannian metric and a symplectic form on the manifold. In the effective IR description of a theory with a moduli space, the massless scalars are analogous to Goldstone bosons, in that they are associated with translations on the moduli space. Unlike conventional Goldstones, which generate a spontaneously broken symmetry, the translations on a supersymmetric moduli space are not necessarily symmetry transformations, and the K{\"a}hler manifold is not necessarily a symmetric space. Nonetheless, the supersymmetric effective IR Lagrangian describing the massless excitations reflects the geometry of the moduli space much like a typical Goldstone Lagrangian reflects the geometry of its associated symmetric space.  

To be explicit, upon Grassmann integration the Lagrangian $\mathcal{L}=\int d^4\theta K$ yields the following component field Lagrangian:
\begin{align} \label{eq:supersymmetric sigma Lagrangian}
\mathcal{L}=h_{i\bar{j}}\left[\partial_\mu\varphi^i\partial^\mu\varphi^{\dagger\bar{j}}+\frac{i}{2}\mathcal{D}_\mu\psi^i\sigma^\mu\bar{\psi}^{\bar{j}}-\frac{i}{2}\psi^i\sigma^\mu\mathcal{D}_\mu\bar{\psi}^{\bar{j}}\right]+\frac{1}{4}
R_{i\bar{j}k\bar{l}}\psi^i\psi^k\bar{\psi}^{\bar{j}}\bar{\psi}^{\bar{l}},
\end{align}
where $\varphi^i, \varphi^{\dagger \bar{j}}$ are the scalar fields parameterizing the vacua. The K{\"a}hler metric $h^{i\bar{j}}$ is related to the K{\"a}hler potential via \begin{align}
h_{i\bar{j}}=\partial_i \bar{\partial}_{\bar j} K,
\end{align}
 The partials are defined as $\partial_i \equiv \frac{\partial}{\partial\varphi^i}$ and $\bar{\partial}_{\bar{j}} \equiv \frac{\partial }{\partial \varphi^{\dagger \bar j}}$. The left and right handed superpartners of the scalars are denoted $\psi^i$ and $\bar{\psi}^{\bar{i}}$, respectively. The tensor $R_{i\bar{j}k\bar{l}}$ is a K{\"a}hler analog of the Riemann tensor and $\mathcal{D}_\mu\psi^i$ is the K{\"a}hler covariant derivative, given by $\partial_\mu\psi^i - i\Gamma^i_{jk}\psi^j\partial_\mu\varphi^k$ with $\Gamma^i_{jk}$ the Christoffel symbol. These terms are defined in terms of the K{\"a}hler metric as \begin{align}
\Gamma^i_{jk}&\equiv h^{i\bar{l}}\partial_j h_{k\bar{l}} \nonumber \\
R_{i\bar{j}k\bar{l}}&\equiv g_{m\bar{j}}\partial_{\bar{l}}\Gamma^m_{ik}
\end{align}
On smooth regions of the moduli space, these tensors are analytic functions of the coordinates. To obtain a polynomial effective Lagrangian for the low-energy dynamics one may pick a vacuum and Taylor expand the Lagrangian about the point corresponding to that vacuum. Since the tensors are analytic, the expansion has a nonzero radius of convergence, and the radius of convergence of the Taylor expansion defines a sort of UV cutoff, albeit one that can be modified by loop corrections. The origin of this UV cutoff is very physical in nature --- typically, the scalar VEVs that parameterize the moduli space give a mass to fields in the fundamental Lagrangian via the Higgs mechanism, and in the low-energy description, these fields are integrated out. The coordinates set the masses of the Higgsed particles that have been integrated out, and so the point about which this Taylor expansion is performed is closely related to the Wilsonian cutoff of this description, although the relationship is typically not a simple equality once quantum corrections are taken into account. 

It is possible for singularities to arise on the moduli space. These singularities typically signify that the Wilsonian cutoff has gone to 0, meaning that additional fields become massless at the singular point and must be incorporated into the Lagrangian to get a sensible effective description. For example, the classical moduli space of $\mathcal{N}=2$ SYM is singular at the origin $\phi=0$. At this point,\footnote{This point does not exist in the quantum moduli space~\cite{Seiberg_1994}.} gauge bosons that acquire a mass from the VEV of $\phi$ become massless, and must be included in the Lagrangian.


\subsection{Anomaly mediation}
\label{sec:AMSB prelims}
\
Anomaly mediated SUSY breaking (AMSB)~\cite{ Giudice_1998, Randall_1999, Pomarol_1999,Jung_2009} is a controlled form of soft SUSY breaking in which the conformal anomaly of a SUSY theory is promoted to a superconformal anomaly. The soft terms are fully calculable in terms of the gravitino mass $m_{3/2}$ and the $\beta$ functions (and anomalous dimensions).

Anomaly mediation can be formulated in terms of the conformal compensator,\footnote{The compensator is sometimes called $\Phi$ in the literature. We avoid this convention because in $\mathcal{N}=2$ theories, the symbol $\Phi$ is conventionally assigned to another object.} $\chi$. This chiral superfield is a spurion for superconformal symmetry. In anomaly mediation, its value is frozen at $\chi = 1 + \theta^2 m_{3/2}$.  Given a generic SUSY Lagrangian \begin{align}
\mathcal{L}=\int d^4\theta K + \int d^2\theta W + h.c., 
\end{align}
we insert powers of the compensator to formally restore Weyl invariance: \begin{align}
\mathcal{L} \to \int d^4\theta \chi^\dagger \chi K + \int d^2\theta \chi^3 W + h.c..
\end{align}
Calculating the resulting $F$ and $D$-terms yields the following SUSY-breaking potential:
\begin{align} \label{eq:treeB}
V_{\text {\bcancel{tree}}} & =m_{3/2}\left(\partial_i W h^{i \bar{j}} \bar{\partial}_{\bar{j}} K -3\,W\right)+ \text{h.c.} +|m_{3/2}|^{2}\left(\partial_i K h^{i \bar{j}} \bar{\partial}_{\bar{j}} K-K\right),  
\end{align}
where $h^{i\bar{j}}$ is the inverse K{\"a}hler metric. This potential vanishes for a classically conformal Lagrangian, but there are also loop-level effects. Couplings and field strengths depend on the renormalization scale $\mu$, but to formally restore Weyl invariance, we need replace $\mu$ with $\frac{\mu}{|\chi|}$. This leads to the following loop-level SUSY-breaking terms:
\begin{align}\label{eq:loopAMSB}
A_{i j k} & =-\frac{1}{2}\,\left(\gamma_i+\gamma_j+\gamma_k\right) m_{3/2} \, \\
\label{eq:softmass} m_i^2 & =-\frac{1}{4}\, \dot{\gamma}_i\, |m_{3/2}|^2 \, , \\ 
\label{eq:gauginomass} m_\lambda & =-\frac{\beta\left(g\right)}{ g} m_{3/2} \, .
\end{align}
Here, $g$ is the gauge coupling, $\beta\left(g\right)$ is the $\beta$ function of the gauge coupling running, and $\gamma_i$ is the anomalous dimension of the chiral superfields.\footnote{We define the anomalous dimension, its derivative, and the $\beta$ function as $\gamma_i=\mu \frac{d}{d \mu} \log Z_i, \dot{\gamma}=\mu \frac{d}{d \mu} \gamma_i$, and $\beta\left(g\right)=\mu \frac{d}{d \mu} g$. } 
$A_{ijk}$ is a trilinear scalar term that appears only if the underlying SUSY theory has a Yukawa term.
In Appendix~\ref{sec:AMSB derivation}, we explicitly derive all of the formulae given in this section.

\subsection{$\mathcal{N}=2$ gauge theories} \label{sec: prelims N=2 main}

We now review the relevant properties of $\mathcal{N}=2$ supersymmetric gauge theories~\cite{Seiberg:1988ur, Seiberg_1994, Seiberg_1994_II,  bilal1996dualityn2susysu2, Lerche_1997, Alvarez_Gaum__1997, Tachikawa_2015}. These remarkable theories exhibit rich, nontrivial physical behaviour, yet are nevertheless sufficiently constrained by extended SUSY that the IR effective action can be determined exactly.

In this work, we focus on theories with gauge group $SU(2)$. Subsections~\ref{sec: prelim N=2 lagrangian}--\ref{sec:N=2 renormalization} summarize general properties of perturbative $\mathcal{N}=2$ gauge theories and apply to all gauge groups. We specialize to gauge group $SU(2)$ in subsection~\ref{sec:Coulomb review} and onward.

\subsubsection{Lagrangian and field content} \label{sec: prelim N=2 lagrangian}
In $\mathcal{N}=1$ language, the minimal $\mathcal{N}=2$ Yang Mills theory consists of a real superfield $V$ together with a chiral superfield $\Phi$ in the adjoint representation. This is the field content of a single $\mathcal{N}=2$ vector multiplet. The Lagrangian can be written \begin{align}
\mathcal{L}= -\frac{i\tau}{8\pi}\Tr( \int d^2\theta \mathcal{W}^\alpha \mathcal{W}_\alpha + h.c. + \int d^4\theta \Phi^\dagger e^V \Phi e^{-V}),
\label{eq:UVLag}
\end{align}
$\mathcal{N}=2$ matter comes in the form of hypermultiplets, pairs of chiral $\mathcal{N}=1$ multiplets $Q, \widetilde{Q}$ that transform in conjugate representations of the gauge group. The unique $\mathcal{N}=2$ invariant form of the superpotential is \begin{align} \label{eq: N=2 hypermultiplet superpotential}
W = \sum_i\left(\sqrt{2}\widetilde{Q}_i\Phi Q_i+M_i \widetilde{Q}_iQ_i\right).
\end{align}
The masses $M_i$ are free parameters, but the trilinear term is fixed by $\mathcal{N}=2$ supersymmetry. The real superfield $V$ containing the gauge boson sits in an $\mathcal{N}=2$ vector multiplet with $\Phi$, and this Yukawa term is related by $\mathcal{N}=2$ SUSY to the kinetic term $Q^\dagger_i e^{V} Q$ for the hypermultiplets. Table~\ref{tab:field symbols} lists the symbols we assign to the fundamental component fields.
\begin{table}[h]
    \centering
    \begin{tabular}{|c|c c c|}
        \hline
        Superfield &  Spin 0 & Spin 1/2 & Spin 1  \\ \hline
       $\mathcal{W_\alpha}$  &  & $\lambda$ & $A_\mu$ \\
       $\Phi$ & $\phi$ & $\psi_\phi$ & \\
       $Q$ & $q$ & $\psi_q$ & \\
       $\widetilde{Q}$ & $\widetilde{q}$ & $\widetilde{\psi}_{\widetilde{q}}$ &\\
       \hline
    \end{tabular}
    \caption{The fundamental fields in an $\mathcal{N}=2$ gauge theory, written as $\mathcal{N}=1$ superfields in the leftmost column. The component fields are listed by spin in the other columns.}   
 \label{tab:field symbols}
\end{table}

The absence of a coupling constant on the Yukawa term in Eq.~\eqref{eq: N=2 hypermultiplet superpotential} results from the holomorphic normalization of the $\mathcal{N}=2$ vector multiplet \cite{Arkani_Hamed_1998}. In $\mathcal{N}=2$ holomorphic normalization, a factor of $\tau$ multiplies the gauge kinetic term \textit{and} the kinetic term for $\Phi$.  In this normalization, the factor of $g$ appearing in the gauge covariant derivative has been absorbed into the field strength normalization of the gauge boson. The gauge invariant kinetic terms for $Q, \widetilde{Q}$ are related by $\mathcal{N}=2$ SUSY to the Yukawa superpotential, and so in canonical normalization, the Yukawa coupling would be proportional to the gauge coupling. In holomorphic normalization, it is simply a numerical constant.

\subsubsection{R-symmetries} \label{sec:R-symmetry review}

For a generic $\mathcal{N}=1$ theory, one may define a $U(1)_R$ symmetry under which the Grassmannian components $\theta$ transform as $U(1)_R:\theta \to e^{i\alpha}\theta$, while the integrand $d\theta$ transforms as $d\theta \to e^{-i\alpha}d\theta$. For this reason, superpotentials and gauge kinetic terms have $R$-charge 2. Given a chiral superfield $\Phi \sim  \phi+\theta\psi+\theta^2F+...$ with overall $R$-charge $r$, its components $(\phi, \psi, F)$ transform with charges $(r, r-1, r-2)$. 

In $\mathcal{N}=2$ theories, the $R$-symmetry group is $SU(2)_R\times U(1)_R$. In $\mathcal{N}=2$ superspace, there are additional Grassmanian directions which are rotated into each other under the
$SU(2)_R$ symmetry group. However, we will work in $\mathcal{N}=1$ superspace, for which the nonabelian $SU(2)_R$ symmetry is not manifest. In this formalism, we can handle $SU(2)_R$ in terms of its action on component fields, largely treating the Grassmann variables as singlets. The diagonal subgroup of $SU(2)_R$ is an exception, for it can be treated manifestly in the $\mathcal{N}=1$ formalism as an additional abelian $R$-symmetry under which the $\mathcal{N}=1$ Grassmann variables are charged. We denote this additional $R$-symmetry by $U(1)_J$.

The $U(1)_R$ and $U(1)_J$ symmetries transform the fundamental $\mathcal{N}=1$ superfields with charges listed in Table~\ref{tab: R, J charges for fundamental fields}. The table also lists the corresponding component field charges for the scalars and fermions. We will not explicitly require the charges for the other component fields. 
\begin{table}[h]
    \centering
    \begin{tabular}{|c|c c c c|c c c c c|}
         \hline
         & $\Phi$ & $Q$ & $\widetilde{Q}$ & $\mathcal{W}_\alpha$ & $\phi$ & $\psi_{\phi}$ & $q/\widetilde{q}$ & $\psi_q/\psi_{\widetilde{q}}$ & $\lambda$ \\
         \hline
      $U(1)_R$   & 2 & 0 & 0 & 1 & 2 & 1 & 0 & -1 & 1\\
      $U(1)_J$   & 0 & 1 & 1 & 1 & 0 & -1 & 1 & 0 & 1 \\ \hline
    \end{tabular}
    \caption{Charges of the fundamental fields under $U(1)_R$ and $U(1)_J$ transformations.}\label{tab: R, J charges for fundamental fields}
\end{table}
Under nonabelian $SU(2)_R$ transformations, the doublets $(\lambda, \psi_\phi)$, $(q, \widetilde{q}^\dagger)$, and $(\widetilde{q}, q^\dagger)$ all transform in the fundamental representation. The other fields are singlets. This is reflected in their neutrality under $U(1)_J$. 

We will later consider $SU(2)_R$ rotations that exchange the components of the doublets. In anticipation of this discussion, we now define a transformation \begin{align}
s = \begin{bmatrix}
0 & 1 \\
-1 & 0
\end{bmatrix},
\end{align}
which swaps the doublets: \begin{align} \label{eq: swap operator}
s:\begin{bmatrix} \lambda \\\psi_{\phi}
\end{bmatrix} \to \begin{bmatrix} \psi_{\phi} \\ -\lambda \end{bmatrix}.
\end{align}
Note the relative minus sign introduced by this transformation. When we swap the doublets with the transformation $s$, it also swaps their $U(1)_J$ charges. We can write a $U(1)_J$ element $j\in SU(2)_R$ as \begin{align}
j = \begin{bmatrix}
e^{i\alpha} & 0 \\
0 & e^{-i\alpha}
\end{bmatrix},
\end{align}
If we act on the doublets first with $s$ then with $j$, we have \begin{align}
js:\begin{bmatrix} \lambda \\ \psi_\phi\end{bmatrix} \to \begin{bmatrix}e^{i\alpha}\psi_{\phi} \\ -e^{-i\alpha}\lambda \end{bmatrix}.
\end{align}
The $\mathcal{N}=1$ Grassmann variables, by contrast, do not flip $J$-charge under $s$. This ensures that the overall Lagrangian remains $J$-invariant. For example, the term $\theta \lambda$ has $J$-charge $1+1=2$. Acting on it with $s$ gives $s:\theta\lambda\to -\theta\psi_\phi$. Since the charge of $\psi_\phi$ becomes $+1$ after the $s$ transformation, the charge of $\theta$ must remain equal to $+1$ to preserve the overall $U(1)_J$ symmetry.

\subsubsection{Perturbative renormalization of $\mathcal{N}=2$ theories} \label{sec:N=2 renormalization}

Quantum corrections to $\mathcal{N}=2$ theories are highly constrained. Perturbatively, the $\beta$ function is 1-loop exact~\cite{Seiberg:1988ur}. This statement is also true of the holomorphic coupling $\tau$ in $\mathcal{N}=1$ theories, but it is less powerful in the $\mathcal{N}=1$ context. This is because the passage to canonical normalization ($V\to gV$) introduces a Jacobian to the path integral measure. The Jacobian induces higher order corrections in the physical gauge coupling. Theories with $\mathcal{N}=2$ supersymmetry evade this subtlety. Since the chiral superfield $\Phi$ sits in a multiplet with the real superfield $V$, it is also rescaled when we pass to canonical normalization. The Jacobians of the two transformations cancel such that the physical gauge coupling coincides with the holomorphic coupling, receiving only 1-loop corrections in perturbation theory. These issues are explained in detail in~\cite{Arkani_Hamed_2000}. 

From~\cite{MARTIN_1998}, the one-loop $\beta$ function for a general supersymmetric gauge theory is:
\begin{align} \label{eq: 1-loop beta function}
    \beta(g) = \frac{g^3}{16\pi^2}\left(\sum_i I(i)-3C(G)\right).
\end{align}
Here $I(i)$ is the Dynkin index of the representation to which the matter multiplets belong, while $C(G)$ is the quadratic Casimir invariant. We normalize our generators so that in the fundamental representation, $I(\Box)=\frac{1}{2}$. If we take, for example, an $\mathcal{N}=2$ $SU(N)$ gauge theory with $N_F$ hypermultiplets in the vectorlike fundamental representation, $C(G)=N$ and we get a $\beta$ function \begin{align}
\beta(g) = \frac{g^3}{16\pi^2}(N_F-2N)\equiv -b_0g^3.
\end{align}
This theory is asymptotically free for $N_F<2N$, conformal for $N_F=2N$, and IR-free for $N_F>2N$. These statements hold nonperturbatively. The asymptotically free theories become strongly coupled at some characteristic scale $|\Lambda|$, which we complexify as $\Lambda\equiv |\Lambda|e^{\frac{i\theta}{4-N_F}}$. In~\cite{csáki2025phasetransitionsunusualvalues}, the $\mathcal{N}=2$-supersymmetric IR effective Lagrangians are written in terms of the rescaled parameters $\Lambda_{N_F}$, related to $\Lambda$ by: \begin{align}\label{eq:LambdaNF}
\Lambda_0 &= \Lambda \nonumber\\
\Lambda_1 &= \sqrt{3}\times 2^{-\frac{4}{3}}\Lambda \nonumber\\
\Lambda_2& = 2^{-\frac{3}{2}}\Lambda \nonumber\\
\Lambda_3& = 2^{-4}\Lambda
\end{align}
We follow this convention.

Anomalous dimensions are also particularly simple in $\mathcal{N}=2$ theories. In holomorphic normalization, the chiral multiplet $\Phi$ (and therefore the real superfield $V$, by $SU(2)_R$ invariance) simply have anomalous dimension $0$~\cite{Arkani_Hamed_2000}. Introducing $N_F$ massless hypermultiplets, we have the superpotential \begin{align} \label{eq: UV yukawa term}
W = \sum_{i=1}^{N_F}\sqrt{2}\widetilde{Q}_i\Phi Q_i.
\end{align}
By the $\mathcal{N}=1$ nonrenormalization theorem~\cite{Seiberg:1994bp}, the superpotential does not run. The Yukawa terms arising from this superpotential couple the matter fields to the fermion $\psi_\phi\in\Phi$. They are related via $SU(2)_R$ to the Yukawa terms that couple those fields to the gaugino $\lambda$. A wavefunction renormalization for $Q$ or $\widetilde{Q}$ would break $SU(2)_R$, and so the anomalous dimensions for $Q$ and $\widetilde{Q}$ are 0.

When we pass to the canonical normalization scheme $\Phi\to g\Phi$, we induce a wavefunction renormalization in $\Phi$ to compensate the running of $g$. We can pick some reference scale with respect to which we can define a bare $\Phi_0$. At arbitrary scale, the Kahler potential for $\Phi$ is  \begin{align}
K = \int d^4\theta Z_\Phi \Tr(\Phi^\dagger_0 \Phi_0)
\end{align}
In this scheme, the superpotential is \begin{align}
W = \sqrt{2Z_{\Phi}} g\widetilde{Q}\Phi_0 Q
\end{align}
It should still satisfy the $\mathcal{N}=1$ nonrenormalization theorem, so we have
\begin{align}
\frac{dW}{dt}& = \sqrt{2}\widetilde{Q}\Phi_0Q\left(\frac{dg}{dt}\sqrt{Z_\phi}+g\frac{d\sqrt{Z_\phi}}{dt}\right) \nonumber \\
&= 0
\end{align}
where $t$ is the logarithm of the renormalization scale.  We then have \begin{align} \label{eq: diff eq for Z}
\frac{1}{\sqrt{Z_\phi}}\frac{d\sqrt{Z}}{dt} & = -\frac{dg}{dt}\frac{1}{g} \nonumber \\
&=-\frac{\beta}{g}.
\end{align}
In Section~\ref{sec:AMSB prelims}, we defined the anomalous dimension as $\gamma \equiv \frac{d\log Z}{dt}$. We have \begin{align}
\frac{d\sqrt{Z}}{dt}=\frac{1}{2\sqrt{Z}}\frac{dZ}{dt}
\end{align}
Hence, we find that in canonical field normalization, \begin{align} \label{eq: anomalous dimension for phi}
\gamma_\Phi= -\frac{2}{g}\beta(g)
\end{align}
This derivation assumed the presence of hypermultiplets. But the result holds in the $N_F=0$ theory as well. The $N_F\neq 0$ theories reduces to the $N_F=0$ theory in the limit of large masses for the hypermultiplets. Take for example the $N_F=1$ theory. We can add a mass term $M \widetilde{Q} Q$ to the superpotential, and since the hypermultiplet field strength does not run, the result Eq.~\eqref{eq: anomalous dimension for phi} still holds. As we decouple the hypermultiplet by taking $M\to \infty$, we recover the $N_F=0$ theory with the relation Eq.~\eqref{eq: anomalous dimension for phi} intact.

\subsubsection{Coulomb phase for $N_F=0$} \label{sec:Coulomb review}
We now focus the discussion to gauge group $SU(2)$. Let us first discuss the effective IR description of the theory with $N_F=0$. As discussed in subsection~\ref{sec: SUSY sigma model review}, the low-energy dynamics are closely related to the geometry of the moduli space of vacua. The metric for $\mathcal{N}=2$ SYM with gauge group $SU(2)$ was determined exactly in~\cite{Seiberg_1994}. 

The $D$-term of $\mathcal{N}=2$ SYM is proportional to $[\phi, \phi^\dagger]$, where $\phi$ is the adjoint scalar in $\Phi$. $D$-flatness is achieved when $\langle\phi\rangle$ lies in the Cartan subalgebra of the Lie algebra of the gauge group. For $SU(2)$, 
\begin{align}
\langle\phi\rangle=a\sigma^3,
\label{eq:phivev}
 \end{align} 
where $\sigma^3$ is the third Pauli matrix. \footnote{We follow the convention in ~\cite{Seiberg_1994_II} where classically, $a=\sqrt{\frac{u}{2}}$.}
This VEV breaks the gauge symmetry down to $U(1)$ and gives a mass to all of the field components that are charged under the surviving $U(1)$. There is a moduli space of vacua and it is classically parameterized by the complex number $a$. Since a $U(1)$ gauge symmetry survives on the moduli space, this space is called the Coulomb branch of the theory.\footnote{For higher gauge groups, the surviving gauge group is $U(1)^r$ with $r$ the dimensionality of the Cartan subalgebra. The Coulomb branch is parameterized by $r$ complex VEVs.}

Since the electrically charged particles have a mass set by the VEV, the classical low-energy theory is a trivial theory of a free $U(1)$ $\mathcal{N}=2$ vector multiplet. Effective interactions are induced in perturbation theory via the virtual exchange of heavy charged particles, and instantons contribute nonperturbative corrections~\cite{Seiberg:1988ur}. The classical theory has a singularity at the point $a=0$, where the Higgs mechanism turns off and the electrically charged particles become massless. In the quantum theory, $a$ is bounded from below over the entire moduli space, and there are no vacua in which the full $SU(2)$ gauge symmetry survives. However, there remain singularities on the moduli space. There are two, and they are related by a $\mathbb{Z}_2$ symmetry of the moduli space. At these points, magnetically charged solitons become massless. To describe how this comes about, we must discuss the metric on the moduli space.
 
In the perturbative ($a\gg \Lambda_0$) region of the moduli space, we can conveniently write the superfield Lagrangian in terms of a single holomorphic function of $A$, the chiral superfield whose lowest component is the VEV $a$ from Eq.~\eqref{eq:phivev}. This function is called the prepotential, denoted by $\mathcal{F}$.  The mass of the fields charged under the surviving $U(1)$ is set by $a$. This is the scale by which the virtual exchange of these particles is suppressed, and so at large $a$ the theory is weakly coupled. The effective Lagrangian in this region can be written in terms of the prepotential $\mathcal{F}$ as\footnote{As is standard in $\mathcal{N}=2$ literature, we use $\Im(z) = (z-z^*)/(2i)$ whether $z$ is a complex number or a complex (super)field. }
\begin{align} \label{eq:EFT Coulomb branch}
\mathcal{L}=\frac{1}{4\pi}\Im\left[\int d^4\theta \frac{\partial\mathcal{F}(A) }{\partial A}A^\dagger-\frac{1}{2}\int d^2\theta \frac{\partial^2\mathcal{F}}{\partial A^2}\mathcal{W}^\alpha\mathcal{W}_\alpha\right],
\end{align}
where $\mathcal{W}$ is the massless $U(1)$ gauge multiplet and $A$ is the $\mathcal{N}=1$ chiral multiplet sitting in the larger $\mathcal{N}=2$ vector multiplet. The metric on this region of the Coulomb branch is given by \begin{align} \label{eq: semiclassical metric}
ds^2  = \Im(\tau)dada^\dagger,
\end{align}
where $\tau = \frac{d^2\mathcal{F}}{d a^2}$.  This $\tau$ is the effective holomorphic $U(1)$ gauge coupling. Since it is a function of $a$, it gives rise to nonrenormalizable effective interactions between the components of $A$ and $\mathcal{W}_\alpha$.

For reasons related to the positivity of the metric~\cite{Seiberg_1994}, Eq.~\eqref{eq: semiclassical metric} cannot be globally valid over the whole moduli space. This tells us that in the strongly coupled region of the moduli space where $a\sim\Lambda_{0}$, the dynamics cannot be described by a local polynomial Lagrangian in $a$. It turns out that the coordinate $u$, equal to $\braket{\Tr(\Phi^2)}$, is a good global coordinate over the entire moduli space, and $a$ can be regarded as a holomorphic function of this coordinate. The moduli space has a $\mathbb{Z}_2$ symmetry which is implemented via the transformation $u\to-u$. 

Introducing another holomorphic function 
$a_D(u)\equiv \frac{\partial\mathcal{F}}{\partial a}$, the exact metric has the form
\begin{align} \label{eq: global metric}
ds^2 = \Im\left(\frac{da_D}{du}\frac{da^\dagger}{du^\dagger}\right)dudu^\dagger.
\end{align}
For $a\gg \Lambda_0$, $u\approx 2a^2$, and one can recover the large $a$ Lagrangian of Eq.~\eqref{eq:EFT Coulomb branch} by taking a large $u$ asymptotic expansion and making the substitution $a\to\sqrt{u/2}$.
In terms of $u$, the effective electric gauge coupling $\tau$ can be written 
\begin{align}\label{eq: effective electric coupling coulomb branch}
\tau = \frac{da_D/du}{da/du} & = \frac{\frac{d}{du}\left(\frac{d\mathcal{F}}{da}\right)}{\frac{da}{du}}\nonumber \\ & = \frac{d^2\mathcal{F}}{da^2}, \\ &
\equiv \frac{\theta_{el}}{\pi}+i\frac{8\pi}{g_{el}^2}
\end{align}
where $\theta_{el}, g_{el}$ are the effective electric theta angle and coupling. The imaginary part of the second line of Eq.~\eqref{eq: effective electric coupling coulomb branch} is the metric in the perturbative region, Eq.~\eqref{eq: semiclassical metric}. It can be defined everywhere on the Coulomb branch, but it is only suitable as a metric in the perturbative large-$a$ region.\footnote{The unusual normalization of $\tau$ is due to the normalization of $a$ in~\cite{Seiberg_1994_II} such that $a=\sqrt{u/2}$.}
The corresponding K{\"a}hler potential is given by \begin{align} \label{eq: generic form of coulomb branch kahler potential}
    K=\frac{1}{4\pi}\text{Im}( a_D a^\dagger).
\end{align}

Eq.~\eqref{eq: global metric} makes manifest that the metric has an $Sp(2, \mathbb{R})$ symmetry under which $(\frac{da_D}{du}, \frac{da^\dagger}{du^\dagger})$ transforms, because the function $\Im(\frac{da_D}{du} \frac{da^\dagger}{du^\dagger}) \equiv \frac{1}{2i}(\frac{da_D}{du}\frac{da^\dagger}{du^\dagger}-\frac{da}{du}\frac{da_D^\dagger}{du^\dagger})$ is a symplectic form on $(\frac{da_D}{du},\frac{da^\dagger}{du^\dagger}) $. Similarly, the Kahler potential Eq.~\eqref{eq: generic form of coulomb branch kahler potential} is invariant under $Sp(2, \mathbb{R})$ if the doublet $(a_D, a^\dagger)$ transforming in the same manner as the doublet $(\frac{da_D}{du},\frac{da^\dagger}{du^\dagger})$.

Taking into account the transformation properties of the gauge kinetic terms, one finds that the true symmetry group of the theory is the discrete $SL(2, \mathbb{Z})$. It has generators $S$ and $T_b$, which act on both the vector $(\frac{da_D}{du}, \frac{da}{du})$ and its conjugate as \begin{align} \label{eq: duality transforms}
T_b & = \begin{bmatrix} 1 & b \\
0 & 1 \end{bmatrix}, \nonumber \\
S & = \begin{bmatrix} 0 & 1 \\
-1 & 0 \end{bmatrix},
\end{align}
where $b$ is an integer.\footnote{The generator $S$ is not to be confused with the $SU(2)_R$ transformation $s$ defined in Eq.~\eqref{eq: swap operator}.} These generators do not act on $u$, and so the vector $(a_D, a)$ and its conjugate also sit in representations of $SL(2, \mathbb{Z})$. 

This $SL(2, \mathbb{Z})$ symmetry lends some context as to why $a$ fails to be a globally valid coordinate on the moduli space. The K{\"a}hler potential and the metric are both single valued over the $u$-plane, but the quantities $(a(u), a_D(u))$, their conjugates, and their derivatives all transform nontrivially under $SL(2,\mathbb{Z})$ while $(u, \bar{u})$ is a singlet. 

Now, since this theory contains a nonabelian gauge group spontaneously broken to $U(1)$, the spectrum of the low-energy theory contains magnetically charged 't Hooft-Polyakov monopoles~\cite{harvey1996magneticmonopolesdualitysupersymmetry}, and, more generally, dyons that have both electric and magnetic charge. It turns out that the transformations in Eq.~\eqref{eq: duality transforms} can be physically interpreted as electromagnetic duality transformations. $S$ implements the duality transform exchanging electric and magnetic degrees of freedom, and $T$ implements the transformation $\theta \to \theta +2\pi b$, under which the electric charge of magnetically charged objects is shifted. Hereafter, we will not be very careful about distinguishing between dyons and monopoles, since a sequence of duality transforms can map one to the other.

The relation between $(a, a_D)$ and the electromagnetic charge vector $(n_m, n_e)$ (where $n_m$ and $n_e$ are the magnetic and electric charges, normalized to be integers) is as follows:
the $\mathcal{N}=2$ supersymmetry algebra admits a central extension, under which dyonic states with charge vector $(n_m, n_e)$ have central charge $(a_Dn_m, a n_e)$. All charged states satisfy the following inequality: \begin{align} \label{eq: BPS bound}
M\geq \sqrt{2}|Z| = \sqrt{2} |a n_e + a_Dn_m|.
\end{align}
Here $Z$ is the central charge of the SUSY algebra and $M$ is the mass of the state. States that saturate this inequality are called BPS states. A consequence of the representation theory of $\mathcal{N}=2$ supersymmetry is that the masses of BPS states are protected from quantum corrections. In the semiclassical region, we argued that $a$ sets the mass of those charged states that are integrated out in the low-energy theory. More generally, the lower bound on the masses of charged states is set by the lightest BPS state which saturates the bound Eq.~\eqref{eq: BPS bound}.

Using the semiclassical description Eq.~\eqref{eq:EFT Coulomb branch}, one can show that the vector $(a_D, a)$ undergoes a monodromy at infinity, with the monodromy an element of the duality group $SL(2, \mathbb{Z})$. This implies that there must be a singularity somewhere in the $u$-plane. Considerations involving the positivity of the metric imply that one is not enough---there must be at least two. These singularities come with their own noncommuting monodromies. 

The singularities occur at $u=\pm \Lambda_0^2$. It can be shown from the monodromy structure that the new massless degrees of freedom at each singularity are dyons with charge vectors $(0, 1)$ and $(1, 1)$, respectively. To get a weakly coupled description of the theory in the neighbourhood of these singularities, one must work in the duality frame appropriate to each dyon. In their respective duality frames, these dyons are treated as fundamental fields charged under the $U(1)$ gauge field associated with the dual field strength tensor. The dynamics at the singularities are therefore described, at the level of renormalizable operators, by a dual $\mathcal{N}=2$ SQED where the massless dyons act like weakly coupled ``electron'' multiplets. In subsection~\ref{sec: review on near-singularity EFT}, we describe this EFT in detail. 

The functions $a_D(u)$ and $a(u)$ can be determined exactly. The solutions, which can be expressed in terms of hypergeometric functions, are given in~\cite{csáki2025phasetransitionsunusualvalues} and we have restated them in Appendix~\ref{sec:Hypergeo}. The arguments that lead to these solutions are quite subtle, and we will not describe the derivation in any depth. The original work~\cite{Seiberg_1994} gives a detailed explanation. Pedagogical reviews from a variety of perspectives can be found in~\cite{Lerche_1997, Tachikawa_2015, Alvarez_Gaum__1997, bilal1996dualityn2susysu2}.

Speaking very broadly, the duality structure on the moduli space admits a powerful complex geometric interpretation that allows one to determine $a(u)$ and $a_D(u)$ given only the monodromy data as input. In particular, $a(u)$ and $a_D(u)$ as well as their derivatives and conjugates can be shown to satisfy a type of second-order ODE known as a Picard Fuchs equation~\cite{klemm1998geometryn2supersymmetriceffective}. For the theory at hand, the Picard Fuchs equation takes the form
\begin{align} \label{eq:picard fuchs}
\frac{d^2f}{du^2}-\frac{f}{p(u)}=0.
\end{align}
The function $p(u)$ encodes the singularities on the $u$ plane. Ref.~\cite{klemm1998geometryn2supersymmetriceffective} lists the Picard Fuchs operators for the theories considered in this paper. The monodromy structure and the asymptotic (perturbative) behaviour of the theory fixes the solution $(a_D(u), a(u))$ uniquely up to duality frame. These solutions are branched, and duality transformations move us from one branch to another. In Appendix~\ref{sec:convexK}, we will make direct use of the Picard Fuchs equation. It can also be converted into a hypergeometric equation, which leads to the solutions given in Appendix~\ref{sec:Hypergeo}.

\subsubsection{Near singularity EFT}\label{sec: review on near-singularity EFT}
We now discuss the effective field theory that describes the dynamics at the singularities where monopoles or dyons become massless.
The dyons come in the form of $\mathcal{N}=2$ hypermultiplets. In general, we denote light dyon/antidyon pairs as $M, \widetilde{M}$, with scalar components $m, \widetilde{m}$.  Near singularities, we can work in a duality frame in which the massless dyons are regarded as weakly coupled fundamental fields. In this frame, the Lagrangian is that of $\mathcal{N}=2$ SQED up to nonrenormalizable K{\"a}hler corrections. 

As an example, let us take the singularity at $u=+\Lambda_0^2$, at which monopoles with charge vector $(1, 0)$ become massless. By the BPS relation Eq.~\eqref{eq: BPS bound}, the monopole mass near the singularity is given by $\sqrt{2}a_D$, and at the singularity, $a_D=0$. This can be captured in a superpotential \begin{align}
    W=\sqrt{2}A_DM\widetilde{M},
\end{align}
with $A_D$ the chiral multiplet whose lowest component is $a_D$. This is the obligatory Yukawa term in equation~\eqref{eq: N=2 hypermultiplet superpotential}. If we regard the pair $(\mathcal{W}_D, A_D)$ as an $\mathcal{N}=2$ vector multiplet (where $\mathcal{W}_D$ is the dual field strength superfield), then at leading order, the Lagrangian is simply a standard $\mathcal{N}=2$ gauge theory Lagrangian of the form described in Section~\ref{sec: prelim N=2 lagrangian}, with the substitutions $(\mathcal{W}, \Phi, Q, \widetilde{Q})\to (\mathcal{W}_D, A_D, M, \widetilde{M})$. 

There are also nonrenormalizable interactions, which we can work out as follows. On the Coulomb branch, the K{\"a}hler potential and the effective gauge coupling can be written in terms of the functions $a_D(u), a(u)$. In the neighbourhood of the monopole singularity, we can power expand in terms of $u$, then invert the series to produce a series expansion for $a$ and $u$ in terms of $a_D$. Subleading terms in this expansion contribute nonrenormalizable couplings to the Lagrangian.

In analogy to the weakly coupled electric prepotential of Eq.~\eqref{eq:EFT Coulomb branch}, we can introduce a dual prepotential $\mathcal{F}_D(a_D)$, defined by the relation \begin{align}
a^\dagger=-\bar\partial\mathcal{F}_D^\dagger,
\end{align}
where $\partial \equiv \frac{\partial}{\partial a_D}, \bar{\partial}\equiv \frac{\partial}{\partial a_D^\dagger}$. The dual complex coupling constant is related to the dual prepotential by \begin{align}
\tau_D = \partial^2 \mathcal{F}_D.
\end{align}
We can write the K{\"a}hler potential in terms of this dual prepotential. Near the singularity, the K{\"a}hler potential includes canonical terms for the monopoles and can be written 
\begin{align} \label{eq:KTotal}
    K_{\text{total}} = \frac{1}{4\pi} \Im{A_D^\dagger\partial\mathcal{F}_D} + M^\dagger e^{V_D}M + \widetilde{M}^\dagger e^{-V_D}\widetilde{M},
\end{align} 
where $V_D$ is the dual $\mathcal{N}=1$ vector multiplet. For the sake of notational concision, we will continue to use the name $K$ exclusively for the monopole independent part of this expression. The metric is then given by 
\begin{align} \label{eq: handy singularity EFT formulae}
h \equiv \partial \bar\partial K= \frac{1}{4\pi}\Im(\tau_D) = \frac{2}{g_{eff}^2},
\end{align}
where $g_{eff}$ is the real effective coupling constant, related to $\tau_D$ by \begin{align}
\tau_D = \frac{\theta_{eff}}{\pi}+i\frac{8\pi}{g_{eff}^2}.
\end{align}

The near singularity expansion of $\mathcal{F}_D$ in terms of $A_D$ is given to next-to-leading order (NLO) in~\cite{csáki2025phasetransitionsunusualvalues},\footnote{The notation of~\cite{csáki2025phasetransitionsunusualvalues} is somewhat different from ours. What we generically call $a_D$ is referred to in their work as $a^{(N_F, k)}$, where $k$ indexes the singularities. Also, what they call $\mathcal{F}^{N_F, k}$ we refer to generically as $\mathcal{F}_D$, related to our $\mathcal{F}$ by a negative sign.} not just for the monopole singularity but for all singularities and all $N_F<4$. Plugging  their expression for $\mathcal{F}_D$ into $K=\frac{1}{4\pi}\text{Im}(A_D^\dagger \partial \mathcal{F}_D)$, we derive the following K{\"a}hler potential,  including monopoles, 
to NLO in $A_D$: 
\begin{align} \label{eq:IR Kahler potential}
K_{\text{total}} = \frac{1}{8\pi^2}\left[2i(A_D \Lambda_0^\dagger-A_D^\dagger\Lambda_0)-\frac{A_D A_D^\dagger}{2}\log\left(\frac{A_D^\dagger A_D}{\Lambda_0\Lambda_0^\dagger}\right)+...\right]+M^\dagger e^{V}M + \widetilde{M}^\dagger e^{-V} \widetilde{M}\nonumber \\
= K + M^\dagger e^{V}M + \widetilde{M}^\dagger e^{-V} \widetilde{M}.
\end{align}
This is the lowest order for which the effective magnetic gauge coupling $g_{eff}$ is nonzero. It is related to the (monopole independent) K{\"a}hler potential straightforwardly:  \begin{align}
\frac{1}{g_{eff}^2} = \frac{h}{2} \equiv \frac{\partial\bar{\partial} K}{2}.
\end{align}
Because of this relationship, the factors of $h$ multiplying the bosonic kinetic terms for $a_D$ and the dual photon are $1/g_{eff}^2$ normalization factors similar to a standard renormalizable Yang-Mills theory. They differ in that subleading corrections to $g_{eff}$ contribute nonrenormalizable interactions to the Lagrangian (see Appendix~\ref{sec: higher corrections to the dual prepotential}).

From Eq.\eqref{eq:IR Kahler potential}, $g_{eff}$ is given by\footnote{This expression is in minor disagreement with that in~\cite{csáki2025phasetransitionsunusualvalues}. They have a $-1$ in the denominator, rather than a $-2$.}
\begin{align}\label{eq: leading order magnetic coupling}
g_{eff}^2 = \frac{32\pi^2}{\log{(|\frac{\Lambda_0}{a_D}|^2)}-2}.
\end{align}
This coupling approaches 0 as $a_D\to 0$. Treating $a_D$ as a complexified version of the renormalization group scale, we can take the derivative of $g_{eff}$ with respect to $\log(|a_D|)$ to find \begin{align}\label{eq: running of effective magnetic coupling}
\frac{\partial g_{eff}}{\partial \log(a_D)}\propto g_{eff}^3,
\end{align}
with a positive coefficient, as expected for the $\beta$ function of a weakly coupled $U(1)$ theory.\footnote{To make $a_D$ more like a renormalization group scale, work in polar coordinates and treat the radial coordinate $|a_D|$ as the scale.} This quantity is perfectly regular at the origin, but notice that $\frac{\partial g_{eff}}{\partial a_D}$ is not. We instead have \begin{align}
\frac{\partial g_{eff}}{\partial a_D}\propto\frac{g_{eff}^3}{a_D},
\end{align}
which diverges at the origin. The presence of nonanalytic behaviour at the origin is to be expected. It reflects the presence of the singularity on the Coulomb branch at this point. Recalling subsection~\ref{sec: SUSY sigma model review}, perturbative calculations can be performed in this theory by Taylor expanding the Lagrangian about some background value of $a_D$. The expansion coefficients play the role of effective couplings. This action is a Wilsonian effective action for which fluctuations above the scale set by $|a_D|$ have been integrated out. In the $a_D\to 0$ limit, all fluctuations have been integrated out, and there are no quantum corrections left to calculate. The action is an exact effective action whose Euler-Lagrange equations calculate VEVs.

On the other hand, the weakly coupled magnetic description of the theory must break down as $a_D$ becomes large. When $\left|\frac{\Lambda_0}{a_D}\right|^2=e^2$, $g_{eff}$ diverges. This is the Landau pole of the low energy EFT.\footnote{The location of the Landau pole can be altered by higher order corrections to the dual prepotential. See Appendix \ref{sec: higher corrections to the dual prepotential}.} At some smaller finite value of $|a_D|$, the coupling is of order 1 and perturbation theory breaks down. The series expansion of $K$ and $a$ in terms of $a_D$ also has some finite radius of convergence, outside of which the weakly coupled EFT must break down.

The effective theory at the other singularity $u=-\Lambda_0^2$ is nearly identical to the one described here. The weakly coupled duality frame is different, meaning that the elementary fields appearing in the theory at that singularity are complicated solitons in the duality frame of the monopole singularity. Expressed in its own duality frame, the other singularity is described by a Lagrangian that is identical up to some differences in the numerical values of the coefficients in Eq.~\eqref{eq:IR Kahler potential}.

\subsubsection{$N_F\neq 0$} \label{sec: prelims on NF neq 0}
We now tersely summarize how this story changes in the presence of massless hypermultiplets. 
$\mathcal{N}=2$ SQCD with gauge group $SU(2)$ is exactly conformal at $N_F=4$, IR free for $N_F>4$, and asymptotically free for $N_F<4$. We restrict our attention to $N_F< 4$. All of these theories' Coulomb branches have the broad structure described in sections~\ref{sec:Coulomb review} and satisfy Picard-Fuchs equations of the form Eq.~\eqref{eq:picard fuchs}. But the global symmetry on the $u$ plane, the number of singularities, and the monodromies associated with those singularities all vary from theory to theory. Some singularities in the $N_F=2, 3$ theories have multiple massless dyon hypermultiplets. Table~\ref{tab:list of singularities vs NF} lists the singularities and some of their basic properties in each case.
\begin{table}[h] 
    \centering
 \begin{tabular}{|c|c|c|c|c|c|}
    \hline
    $N_F$ & Massless $(n_m, n_e)$  & Number of light dyons & Global symmetry on Coulomb branch\\
    \hline
    0 & (0, 1)  & 1 & $\mathbb{Z}_2$\\
      & (1, 1) &  1 &  \\
     \hline
      1 & (1, -1) & 1 & $\mathbb{Z}_3$\\
        & (1, 1) & 1& \\
        & (1, 0)  & 1 &  \\
       \hline
       $2$ & (1, 0) & 2 & $\mathbb{Z}_2$\\
        & (1, 1)&  2 & \\
      \hline
        $3$ & (1, 0) & 1 & No global symmetry\\
        &(2, -1) & 4 & \\
       \hline
\end{tabular}
    \caption{A list of singularities, dyon multiplicities, and symmetries on the Coulomb branch. }\label{tab:list of singularities vs NF}
\end{table} 

In theories with $2$ or more flavours, it is possible for the squarks to have nonzero VEVs. This branch of the moduli space is called the Higgs branch. Suppressing colour indices, the Higgs branch is defined by the VEV configuration \begin{align} \label{eq: higgs branch VEVs}
\vec{Q} & = (B, 0, ...) ,\nonumber\\
\vec{\widetilde{Q}} &= (0, B, ..), \nonumber\\
\Phi& = 0,
\end{align}
where $\vec{Q}$ is a vector whose components are the different flavours of $Q$. This is in fact the exact expression for a $U(1)$ theory. For a nonabelian theory, the flatness conditions are satisfied by the constraints \begin{align}
Q_{ia}^\dagger T_{k}^{ab} Q^i_{b} & =\widetilde{Q}_{ia}T_{k}^{ab}\widetilde{Q}^{i \dagger}_{b} \nonumber\\
\widetilde{Q}_{ia} T_k^{ab}Q^{i}_{b}&=0,
\end{align}
where $T_k$ is the $k$th generator of the gauge group, $i$ is a flavour index, and $a, b$ are colour indices.

On the Higgs branch the gauge symmetry is broken completely. The full $\mathcal{N}=2$ vector multiplet is given a mass set by $B$. The classical Higgs branches are unmodified by quantum corrections~\cite{Seiberg_1994_II}. The $N_F=2$ theory has a very simple Higgs branch: a pair of flat cones, $\mathbb{C}^2/\mathbb{Z}_2$. The $N_F=3$ Higgs branch is a more complicated Ricci-flat K{\"a}hler manifold derived in~\cite{Antoniadis_1997}.

The Higgs branch makes contact with the Coulomb branch at the singularities with multiple light dyons shown in Table~\ref{tab:list of singularities vs NF}. The effective scalar potentials are minimized at any $a_D$ if the dyons are set to 0. But when $a_D=0$, the dyons can get a VEV that is given, up to symmetries, by 
\begin{align}
\vec{m} = (B, 0, 0, ...) \nonumber\\
\vec{\widetilde{m}} = (0, B, 0, ...).
\end{align}
The resemblance to Eq.~\eqref{eq: higgs branch VEVs} is not coincidental. This is the Higgs branch, expressed in variables appropriate to the near singularity region.

\subsubsection{Explicitly breaking $\mathcal{N}=2$ down to $\mathcal{N}=1$}\label{sec:N=1-preserving deformation in SW}
In~\cite{Seiberg_1994, Seiberg_1994_II} the authors explicitly break $\mathcal{N}=2$ SUSY down to $\mathcal{N}=1$ by adding a bare mass to the adjoint chiral multiplet, \begin{align} \label{eq: SW deformation superpotential UV}
\Delta W = \mu \Tr(\Phi^2).
\end{align}
By holomorphy, this term will appear in the effective Lagrangian for the massless dyons near a singularity~\cite{Seiberg_1994}. In terms of the local degrees of freedom, it has the form \begin{align} \label{eq: N=1 preserving mass deformation superpotential}
\Delta W_{eff}=\mu U(A_D),
\end{align}
where $U(A_D)$ is the chiral superfield whose lowest component is $u$. It can be written as a power series in $A_D$, of the form \begin{align}
U(A_D) = c_0\Lambda_{N_F}^2+c_1\Lambda_{N_F}A_D+c_2 A_D^2 + ... \end{align}

Eq.~\eqref{eq: N=1 preserving mass deformation superpotential} gives rise to the following scalar potential:
\begin{align} \label{eq: N=1 preserving scalar potential}
V = |a_D|^2(|m_n|^2+|\widetilde{m}_n|^2)+ \frac{g_{eff}^2}{2}\left|\sqrt{2}m_n\widetilde{m}^n+\mu\frac{du}{da_D}\right|^2 + \frac{g_{eff}^2}{2}(|m_n|^2-|\widetilde{m}_n|^2)^2 
\end{align}
Minimizing this potential, one finds that a monopole condensate forms. The vacuum field configuration is given by 
\begin{align}\label{eq:N=1 VEV configuration, prelims}
a_D &= 0 \nonumber \\
m_n^\dagger m^n &=\widetilde{m}_n^\dagger\widetilde{m}^n\nonumber \\
\widetilde{m}_nm^n & = -\frac{\mu}{\sqrt{2}}\frac{du}{da_D}|_{a_D=0}.
\end{align}
The condensate gives the theory a mass gap---in particular, the local dual photon acquires a mass. This dual Meissner effect implies confinement of electric charge. A nonrenormalization argument suggests that the condensate given in the last line of Eq.~\eqref{eq:N=1 VEV configuration, prelims} persists as $\mu$ becomes arbitrarily large~\cite{Seiberg_1994}.

At the theories with multiple massless dyons (see Table~\ref{tab:list of singularities vs NF}), there is a space of vacua satisfying Eq.~\eqref{eq:N=1 VEV configuration, prelims}. Up to symmetries, it is given by the following configuration:
 \begin{align} \label{eq: deformed higgs vev}
\vec{m} & = (C, 0,...) \nonumber \\
\vec{\widetilde{m}} & = (-\frac{D}{C}, B, ...),
\end{align}
with $\left|\frac{D}{C}\right|^2 + |B|^2 =  |C|^2$, and $D=\frac{\mu}{\sqrt{2}}\frac{du}{da_D}|_{a_D=0}$. This is a deformation of the Higgs branch. In~\cite{Seiberg_1994_II}, the authors introduce superpotentials desribing the deformation of the Higgs branch when the SUSY-breaking parameter $\mu$ is large compared to the confinement scale. They demonstrate that in the $\mu\to\infty$ limit, one recovers the moduli space of the equivalent $\mathcal{N}=1$ theory. We list these superpotentials in Appendix~\ref{sec: large SW-deformation Higgs branch superpotentials}

We will frequently compare the theories with this $\mathcal{N}=1$-preserving deformation to the theories with anomaly mediation. As a shorthand, we will refer to theories with an $\mathcal{N}=1$-preserving mass as SW-deformed theories.

\section{$\mathcal{N}=2$ SYM with AMSB} \label{sec:AMSB main body}

In this section, we apply AMSB to $\mathcal{N}=2$ SYM with gauge group $SU(2)$. Most of the derivations here generalize straightforwardly to $N_F\neq0$, which we will discuss in Section~\ref{sec: SQCD with AMSB}.

In Section~\ref{sec:UVAMSB}, we introduce AMSB to the fundamental Lagrangian. At the perturbative level, AMSB simply contributes an $\mathcal{N}=1$-preserving mass term for the adjoint multiplet $\Phi$, up to an $SU(2)_R$ rotation that swaps the gaugino with the fermionic component of $\Phi$. On its face, this would suggest that in this theory with AMSB is identical to the SW deformation, the $\mathcal{N}=1$-preserving derivation considered in the original paper~\cite{Seiberg_1994} and described in Section~\ref{sec:N=1-preserving deformation in SW}. 
This turns out not to be the case. Even in the UV, instanton corrections to the $\beta$ function produce a mass splitting among the components of $\Phi$. Supersymmetry is broken completely, although in the far UV where the coupling is small, the mass splitting is negligible. 

In Section~\ref{sec:coulomb branch AMSB}, we apply AMSB to the Coulomb branch of the theory. We find that AMSB lifts the moduli space and the theory flows to the singularities where monopoles and dyons become massless. There, we must work with the effective dual SQED described in Section~\ref{sec: review on near-singularity EFT}. In Section~\ref{sec:singularityEFT}, we introduce AMSB to this effective theory and solve for the vacuum exactly. The spectrum and dynamics in the theory with AMSB differ from the $\mathcal{N}=1$ case, with the AMSB theory breaking all supersymmetry. However, the vacua themselves are quite similar to those obtained in the SW-deformed theory. The Coulomb branch moduli are stabilized at the singularity, and the light dyons condense, with condensate values that are consistent with the perturbative UV description.

\subsection{AMSB in the UV} \label{sec:UVAMSB}

In the absence of mass terms, the Lagrangian is dimensionless and so the tree-level AMSB terms of Eq.~\eqref{eq:treeB} vanish. We must consider loop corrections. To apply the formulae of Section~\ref{sec:AMSB prelims}, we must work in the canonical normalization where $V, \Phi \to gV, g\Phi$. As discussed in Section~\ref{sec: prelim N=2 lagrangian}, changing our normalization from holomorphic to canonical does not alter the $\beta$ function, but it does introduce an anomalous dimension for $\Phi$, given in Eq.~\eqref{eq: anomalous dimension for phi}. 

In terms of the $\beta$ function, the loop terms from Eq.~\eqref{eq:loopAMSB} are \begin{align} \label{eq:UV AMSB SYM}
m_\Phi^2&  =  -b_0g\beta(g)|m_{3/2}|^2, \nonumber\\
& = \frac{\beta(g)^2 |m_{3/2}|^2}{g^2}, 
\nonumber\\
m_\lambda & = -\frac{\beta(g)}{g} m_{3/2}. 
\end{align}
The second line of this equation follows from the generic form of the 1-loop $\beta$ function~\cite{Seiberg:1988ur} \begin{align} \label{eq: generic perturbative beta}
\beta = -b_0g^3.
\end{align}

We now demonstrate that up to an $SU(2)_R$ rotation, these masses are those one would expect from an $\mathcal{N}=1$-preserving mass term for $\Phi$, the SW deformation described in Section~\ref{sec:N=1-preserving deformation in SW}.
Let $M=  -\frac{\beta(g)m_{3/2}}{g}$. If we perform the swap operation defined in equation~\eqref{eq: swap operator}, we exchange the gaugino and the fermion in the adjoint chiral multiplet. Then $\lambda$ lies within $\Phi$ and picks up a negative sign. Let us then add an $\mathcal{N}=2$ breaking mass term to the Lagrangian: \begin{align}\label{eq:amsb superpotential in canonical normalization}
W =  M \Tr{ \Phi^2 } \nonumber  \\
= \frac{M}{2}\Phi_a\Phi^a ,
\end{align}
where $a$ is a colour index. Expanding in Grassman components and integrating over $\mathcal{N}=1$ superspace, we get the mass terms
\begin{align} \label{eq: mass term in superpotential expansion}
\mathcal{L}_{\text{Mass}}= -|M\phi_a|^2 +\frac{M}{2}\lambda_a^2.
\end{align}
Since $|m_{\phi}|^2=|m_\lambda|^2$ in Eq.~\eqref{eq:UV AMSB SYM}, the superpotential Eq.~\eqref{eq:amsb superpotential in canonical normalization} reproduces the AMSB masses.

As a cross check, we can verify that this superpotential satisfies the nonrenormalization theorem. We compute \begin{align}
\frac{dW}{dt} & =-\frac{d}{dt}\left(\frac{\beta(g)m_{3/2}}{g}Z_\Phi \Tr(\Phi^2)\right)\nonumber \\ &
= m_{3/2}\Tr(\Phi^2)\left(\frac{\beta^2}{g^2}Z_\Phi -\frac{Z_\Phi}{g}\frac{d\beta}{dt}-\frac{\beta}{g}(2\sqrt{Z}_\Phi) \frac{d\sqrt{Z_\Phi}}{dt}\right) ,
\end{align}
then subbing in the relation Eq.~\eqref{eq: diff eq for Z} and the 1-loop $\beta$-function Eq.~\eqref{eq: 1-loop beta function}, we get \begin{align}
&\frac{\beta m_{3/2}\Tr(\Phi^2)Z_\Phi}{g}\left(\frac{\beta}{g} -\frac{1}{\beta}\frac{d\beta}{dt}-\frac{2}{\sqrt{Z_\Phi}} \frac{d\sqrt{Z_\Phi}}{dt}\right) \nonumber \\&
= \frac{\beta m_{3/2} \Tr(\Phi^2)Z_\Phi}{g}\left(\frac{\beta}{g}-\frac{1}{\beta}\frac{d\beta}{dt}+\frac{2\beta}{g}\right) \nonumber \\&
=-\frac{\beta m_{3/2} \Tr(\Phi^2)Z_\Phi}{g}\left(-b_0g^2+3b_0g^2-2b_0g^2\right) \nonumber \\&
=0.
\end{align}

The superpotential in Eq.~\eqref{eq:amsb superpotential in canonical normalization} is canonically normalized. whereas the usual treatment of the $\mathcal{N}=1$-preserving superpotential Eq.~\eqref{eq: N=1 preserving mass deformation superpotential} considered in the SW-deformed theory is typically treated in holomorphic normalization, and adapting it to canonical normalization would be a rather technical endeavor (see Appendix~\ref{sec:IR rescaling}). Later in our analysis, it will be useful to have an expression for the anomaly mediated result in holomorphic normalization so that we can compare AMSB to the SW deformation in a direct, one-to-one manner. We conclude this subsection with a derivation of the holomorphically normalized superpotential that gives rise to Eq.~\eqref{eq:amsb superpotential in canonical normalization} when the fields are canonically normalized.

We can introduce a holomorphically normalized mass term $\Delta W =Cm_{3/2}\Tr(\Phi^2)$ by hand, and solve for $C$ by canonically normalizing and matching onto Eq.~\eqref{eq:amsb superpotential in canonical normalization}. Canonically normalizing the theory by transforming $\Phi\to g\Phi$ gives $\Phi$ an anomalous dimension $\gamma$, as in Section~\ref{sec:N=2 renormalization}, so that the overall superpotential does not run. Under this rescaling, the superpotential becomes \begin{align} 
 W = Cm_{3/2} Z_{\Phi}g^2 \Tr(\Phi^2),
\end{align}
while the K{\"a}hler term for $\Phi$ is $Z_{\Phi}\Tr(\Phi^\dagger \Phi)$ (neglecting the gauge field). If we are performing a perturbative calculation and we want a normalized propagator for $\Phi$, we must divide this K{\"a}hler term by the field strength renormalization. Then the running of the theory is captured in the running of the parameters in the superpotential,  and our superpotential becomes \begin{align}\label{eq: AMSB superpotential, not quite done}
 W=Cm_{3/2}g^2 \Tr(\Phi^2)
\end{align}
Plugging Eq.~\eqref{eq: 1-loop beta function} into Eq.~\eqref{eq:amsb superpotential in canonical normalization}, we get
 \begin{align} \label{eq: AMSB N=1 mass}
M = b_0g^2 m_{3/2}.
\end{align}
From Eq.~\eqref{eq: AMSB superpotential, not quite done}, we can identify the constant $C$ with the quantity $b_0$. We thus conclude that in holomorphic normalization, the effect of AMSB is to add a superpotential \begin{align} \label{eq: AMSB perturbative superpotential}
 W = b_0m_{3/2}\Tr(\Phi^2).
\end{align}

For gauge group $SU(2)$ with $N_F=0$, the constant $b_0$ is equal to $\frac{1}{4\pi^2}$. 
If we perform an $SU(2)_R$ swap and identify $\mu = b_0m_{3/2}$, the perturbative AMSB superpotential is identical to the SW deformation.
The emergence of an $\mathcal{N}=1$-preserving mass depends only on the fact that the $\beta$ function takes the form of Eq.~\eqref{eq: generic perturbative beta}. 

This result came as a surprise to us. AMSB typically breaks all supersymmetry and the partial preservation of SUSY here arises in a rather slick manner. More surprisingly, we will find that in the IR description with small $m_{3/2}$, the theory is \textit{not} quite the same as the SW-deformed theory. We will find in Section~\ref{sec:singularityEFT} that while the two theories are similar in some respects, they are not equivalent. In particular, the effective low-energy dynamics do \textit{not} preserve $\mathcal{N}=1$ SUSY, in stark contrast to the situation in the UV. 
As bizarre as this behaviour is, there is a very good reason to expect that complete SUSY breaking occurs in the IR. We discuss it in the next subsection, Section~\ref{sec:instantons}.

\subsection{SUSY-breaking from instantons}\label{sec:instantons}

In the previous subsection, we found that anomaly mediation in perturbation theory merely adds an $\mathcal{N}=1$-preserving mass to the adjoint chiral multiplet $\Phi$, up to an $SU(2)_R$ rotation. But perturbation theory is not the whole story. Even in the UV, we can already see how nonperturbative effects break the surviving $\mathcal{N}=1$ SUSY. 

In $\mathcal{N}=2$ theories, it is possible to extend the $\beta$ function to incorporate instanton corrections~\cite{Seiberg:1988ur}. These instantons also induce a $\beta$ function for the theta angle. These $\beta$ functions are given by
\begin{align}
\beta_g &= Ag^3 + B\cos\theta g^3e^{-\frac{8\pi^2}{g^2}} + O\left(e^{-\frac{16\pi^2}{g^2}}\right) \\
\beta_\theta &= C\sin\theta e^{-\frac{8\pi^2}{g^2}}+O\left(e^{-\frac{16\pi^2}{g^2}}\right).
\end{align}
where the exponentially small terms are instanton corrections. These are thus heavily suppressed in the UV and negligible in perturbative calculations. Regardless, let us recalculate the loop masses from anomaly mediation with the leading-order instanton correction incorporated.

For simplicity, let us set the $\theta$-angle to 0. Then from Eq.~\eqref{eq: anomalous dimension for phi}, we have \begin{align}
\dot\gamma_\Phi & = \frac{2}{g^2}\beta^2 -\frac{2}{g}\frac{d\beta}{d\log\mu}\nonumber \\
& = -g^4\left(4A^2 + 8ABe^{-\frac{8\pi^2}{g^2}}\right) -32\pi^2AB g^2e^{-\frac{-8\pi^2}{g^2}}
\end{align}
The resulting AMSB masses are
\begin{align}
m_\lambda & =  -m_{3/2}g^2\left[A + Be^{-\frac{8\pi^2}{g^2}}\right] \nonumber \\
m_\phi^2  & =  |m_{3/2}|^2\left[g^4\left(A^2+2ABe^{-\frac{8\pi^2}{g^2}}\right)+8g^2AB\pi^2 e^{-\frac{8\pi^2}{g^2}}\right]
\end{align}
With $B=0$, we get that $m_{\phi}^2 = |m_\lambda|^2$ as needed to preserve SUSY. But with nonzero $B$, the $g^2$ term in the scalar mass breaks the relationship between the masses, hence breaking SUSY down to $\mathcal{N}=0$. Where perturbation theory is reliable, this mass splitting is negligible. But as we flow to the IR, it becomes larger and larger. As we discussed in Section~\ref{sec: prelims N=2 main}, the effective IR description of the theory deviates strongly from the perturbative description which is valid when $a\gg\Lambda_0$. The strong coupling effects in $\mathcal{N}=2$ theories arise from instantons~\cite{Lerche_1997}. We should therefore anticipate that the differences between the anomaly mediated and SW-deformed theories will be very pronounced in the IR, in contrast to the UV.

\subsection{AMSB on the Coulomb branch}\label{sec:coulomb branch AMSB}

We now apply anomaly mediation to the effective theory on the Coulomb branch described in Section~\ref{sec:Coulomb review}. Since AMSB contributes a mass term for $\phi$ in the perturbative regime, one might expect that the Coulomb branch is lifted and the moduli roll to the singularities. We will see that this is indeed the case. 

We will perform this derivation at tree level. In the UV, there was no tree-level AMSB potential, but on the Coulomb branch, the tree-level AMSB potential is nonvanishing. This is because the exact Coulomb branch metric explicitly encodes quantum effects that are absent at tree level in the UV Lagrangian. The effective action on the Coulomb branch is a Wilsonian action and so it does still run nontrivially, but it is safe to ignore this running --- it cannot affect our conclusions. We will explain why this is the case at the end of the section.

The Coulomb branch has no superpotential, so the supersymmetric action consists only of a K{\"a}hler potential and a gauge kinetic term. The tree level AMSB potential is then given by the $|m_{3/2}|^2$ term in equation~\eqref{eq:treeB}, that is, \begin{align} \label{eq:VAMSB when W=0}
V_{\text{AMSB}} = |m_{3/2}|^2(\partial K\bar\partial K h^{-1} -K),
\end{align}
where $h$ is the (single component) K{\"a}hler metric and the partials are with respect to whatever operator is taken to be an elementary field. On the Coulomb branch, we assume that this potential is the dominant AMSB contribution, neglecting loop-level terms.\footnote{In Appendix~\ref{sec:IR rescaling}, we do consider loop corrections to the near singularity EFT. They merely reinforce the conclusion that the moduli are lifted leaving the singularities as minima.} 
The K{\"a}hler potential on the Coulomb branch is given by \begin{align} \label{eq: coulomb branch Kahler potential}
K = \frac{i}{8\pi}\left[(a_D(u))^\dagger a(u)-a_D(u)(a(u))^\dagger\right].
\end{align}
The definitions of $a$, $a_D$, and $u$ are reviewed in Section~\ref{sec:Coulomb review}.

The literature around $\mathcal{N}=2$ theories tends to be concerned with the near singularity EFTs, in which some linear combination of $a_D$ and $a$ is treated as an elementary field. But at a generic point on the moduli space, the global coordinate $u$ is a more natural choice of elementary field, since $K$ is analytic everywhere in $u$ except at the singularities. At a generic point $u_0$, $K$ can be Taylor expanded in $u-u_0$, and the Taylor expansion is convergent. The dynamics of the fluctuations $u-u_0$ are then well approximated by a finite order polynomial Lagrangian, with the coupling strengths set by the coefficients of the Taylor expansion. So, the SUSY-breaking potential can be written 
\begin{align} \label{eq:CoulombAMSB}
V_{\text{AMSB}}(u, \bar{u}) = |m_{3/2}|^2\left[\frac{1}{8\pi i}\frac{(\partial a_Da^\dagger- \bar{a_D} \partial{a})(\bar\partial a^\dagger a_D - a\bar{\partial}a_D^\dagger)} {(\bar\partial a_D^\dagger \partial a - \partial a_D\bar\partial a^\dagger)}+\frac{1}{8\pi i}(a^\dagger a_D - aa_D^\dagger)\right]
\end{align}
where the partials $\partial, \bar{\partial}$ are with respect to $u, \bar{u}$. The vacua of the theory lie at the minima of this potential. The functions $a(u), a_D(u)$ can be written in terms of hypergeometric functions and are listed in \cite{csáki2025phasetransitionsunusualvalues} and Appendix~\ref{sec:Hypergeo}. 

One may worry that $u$ is canonically a dimension 2 parameter. This is not a problem. Suppose we define $w\equiv \frac{u}{\Lambda_0}$ and $ \widetilde{K}(w(u)) = K(u)$. With respect to these variables, the AMSB potential Eq.~\eqref{eq:VAMSB when W=0} in  \begin{align}
V_{\text{AMSB}} &=|m_{3/2}|^2\left(\left|\frac{\partial \widetilde{K}}{\partial w}\right|^2\left(\frac{\partial^2\widetilde{K}}{\partial w\partial w^\dagger}\right)^{-1} -\widetilde{K}\right) \nonumber \\
&=  |m_{3/2}|^2\left(\left|\frac{\partial \widetilde{K}}{\partial u}\right|^2\left|\frac{du}{dw}\right|^2\left(\frac{\partial^2\widetilde{K}}{\partial u\partial u^\dagger}\right)^{-1} \left|\frac{du}{dw}\right|^{-2}-\widetilde{K}\right) \nonumber \\
&= |m_{3/2}|^2\left(\left|\frac{\partial K}{\partial u}\right|^2\left(\frac{\partial^2K}{\partial u\partial u^\dagger}\right)^{-1} -K\right)
\end{align}
where the last line simply follows from the fact that $\widetilde{K}(w(u))=K(u)$. We therefore proceed to work with $u$ itself.

\begin{figure}[h]
    \centering
    \includegraphics[width=0.5\linewidth]{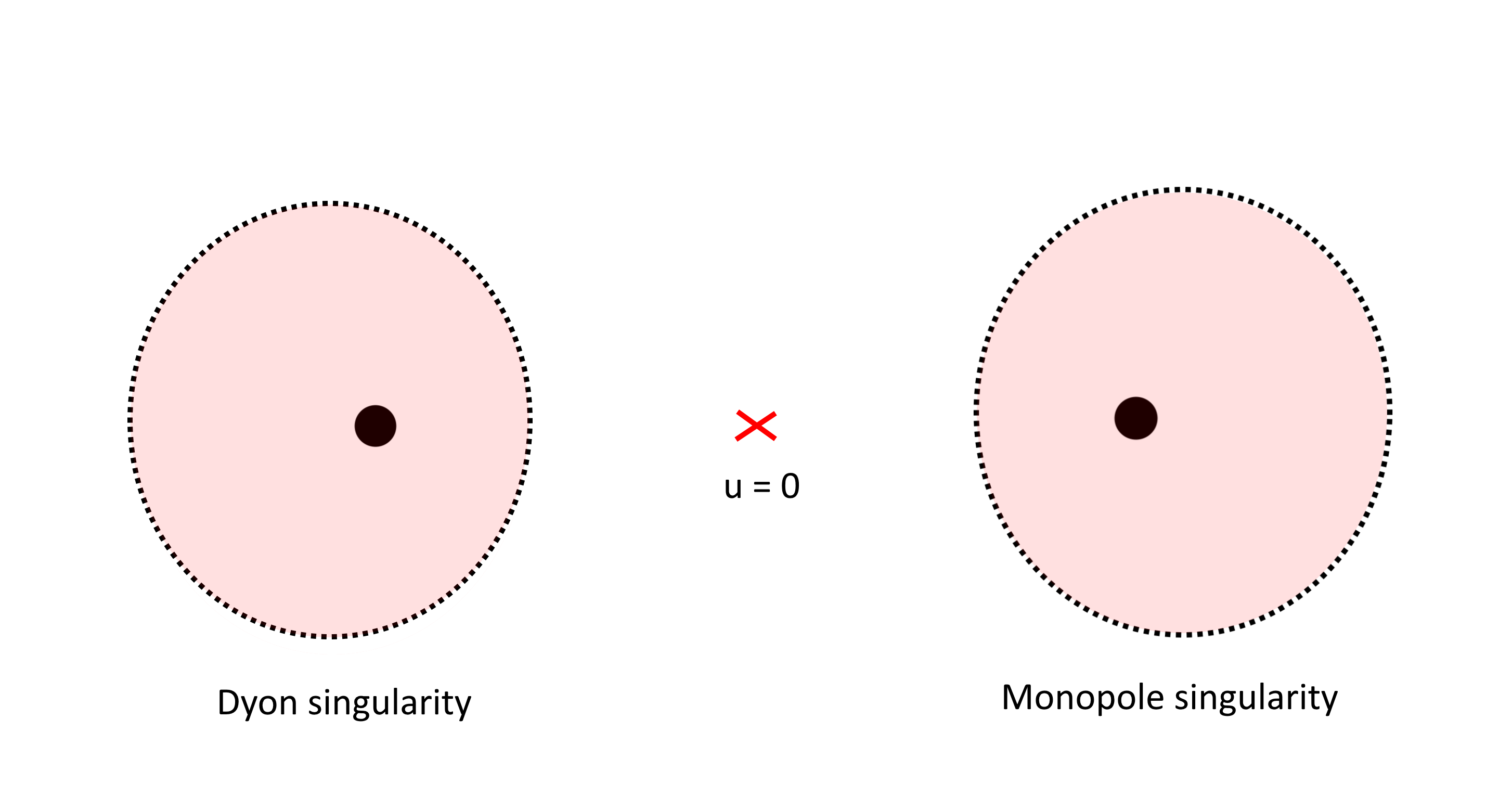}
    \caption{A sketch of the Coulomb branch. The black dots are the singularities where monopoles or $(1, 1)$-dyons become massless. The dashed ellipses represent curves along which the minimum BPS mass is constant and equal to $m_{3/2}$. This mass acts as a cutoff for the family of EFTs obtained by Taylor expanding the action about a point lying on the curve. Within the red shaded region, Eq.~\eqref{eq:CoulombAMSB} cannot be trusted and the light dyons must be incorporated in the calculation. The area of this region increases when we increase $m_{3/2}$. The origin of the $u$-plane, marked with a red x, will turn out to be a saddle point of the AMSB potential. When $m_{3/2}$ becomes larger than the cutoff at the origin, the two regions of incalculability will merge into one large blob enclosing both singularities.}
    \label{fig:coulomb with equicutoffs}
\end{figure}

How large can $m_{3/2}$ become before this description of the theory breaks down? The functions $a$ and $a_D$ set the mass scale of the lightest degrees of freedom that have been integrated out. At the singularities, some of these degrees of freedom become massless and must be incorporated into the Lagrangian for the theory to make sense. We can imagine cutting out closed curves around the singularities along which the minimum dyon mass is constant. In the region outside of the curves, the cutoff of the theory is always greater than this minimum mass scale. Equation~\eqref{eq:CoulombAMSB} is valid in this region as long as $m_{3/2}$ is no greater than the aforementioned minimum dyon mass. Figure~\ref{fig:coulomb with equicutoffs} presents a sketch of what this might look like. We can get arbitrarily close to the singularities as long as we are willing to entertain arbitrarily small values of $m_{3/2}$. But as $m_{3/2}$ grows, so too does the region around the singularity within which equation~\eqref{eq:CoulombAMSB} cannot be trusted.

Based on the results in the Section~\ref{sec:UVAMSB}, one might expect that the Coulomb branch is lifted by AMSB and the modulus rolls to the singularities. This would mean that the potential, Eq.~\eqref{eq:CoulombAMSB}, has no local minima on its domain of analyticity. 
In Appendix~\ref{sec:convexK}, we prove that this is indeed the case if $K$ is a convex function.\footnote{Here `convex' means real convex, not plurisubharmonic. Real convexity is not preserved by K{\"a}hler transformations, so at first glance this claim seems vacuous. But the anomaly mediation formulae in Section~\ref{sec:prelim} are not K{\"a}hler invariant either, so there is a preferred K{\"a}hler gauge. This is the one in which convexity is relevant. We expand on this point in Appendix~\ref{sec:convexK}.} Ref.~\cite{D_Hoker_2022} presents evidence that $K$ is a convex function for all $SU(N)$ gauge groups, suggesting that indeed, Eq.~\eqref{eq:CoulombAMSB} is minimized at the singularities.

\begin{figure}[h]
    \centering
    \includegraphics{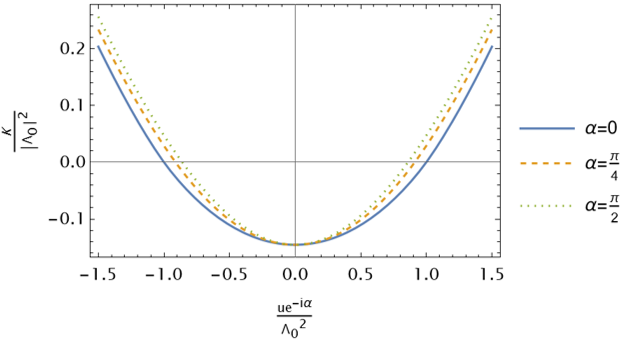}
    \caption{Representative slices of the nondimensionalized K{\"a}hler potential on the Coulomb branch in the $N_F=0$ theory. Rephasing $\Lambda_0$ simply rotates $K$ in the complex $u$-plane, while rescaling $|\Lambda_0|$ dilates the $u$-plane and rescales $K$ by a constant factor. Under these transformations, the qualitative shape of the function remains the same. In particular, it remains convex. Each curve shows the function evaluated on a straight line through the origin of the $u$-plane, on an angle given by $\alpha$. Changing $\alpha \to \alpha+\pi$ simply takes the plot to its mirror image, and the $\mathbb{Z}_2$ symmetry on the $u$-plane means that curves of angle $\alpha=\frac{\pi}{2}+\delta\alpha$ are equivalent to curves of angle $\alpha=\frac{\pi}{2}-\delta\alpha$.}
    \label{fig:Kahler potential on Coulomb branch for NF=0}
\end{figure}

\begin{figure} [h]
    \centering
    \includegraphics{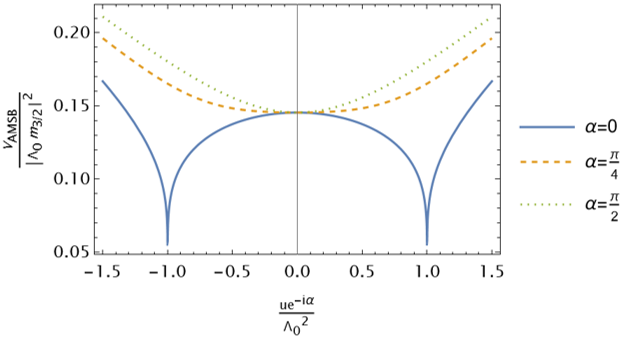}
    \caption{This plot shows representative slices of the nondimensionalized AMSB potential on the $N_F=0$ Coulomb branch, along the same angles as in Fig.~\ref{fig:Kahler potential on Coulomb branch for NF=0}. On angles $\alpha\neq 0$, it appears as though there is a local minimum at the origin, but this is not the case. Rather, these stationary points are saddle points of $V_{\text{AMSB}}$, since along the real axis $\alpha=0$, the origin is a local maximum. The origin is the global minimum of $K$, and the existence of saddle points in $V_{\text{AMSB}}$ at extrema of $K$ is a prediction of our analysis in Appendix~\ref{sec:convexK}.}
    \label{fig:VAMSB on Coulomb branch for NF=0}
\end{figure}

Figs.~\ref{fig:Kahler potential on Coulomb branch for NF=0} and~\ref{fig:VAMSB on Coulomb branch for NF=0} show representative slices of the nondimensionalized K{\"a}hler and AMSB potentials, and they show the expected behaviour. The K{\"a}hler potential is convex everywhere, and the AMSB potential's only local minima lie at the singularities.

The functions are nondimensionalized as follows: from Appendix~\ref{sec:Hypergeo}, the K{\"a}hler potential depends only on the combination $u/\Lambda_0^2$ and its conjugate, up to an overall factor of $|\Lambda_0|^2$. The function $\frac{K(u/\Lambda_0^2)}{|\Lambda_0|^2}$ is thus $\Lambda_0$ independent. We can then consider the nondimensionalized AMSB potential 
\begin{align} \label{eq:coulomb branch AMSB, nondimensionalized}
\left(\left|\frac{\partial (K/|\Lambda_0|^2)}{\partial (\Lambda_0^2u)}\right|^2\left(\frac{\partial^2(K/|\Lambda_0|^2)}{\partial(\Lambda_0^2 u)\partial(\Lambda_0^{\dagger 2} u^\dagger)}\right)^{-1} -\frac{K}{|\Lambda_0|^2}\right) = \frac{V_{\text{AMSB}}}{|m_{3/2}\Lambda_0|^2}.
\end{align}
We see that changes in $m_{3/2}$ and $\Lambda_{N_F}$ scale and rotate the potentials in the $u$-plane, but they do not change the convexity of $K$ or the fact that $V_{\text{AMSB}
}$ is minimized exclusively at the singularities. The conclusions we have drawn from Figs.~\ref{fig:Kahler potential on Coulomb branch for NF=0} and~\ref{fig:VAMSB on Coulomb branch for NF=0} are independent of $m_{3/2}$ and $\Lambda_0$.


At the singularities, the Wilsonian cutoff of the effective action goes to 0 and the potential Eq.~\eqref{eq:CoulombAMSB} cannot be trusted. To proceed with our analysis of the vacua, we must work with the weakly coupled effective actions, reviewed in Section~\ref{sec: review on near-singularity EFT}, which incorporate the massless dyons.

Before we move forward with this analysis, we comment on loop-level contributions to $V_{\text{AMSB}}$ on the Coulomb branch. The Coulomb branch is described by a pure supersymmetric sigma model. Taylor expanding about any particular $\mathcal{N}=2$ vacuum $U=p$, the resulting interaction terms are all nonrenormalizable, and nonrenormalizable couplings do not contribute to the running of marginal couplings~\cite{Weinberg:1973xwm}. Therefore, the loop level AMSB terms in Eq.~\eqref{eq:loopAMSB} do not arise on the Coulomb branch. The nonrenormalizable couplings can run, however. This running affects the anomaly mediated theory, but it does not change the vacuum, for reasons we will now explain.

Within the radius of convergence of the Taylor expansion about $U=p$, we have some effective nonrenormalizable K{\"a}hler potential $K_{\text{eff}}(p)$. Introducing loop corrections, the subleading coefficients in the series $K_{\text{eff}}(p)$ will run with the energy scale $\sigma$, so that $K_{\text{eff}}(p, \sigma=0) = K_{\text{eff}}(p)\neq K_{\text{eff}}(p, \sigma \neq 0)$. Plugging this effective Kahler potential into Eq.~\eqref{eq:VAMSB when W=0}, the resulting SUSY-breaking terms would also run according to the running of $K_{\text{eff}}(p, \sigma)$. But in vacuum, $\sigma=0$ and these AMSB terms  simply reduce to the Taylor expansion of the global potential Eq~\eqref{eq: coulomb branch Kahler potential}. 

One may wonder if additional terms arise by the substitution $\sigma\to \sigma/\sqrt{\chi^\dagger \chi}$, in analogy to the perturbative UV theory where the scalar mass, for example, arises from the addition of an $F$-term to the field strength normalization $Z$ via the substitution $Z(\sigma)\to Z(\sigma/\sqrt{\chi^\dagger \chi})$. This does not happen on the Coulomb branch, however. We can see this by dimensional analysis. The K{\"a}hler potential is dimension 2. The expansion of the K{\"a}hler potential takes the generic form \begin{align}
    K_{\text{eff}}\sim K|_p + \partial K|_p\delta U + \bar{\partial} K|_p \delta U^\dagger +\partial \bar{\partial} K|_p \delta U^\dagger \delta U + \partial \partial K|_p \delta U\delta U\nonumber \\+\bar{\partial} \bar{\partial} K|_p\delta U^\dagger\delta U^\dagger +\partial^2\bar{\partial}K|_p\delta U^\dagger \delta U^2 + ... ,
\end{align} with $\delta U$  the perturbation of $U$ about the expansion point. The 0th and 1st order terms vanish under Grassmann integration, but they do affect the AMSB potential. The coefficient $\bar{\partial}^n
\partial^m K|_p$ of $\delta U^{\dagger n} \delta U^m$ must have dimension $2-n-m$, and the dimensionful coefficients available are $\Lambda_0$, $p$, and their conjugates. The coefficient will therefore generically take the form \begin{align}  \label{eq: dimensional analysis of nth order coefficient in local Kahler potential expansion}
\bar{\partial}^n
\partial^m K|_p \sim \sum_{k, q}c^{kq}\left(\frac{p}{\Lambda_0}\right)\Lambda_0^kp^q 
\end{align}
where $c^{k, q}$ is a dimensionless function of the parameters, the sum is over all $k, q$ satisfying $k+q=2-n-m$, and for the sake of notational concision we are not bothering to distinguish between $(p, \Lambda_0)$ and $(p^\dagger, \Lambda_0^\dagger)$, both of which can appear in Eq.~\eqref{eq: dimensional analysis of nth order coefficient in local Kahler potential expansion}.

When we renormalize a particular $n$-point interaction, the counterterms will generically generate higher-dimensional interactions. But the series prior to renormalization already includes all terms consistent with the symmetries of the theory, and so a given higher-dimensional counterterm will simply contribute to the running of an already-existing term in the series. Overall then, the effect of renormalization is to modify the dimensionless coefficients $c^{kq}\left(\frac{\Lambda_0}{p}\right)\to c^{kq}\left(\frac{\Lambda_0}{p}, \frac{\sigma}{\Lambda_0}, \frac{\sigma}{p}\right)$. Since these renormalized coefficients are dimensionless, spuriously restoring superconformal invariance introduces no new powers of the conformal compensator, other than those already appearing at tree level.  Therefore, loop level AMSB on the Coulomb branch is completely captured by simply plugging $K(p, \sigma)$ into Eq.~\eqref{eq:VAMSB when W=0}. This can alter the dynamics at finite energy, but it cannot alter the value of $V_{\text{AMSB}}$ in vacuum.  

We now proceed to Section~\ref{sec:singularityEFT}, where we will couple AMSB to the EFTs at the singularities.


\subsection{AMSB at the singularities}  \label{sec:singularityEFT}

We now apply AMSB to the effective theories at the singularities on the moduli space, where dyons become massless. These EFTs are reviewed in Section~\ref{sec: review on near-singularity EFT}. We will show that the vacua are qualitatively very similar to those in the SW-deformed theory, in that the Coulomb branch moduli are fixed at the singularities and the massless dyons condense. However, the spectrum and the dynamics are \textit{not} the same as the SW-deformed theory. The anomaly mediated theory breaks all SUSY when applied to the low-energy theory, in contrast to the perturbative UV results of Section~\ref{sec:UVAMSB}.

Here, we work in holomorphic normalization and at tree level. In Appendix~\ref{sec:IR rescaling}, we change to canonical normalization and calculate loop corrections, verifying that they do not alter the conclusions of this section. 

In holomorphic normalization, the supersymmetric scalar potential of the $\mathcal{N}=2$ theory near the singularity is given by \begin{align}
     V_{\textit{SUSY}}  =|a_D|^2(|m|^2+|\widetilde{m}|^2)+ g_{eff}^2|m\widetilde{m}|^2 + \frac{g_{eff}^2}{2}(|m|^2-|\widetilde{m}|^2)^2.
\end{align}
The first term lacks a factor of $g_{eff}$ because it comes from the Yukawa term in the superpotential. Prior to turning on AMSB, the potential is minimized by simply setting $m, \widetilde{m}=0$. Then $a_D$ can take on any value. This is the guise through which the Coulomb branch appears in this description of the theory.

Applying Eq.~\eqref{eq:treeB} to the effective Lagrangian described in Section~\ref{sec: review on near-singularity EFT}, we arrive at the tree-level AMSB potential:
\begin{align} \label{eq: general form for IR AMSB potential}
  V_{\text{AMSB}}  =|a_D|^2(|m|^2+|\widetilde{m}|^2)+ \frac{g_{eff}^2}{2}|\sqrt{2}m\widetilde{m}+m_{3/2}^\dagger\partial K|^2+\frac{g_{eff}^2}{2}(|m|^2-|\widetilde{m}|^2)^2\nonumber\\-|m_{3/2}|^2K -\sqrt{2}(m_{3/2}a_Dm \widetilde{m}+h.c.).
\end{align}
where we reiterate that the $K$ here refers to the dyon independent part of the broader K{\"a}hler potential. The partials $\partial, \bar{\partial}$ are with respect to $a_D, a_D^\dagger$.
Unlike on the Coulomb branch, the theory here contains renormalizable couplings and so loop level AMSB contributions may arise. 

In the appropriate limit, this potential closely resembles that of the SW-deformed theory considered in~\cite{Seiberg_1994}. In that theory we have the scalar potential given in Eq.~\eqref{eq: N=1 preserving scalar potential}, which we restate here for the purpose of comparison:

\begin{align} \label{eq:N=1 potential again (monopole EFT w SW deformation)}
V_{\mathcal{N}=1} = |a_D|^2(|m|^2+|\widetilde{m}|^2)+ \frac{g_{eff}^2}{2}\left|\sqrt{2}m\widetilde{m}+\mu\frac{du}{da_D}\right|^2 + \frac{g_{eff}^2}{2}(|m|^2-|\widetilde{m}|^2)^2.
\end{align} 

If we drop the second line of the AMSB potential Eq.~\eqref{eq: general form for IR AMSB potential}, we are left with
\begin{align}\label{eq: truncated IR potential}
V_{\text{AMSB}}  \sim |a_D|^2(|m|^2+|\widetilde{m}|^2)+ \frac{g_{eff}^2}{2}|\sqrt{2}m\widetilde{m}+m_{3/2}^\dagger\partial K(a_D=0)|^2+\frac{g_{eff}^2}{2}(|m|^2-|\widetilde{m}|^2)^2.
\end{align}
The resemblance to Eq.~\eqref{eq:N=1 potential again (monopole EFT w SW deformation)} is clear. If we assume that $|a_D|\ll |\Lambda|$, as it must be for this description to apply, this truncation should be reasonable. This is because the terms in the second line of Eq.~\eqref{eq: general form for IR AMSB potential} are proportional to $|m_{3/2}^2a_D\Lambda_0|$ and $|m_{3/2}a_D|$, respectively, while the expression $m_{3/2}^\dagger\partial K$ is proportional to $m_{3/2}\Lambda_0$. Of course in this limit, we could also ignore the first term in the potential. Regardless, the point is that the leading order terms in Eq.~\eqref{eq: general form for IR AMSB potential} closely resemble those of Eq.~\eqref{eq:N=1 potential again (monopole EFT w SW deformation)}. The key difference is the replacement of $\partial u|_{a_D=0}$ with $\partial K|_{a_D=0}$. Then in this limit, the monopoles condense with a VEV given by \begin{align} \label{eq: monopole VEV}
    \braket{m\widetilde{m}}=-\frac{m_{3/2}^\dagger\partial K|_{a_D=0}}{\sqrt{2}}
\end{align}
The scalar $a_D$ is pinned to 0, and the $D$-term $\propto (|m|^2-|\widetilde{m}|^2)^2$ imposes that $|m|=|\widetilde{m}|$. 

We can make this heuristic argument much more rigorous. We will momentarily show analytically that this is the correct global minimum to all orders in $a_D$, as long as $m_{3/2}$ is kept below the UV cutoff of the effective monopole theory.  But before we proceed, we comment on a subtlety of the interpretation of Eq.~\eqref{eq: general form for IR AMSB potential}. It may give pause that upon setting $a_D$ to 0, the effective coupling $g_{eff}$ also goes to 0, causing the entire potential to vanish. This is also true for the potential in the SW-deformed theory. Taken at face value, this suggests that the monopoles can take on \textit{any} VEV as long as $a_D=0$.  However, $g_{eff}$ grows extremely rapidly as a function of $a_D/\Lambda_0$, as discussed in Section~\ref{sec: review on near-singularity EFT}. In particular, $\partial g_{eff}\to\infty$ at the origin. Because of this property, there will always be some finite open interval $|a_D|\in (0, |a_D|_{\text{max}})$ within which the numerical magnitude of $g_{eff}$ is far greater than that of $|a_D|$. The precise value of $|a_D|_{\max}$ depends on $\Lambda_0$, but at any finite $\Lambda_0$, such an upper bound exists. The existence of this upper bound means that if we solve the equations of motion for $m, \widetilde{m}$ at some fixed $a_D$ then adiabatically bring $a_D$ down to 0, the terms in Eq.~\eqref{eq: general form for IR AMSB potential} which are proportional to $g_{eff}$ will dominate the equations of motion and determine the monopole VEVs in the $a_D\to 0$ limit. The $\mathcal{N}=2$ term imposes the $D$-constraint. The $\mathcal{N}=2$ breaking term causes the monopoles to condense.

We now proceed with our derivation. Let us first argue that the $D$-term constraint $|m|^2=|\widetilde{m}|^2$ holds independently of the values of $m\widetilde{m}$ and $a_D$. Only the first two terms of Eq.~\eqref{eq: general form for IR AMSB potential} depend separately on $m$ and $\widetilde{m}$, with the rest depending only on the product $m\widetilde{m}$. These first two terms are positive definite. The second term is 0 if and only if $|m|=|\widetilde{m}|$. Let us write $m\widetilde{m}=v$, being agnostic for the moment about the precise value of $v$. Then we can write the first term as \begin{align}
|a_D|^2\left(|m|^2+\frac{|v|^2}{|m|^2}\right).
\end{align}
Minimizing this expression with respect to $|m|$, we find that $|m|^2=|v|$. Since $|m|^2|\widetilde{m}|^2=|v|^2$, we see that $|\widetilde{m}|^2=|m|^2$. Because this is true independently of the values of $v$ and $a_D$, we can justify dropping the $|m|^2-|\widetilde{m}|^2$ term moving forward. We will still write $m$ and $\widetilde{m}$ as separate fields, as the $D$-flatness condition does not constrain their relative phases. 

Let us now show that the vacuum indeed lies at $m\widetilde{m}=-\frac{m_{3/2}^\dagger}{\sqrt{2}}\partial K$, with $a_D=0$. This is a somewhat delicate calculation, because the potential is nonanalytic at $a_D=0$ and behaves nonphysically at large $a_D$. These features make the equations of motion somewhat difficult to work with. Fortunately, it is possible to show algebraically that the potential is nonnegative for all $a_D$ in the radius of convergence of $K$, and that its global minimum lies at $a_D=0$. We will use equations of motion to determine the monopole condensate, but they will be easier to interpret with this algebraic result in hand. 

To this end, we perform a change of variables, \begin{align} m\widetilde{m}\to\frac{1}{\sqrt{2}}\left(\Delta -m_{3/2}^\dagger\partial K\right).
\end{align}
Plugging this into equation~\eqref{eq: general form for IR AMSB potential} and imposing the $D$-term constraint, we obtain an alternate form of the potential that will prove to be convenient. We have \begin{align} \label{eq: V(delta)}
V_{\text{AMSB}} = \sqrt{2}|a_D|^2 |\Delta-m_{3/2}^\dagger\partial K|^2 +g_{eff}^2|\Delta|^2 - |m_{3/2}|^2 K + 2\Re(m_{3/2}a_D(m_{3/2}^\dagger \partial K-\Delta)).
\end{align}
At this point, it is useful to consult the relations in Eq.~\eqref{eq: handy singularity EFT formulae}. We can expand the K{\"a}hler potential in terms of the dual prepotential, $K=\frac{1}{4\pi}\Im(a_D^\dagger \frac{\partial \mathcal{F}_D}{\partial a_D})$. Then with a bit of straightforward algebra, the last term in Eq.~\eqref{eq: V(delta)} can be written \begin{align}
2\Re(|m_{3/2}|^2a_D\partial K-m_{3/2}\Delta) & =2 \Re(a_D \frac{|m_{3/2}|^2}{8\pi i}(-\bar{\partial}\mathcal{F}^\dagger_D+a_D^\dagger\partial^2\mathcal{F}_D)-m_{3/2}a_D\Delta)) \\
& =|m_{3/2}|^2\left(K+|a_D|^2 \frac{Im(\tau)}{4\pi}\right)-2\Re(m_{3/2}a_D\Delta).
\end{align}
Plugging this back into Eq.~\eqref{eq: general form for IR AMSB potential} we have \begin{align}
V_{\text{AMSB}}=\sqrt{2}|a_D|^2 |\Delta-m_{3/2}^\dagger\partial K|^2 +|m_{3/2}|^2\left(|a_D|^2 \frac{Im(\tau)}{4\pi}\right)+g_{eff}^2|\Delta|^2-2\Re(m_{3/2}a_D\Delta).
\end{align}

Most of the terms in this expression are nonnegative, but the last term is unbounded from below. Nevertheless, we can show that despite the presence of this term in the Lagrangian, the potential is nonnegative everywhere on the domain where the near singularity description is applicable. Consider a case in which the phase of $\Delta$ is such that $m_{3/2}a_D\Delta$ is a real number. Then the final two terms are a quadratic in $|\Delta|$, given by \begin{align}
Q(|\Delta|)=g_{eff}^2|\Delta|^2-2|m_{3/2}a_D\Delta|
\end{align}
The minimum of $Q$ lies at $|\Delta|=\frac{2|m_{3/2}a_D|}{g_{eff}^2}$. Evaluated at the vertex, the quadratic takes the value \begin{align}
Q_{min}& =-\frac{2{|m_{3/2}}a_D|^2}{g^2}\\
&=-\frac{|m_{3/2}a_D|^2}{4\pi}\Im{\tau}.
\end{align}
We thus have an inequality, $\frac{|m_{3/2}^2a_D|^2}{4\pi}\Im{\tau}\geq |Q(|\Delta|)| $. But the left hand side of this inequality is none other than the second term in the potential, equation~\eqref{eq: V(delta)}. When the bound is saturated, the potential is \begin{align}
V_{\text{AMSB}}& = |a_D|^2\sqrt{2}|\Delta -m_{3/2}^\dagger \partial K|^2 \nonumber \\
&=\sqrt{2}|a_D|^2\left|\frac{2m_{3/2}^\dagger a_D^\dagger}{g_{eff}^2}-m_{3/2}^\dagger\partial K\right|^2,
\end{align}
which is positive. We derived this bound assuming that $m_{3/2}a_D\Delta$ is real, but it holds in general. To see this, note that under rephasing of this term, the quadratic $Q(\Delta)$ remains real and can simply be written \begin{align}
Q(\Delta) = g_{eff}^2|\Delta|^2-2\alpha|m_{3/2}a_D\Delta|
\end{align}
with $\alpha<1$. Then the upper bound on $|Q|$ becomes $\frac{|m_{3/2}a_D\alpha|^2}{g_{eff}^2}$ and the inequality only becomes more strict.

It follows that the potential of equation~\eqref{eq: general form for IR AMSB potential} is nonnegative everywhere in the domain of validity of the EFT. Since the potential vanishes when we take $a_D=0$, we see that this point is a global minimum.

Let us finally determine whether the monopole VEV in Eq.~\eqref{eq: monopole VEV} is the right one. Once more, we note that sending $a_D\to 0$ sends the potential to 0 regardless of the value of $m\widetilde{m}$. But we can calculate the limiting behaviour of $\braket{m\widetilde{m}}$ as $a_D$ is adiabatically brought to 0 from some finite value. Taking the derivative of the potential Eq.~\eqref{eq: general form for IR AMSB potential} with respect to $a_D$ (and again ignoring the $D$-term), we have \begin{align}
\partial V_{\text{AMSB}}  = & a_D^\dagger (|m|^2+|\widetilde{m}|^2)+\frac{g_{eff}^4}{16\pi^2a_D}|\sqrt{2}m\widetilde{m} + m_{3/2}^\dagger \partial K|^2 \nonumber\\& +\frac{g_{eff}^2}{2}\left[(m_{3/2}^\dagger\partial^2K)(\sqrt{2}m^\dagger\widetilde{m}^\dagger+m_{3/2}\bar\partial K)
+(\sqrt{2}m\widetilde{m}+m_{3/2}^\dagger \partial K)(m_{3/2}\partial\bar{\partial}K) \right]\nonumber \\ &-|m_{3/2}|^2\partial K - \sqrt{2}m_{3/2}m\widetilde{m},\nonumber \\
= & a_D^\dagger (|m|^2+|\widetilde{m}|^2)+\frac{g_{eff}^4}{16\pi^2a_D}|\sqrt{2}m\widetilde{m} + m_{3/2}^\dagger \partial K|^2 +\frac{g_{eff}^2}{2}m_{3/2}^\dagger \partial^2 K (\bar{\partial} K).
\end{align}
As $a_D$ approaches 0, only the term inversely proportional to $a_D$ survives. We have\begin{align}
\partial V_{\text{AMSB}} \to \frac{g_{eff}^4}{a_D}|\sqrt{2}m\widetilde{m} + m_{3/2}^\dagger \partial K|^2.
\end{align}
Setting $\partial V_{\text{AMSB}}$ to 0, we recover Eq.~\eqref{eq: monopole VEV}.

To summarize, we have shown that the vacuum field configuration is at \begin{align}\label{eq: monopole vacuum summary}
m\widetilde{m} &=-\frac{m_{3/2}^\dagger}{\sqrt{2}}\partial K|_{a_D=0}, \nonumber \\
|m|^2&=|\widetilde{m}|^2, \nonumber \\
a_D&=0.
\end{align}
This is true to all orders in the expansion of $K$ in $a_D$, as long as $m_{3/2}$ is below the cutoff of the effective description.

Let us now briefly consider the other singularity in the $N_F=0$ theory. At this singularity, a dyon hypermultiplet of charge $(n_m, n_e)=(1, 1)$ becomes massless, and so the local singlet scalar is $a_D+a$. The effective K{\"a}hler potential at this singularity has the same form as that given in Eq.~\eqref{eq:IR Kahler potential}, but the coefficients are different and $A_D$ is replaced with $A+A_D$. In terms of the weakly coupled degrees of freedom in the dyon's duality frame, the scalar potential again takes the form of Eq.~\eqref{eq: general form for IR AMSB potential}. All of the arguments we have made in this section carry over directly---we find another global minimum at the $(1, 1)$ singularity with a dyon condensate proportional to the derivative of the K{\"a}hler potential with respect to the locally weakly coupled singlet scalar. 

On one final note, we address the possibility that loop level AMSB might alter the vacua that we have found here. This does not occur. In Appendix~\ref{sec:IR rescaling}, we canonically normalize the theory and work out the leading order loop corrections. The corrections are proportional to $g_{eff}$ and so they vanish at the origin. Away from the origin their leading order effect is to add an $\mathcal{N}=1$-preserving mass to $a_D$ and the gaugino, much like in the UV theory. This simply drives the value of $a_D$ back towards the origin, where the loop corrections vanish. The loop corrections do alter the physical spectrum, however. As we show in Appendix~\ref{sec:IR rescaling}, canonical normalization introduces a finite IR cutoff on $a_D$, below which no dynamics can occur. Above this cutoff, loop corrections are finite and nonzero.

\section{Comparing AMSB to the SW deformation for $N_F=0$} \label{sec: Nf=0 theory comparison}
Over the course of Section~\ref{sec:AMSB main body}, we have frequently compared the theory with AMSB to the SW-deformed theory. Since the theories are identical at the level of perturbation theory when $m_{3/2}$ is large, the well understood SW deformation provides a useful benchmark for the subtler anomaly mediated theory. In particular, the vacuum structure of the $\mathcal{N}=1$ SYM recovered in the large-$\mu$ limit is well understood, and it is widely believed that this limit is continuously connected to the small $\mu$ limit~\cite{Seiberg_1994_II}. The core difficulty with the AMSB research programme initiated in~\cite{Murayama_2021} is that it is generally unclear to what extent the small $m_{3/2}$ results can be extrapolated to large $m_{3/2}$, due to the relative poverty of analytic techniques to determine the vacuum when $m_{3/2}$ is near or above the strong coupling scale. But in this theory, we have a robust understanding of the picture one would expect when $m_{3/2}\gg\Lambda_0$. From Eq.~\eqref{eq:UV AMSB SYM}, $\phi$ and $\lambda$ decouple, leaving us with the field content of $\mathcal{N}=1$ SYM, only the massless gaugino is $\psi_\phi$, rather than $\lambda$. Up to the $SU(2)_R$ gaugino exchange, the SW-deformed theory has the same decoupling limit. 

In this section, we will compare the two theories along the flows of their SUSY-breaking parameters ($\mu$ and $m_{3/2}$) from $0$ to $\infty$. We assume that in the decoupling limits $m_{3/2}, \mu\to\infty$, there exists no low-energy experiment which might distinguish AMSB from the SW deformation. 
We will now examine how this assumption may constrain the behaviour of the flow $m_{3/2}\to\infty$. The outcome of this discussion is ultimately somewhat inconclusive, but it lays the groundwork for an analogous but far more conclusive discussion of the $N_F>0$ theories in Section~\ref{sec: discussion NF nonzero}.

Let us begin with a comparison of the two theories in the limit $m_{3/2},\mu \ll \Lambda_{0}$. The AMSB ground states we derived in Section~\ref{sec:singularityEFT} are very similar to those of the SW-deformed theory. Nevertheless, there are subtle differences between the two theories, some of which we have already touched on and some of which we will introduce in this section. As foreshadowed by the instanton calculation in Section~\ref{sec:instantons}, the low-energy theory with AMSB is not supersymmetric, as the scalar interactions in Eq.\eqref{eq: general form for IR AMSB potential} do not come with fermionic terms into which they can be rotated under SUSY transformations. The anomaly mediated scalar potential also differs from the SW-deformed scalar potential even at the level of renormalizable operators. For example, the trilinear term $\propto\Re(a_D m \widetilde{m})$ is not present in the SW-deformed theory.

Nevertheless, the vacua that we have found in Section~\ref{sec:singularityEFT} are qualitatively very similar to those in the SW-deformed theory. Both singularities give rise to the dyon condensates given in Eq.~\eqref{eq: monopole VEV}. The condensate values are given in Table~\ref{Table:monopole condensates, NF=0}, alongside those for the SW-deformed theory. We calculated them using the series expansions of $K$ and $U$ given in~\cite{csáki2025phasetransitionsunusualvalues}. 

\begin{table}[h] 
    \centering
 \begin{tabular}{|c|c|c|}
    \hline
    Massless $(n_m, n_e)$ & $
    \braket{m
    \widetilde{m}}$ (AMSB) & $\braket{m\widetilde{m}}$ ($\mathcal{N}=1$)  \\
    \hline
   (0, 1) & $-m_{3/2}^\dagger\frac{i\Lambda_0^\dagger}{4\pi^2\sqrt{2}}$ & $-\sqrt{2}i\mu\Lambda_0$   \\
     (1, 1) & $m_{3/2}^\dagger\frac{\Lambda_0^\dagger}{4\pi^2\sqrt{2} }$ &  $-\sqrt{2}\mu\Lambda_0$  \\ \hline
\end{tabular}
    \caption{Dyon condensates at the singularities, for both AMSB and the SW-deformed theory theory. Here $m\widetilde{m}$ generically refers to the light dyon pair at a given singularity. }
    \label{Table:monopole condensates, NF=0}
\end{table} 

It immediately stands out that the AMSB condensates are proportional to $\Lambda_0^\dagger$ while the SW deformation condensates are instead proportional to $\Lambda_0$. However, this is consistent with what would be expected from the perturbative UV calculation. In Section~\ref{sec:UVAMSB}, we needed to perform an $SU(2)_R$ rotation in order to formally replace the now massive gaugino with the massless fermion associated with $\Phi$. If we act on the doublets $(m, \widetilde{m}^\dagger), (\widetilde{m}, m^\dagger)$ with the swap matrix $s$ in Eq.~\eqref{eq: swap operator}, then \begin{align}
s:m\widetilde{m}\to \widetilde{m}^\dagger m^\dagger 
\end{align}
After this rotation, instead of the entry in Table~\ref{Table:monopole condensates, NF=0}, we have \begin{align}
m\widetilde{m}=\frac{i}{4\pi^2} \frac{m_{3/2}\Lambda_0}{\sqrt{2}},
\end{align}
and the AMSB condensates are related to the SW condensates up to an identification 
\begin{align}
-\frac{m_{3/2}}{8\pi^2} \sim \mu
\end{align}
This is not the value we would obtain if we were to treat the perturbative AMSB superpotential Eq.~\eqref{eq: AMSB perturbative superpotential} as though it were exact. Eq.~\eqref{eq: AMSB perturbative superpotential} matches the standard SW-deformed superpotential Eq.~\eqref{eq: N=1 preserving mass deformation superpotential} if we identify $\mu$ with the quantity $\frac{m_{3/2}}{4\pi^2}$. 

Let us now discuss the decoupling limit in more detail, starting with the SW-deformed theory. When $\mu$ is taken to infinity, $\Phi$ decouples from the theory and the field content is that of $\mathcal{N}=1$ SYM. This theory has an anomalous $U(1)_R$ symmetry that rotates the gaugino. Under quantization, the complexified strong coupling scale has $R$-charge $2/3$~\cite{Tachikawa_2015}. A gaugino condensate forms, breaking the symmetry down to $\mathbb{Z}_2$. This symmetry permutes two degenerate vacua, which have fermion condensates $\braket{\Tr(\lambda\lambda)}=\pm 16\pi^2\Lambda_{\text{SYM}}^3$, where $\Lambda_{\text{SYM}}$ is the strong coupling scale of $\mathcal{N}=1$ SYM. These vacua are strongly believed to be continuously connected to the vacua found in the small $\mu$ limit. The $\mathcal{N}=1$ strong coupling scale can be related to the $\mathcal{N}=2$ scale via a relation between the gaugino condensate and $\braket{u}$. From~\cite{Konishi_2003,Finnell_1995}, we have  \begin{align} 
\label{eq: gaugino condensate in terms of phi}
\frac{\braket{\Tr(\lambda\lambda)}}{16\pi^2} 
= \frac{\mu}{2}\braket{\Tr\phi^2}.
\end{align}
Since $\Tr \phi^2\equiv u$, we may write \begin{align}
{\frac{\braket{\Tr(\lambda\lambda)}}{8\pi^2}} = \pm\mu\Lambda_{0}^2 = \pm\Lambda_{\text{SYM}}^3
\end{align}
These relations are valid at any $\mu$~\cite{Konishi_2003}. When $\mu$ is large compared to $\Lambda_0$, we can integrate out $\Phi$ and $\mu$ is the resulting cutoff. Keeping $\Lambda_{SYM}$ fixed, $\Lambda_0\to 0$ as $\mu\to\infty$. Then $\frac{\Lambda_{\text{SYM}}}{\mu}=\left(\frac{\Lambda_0}{\mu}\right)^{2/3}\ll 1$, and we have a wide range over which the dynamics are those of $\mathcal{N}=1$ SYM.

In the theory with AMSB, the decoupling limit should leave us with an analogous gaugino condensate, only with the formal swapping of the $\mathcal{N}=2$ gauginos and the replacement $\mu\to \frac{m_{3/2}}{2^{3/2}\pi^2}$. Applying Eq.~\eqref{eq: gaugino condensate in terms of phi} to the supersymmetric Lagrangian obtained when considering only perturbative AMSB, we find
\begin{align} \label{eq: gaugino condensate for AMSB}
\frac{\braket{\Tr(\psi_\phi \psi_\phi)}}{8\pi^2} = \pm\frac{m_{3/2}}{4\pi^2} \Lambda_0^2 = \pm\Lambda_{\text{SYM}}^3,
\end{align}
In the SW-deformed theory, Eq.~\eqref{eq: gaugino condensate in terms of phi} can be derived classically simply by insisting that the $F$-terms in the fundamental Lagrangian vanish in the vacuum, which implies that the vacuum energy is 0. Eq.~\eqref{eq: gaugino condensate in terms of phi} is exact, but the argument for this depends on supersymmetry~\cite{Konishi_2003}. It is therefore unclear whether Eq.~\eqref{eq: gaugino condensate for AMSB} can be considered an exact relation in the anomaly mediated theory. However, when $m_{3/2}\gg \Lambda_0$, the perturbative, $\mathcal{N}=1$-preserving approximation of AMSB is reasonable, and in either limit, we appear to have a vacuum energy of exactly 0. If we integrate out $\phi$ and $\lambda$, we should expect the vacuum structure to be that of $\mathcal{N}=1$ SYM. We should therefore expect Eq.~\eqref{eq: gaugino condensate for AMSB} to hold at least asymptotically as $m_{3/2}\to \infty$.

Having introduced the gaugino condensate, we now perform a $U(1)$ spurion analysis on the theory. Table~\ref{tab: R, J charges for fundamental fields} 
from Section~\ref{sec:R-symmetry review} lists the $R, J$ charges for fields in the fundamental Lagrangian. Table~\ref{tab:EFT U(1) charges} lists spurious $R, J$ assignments for quantities relevant to the dyonic EFT.  Table~\ref{tab:N=1 condensate charge assignment} lists the condensate charge assignments for the SW-deformed theory, confirming that the charges of the gaugino and monopole bilinears match the charges of their VEVs. Finally, Table~\ref{tab:AMSB condensate spurions with and without m32 charge involution} shows the analogous charges in the theory with AMSB. 

\begin{table}[h]
    \centering
    \begin{tabular}{|c|c c c c c c c c|}
    \hline
       & $U(A_D)$ & $A_D$ & $M, \widetilde{M}$& $\Lambda_{N_F}$ & $\chi$ & $m_{3/2}$ & $\mu$ & $\partial K$ \\
       \hline
       $U(1)_R$ &4 &2 &0 & 2& 0 & -2 & -2 & -2 \\
        $U(1)_J$ & 0 & 0 & 1 & 0 & 0 & -2 & 2 & 0  \\
        \hline
    \end{tabular}
    \caption{$R, J$ charges that restore spurious $U(1)_{R, J}$ invariance to the near singularity EFTs. Here $\chi$ is the conformal compensator, and $\mu$ is the $\mathcal{N}=1$-preserving mass. The charge of $\Lambda_{N_F}$ follows directly from the charge for $U(A_D)$, while the charges for other spurions follow from demanding $U(1)_{R, J}$ invariance of the superspace Lagrangian.}
    \label{tab:EFT U(1) charges}
\end{table}

\begin{table}[h]
    \centering
    \begin{tabular}{|c|c c c c|}
    \hline
         & $\braket{\lambda^\alpha\lambda_{\alpha}}$   & $\braket{\mu U}$ & $\braket{m\widetilde{m}}$ &  
          $\braket{\mu \partial U}$ 
     \\
         \hline
        $U(1)_R$ & 2 & 2 & 0 & 0  \\
        $U(1)_J$ & 2 & 2 & 2 & 2  \\
        \hline
    \end{tabular}
     
    \caption{
    Condensate charges in the SW-deformed theory. The expectation value in the second column is equal to the gluino condensate in the first column. The expectation value in the last column is proportional to the monopole condensate in the third. It is therefore sensible that these pairs of columns have matching charges.}
    \label{tab:N=1 condensate charge assignment}
    
\end{table}

\begin{table}[h]
    \centering
    \begin{tabular}{|c|c c c c|}
    \hline
         & $\braket{\psi_{\phi}^\alpha\psi_{\phi\alpha}}$   & $\braket{m_{3/2} U}$ & $\braket{m\widetilde{m}}$ & 
          $\braket{m_{3/2}^\dagger \partial K}$ 
     \\
         \hline
         $U(1)_R$ & 2 & 2 &  0 & 0  \\
         $U(1)_J$ & -2 & -2 & 2 &  2 \\
         $U(1)_J$ composed with $s$ & 2 &  2 &-2 & -2\\
         \hline
    \end{tabular}
     
    \caption{This table shows the condensate charges in the AMSB case, calculated based on the entries of Table~\ref{tab:EFT U(1) charges}. In the vacuum, the first column is equal to the second, and the third column is equal to the fourth. In the second row, no $SU(2)_R$ transformation has been performed. The charges of column 1 match those of column 2 and likewise for columns 3 and 4, although the gaugino condensate's $J$-charge is opposite that of the SW-deformed theory. The third row shows the charges after the $SU(2)_R$ rotation $s$ reverses the charges under $U(1)_J$. Crucially, the third column is \textit{not} equal to the fourth after performing the $s$ transformation, since $s:m\widetilde{m}\to m^\dagger \widetilde{m}^\dagger$. When this is accounted for, it becomes apparent that the left and right sides of Eq.~\eqref{eq: monopole VEV} have different $J$-charges after the $s$ transformation is carried out.}
    \label{tab:AMSB condensate spurions with and without m32 charge involution}
\end{table}
We see from the second row of Table~\ref{tab:AMSB condensate spurions with and without m32 charge involution} that the charges of the composite operators match the charges of their VEVs, although the $J$-charges are opposite those of the SW-deformed theory. Let us now consider the superpotential associated with perturbative AMSB given in Eq.~\eqref{eq: AMSB perturbative superpotential}. Up to a multiplicative constant, it is given by \begin{align}
 W \propto m_{3/2} U
\end{align}
If we want to preserve spurious $U(1)_J$ symmetry, the superpotential must have a charge of $+2$, but $ W$ has a $J$-charge of $-2$ if we naively apply the charge assignments in Table~\ref{tab:EFT U(1) charges}. This makes sense---under the swap operation in Eq.~\eqref{eq: swap operator}, the $U(1)_J$ charges of elements in an $SU(2)_R$ doublet are exchanged. There is no clear way to interpret $m_{3/2}$ as an element of an $SU(2)_R$ doublet, but at the level of $U(1)_J$ symmetry, we can certainly reverse its charge assignment. Performing this $J$-charge reversal, $ W$ now has $J$-charge $+2$.

The $J$-charges of the condensates under this $s$ rotation are listed in the third row of Table~\ref{tab:AMSB condensate spurions with and without m32 charge involution}. The charges now match those of Table~\ref{tab:N=1 condensate charge assignment}, but this does not mean that the charges in the theory match those of the SW-deformed theory. The fourth column of Table~\ref{tab:AMSB condensate spurions with and without m32 charge involution} is \textit{no longer} the value of $\braket{m\widetilde{m}}$ under the $SU(2)_R$ swap. Instead, we have \begin{align}
    \braket{m\widetilde{m}} =-\frac{m_{3/2}}{
    \sqrt{2}}\bar\partial K
\end{align}
The left hand side of this equation has $J$-charge $-2$, while the right hand side has $J$-charge $+2$.  There is no assignment of spurious $U(1)_J$ charges that simultaneously makes the SUSY-preserving AMSB superpotential Eq.~\eqref{eq: AMSB perturbative superpotential} and the equation for the monopole condensate Eq.~\eqref{eq: monopole VEV} invariant under $U(1)_J$. 

This lack of $U(1)_J$-invariance is an indication that in the anomaly mediated theory, the relative phases between quantities can take on values which are not possible in the SW-deformed theory. In particular, this means that the relative phase between the gaugino condensate Eq.~\eqref{eq: gaugino condensate for AMSB} and the monopole condensate Eq.~\eqref{eq: monopole VEV} can take on values which are not permitted in the SW-deformed theory. As we have already discussed, the relative magnitude between condensates is also different.  

However, Eq.~\eqref{eq: gaugino condensate for AMSB} was derived with the UV Lagrangian, and is most trustworthy when $m_{3/2}\gg \Lambda_0$. Eq.~\eqref{eq: monopole VEV} was derived using the effective IR description, which is valid only for $m_{3/2}\ll \Lambda_0$. This may indicate that Eq.~\eqref{eq: monopole VEV} must break down as $m_{3/2}$ exceeds the cutoff of the IR EFT. However, it is possible that both equations could hold at large $m_{3/2}$ if the difference in condensates is not observable at energies $\ll m_{3/2}$. The role of the monopole condensate is rather opaque when $m_{3/2}$ or $\mu$ exceed $\Lambda_0$, since the effective abelian description is no longer applicable. But in the SW-deformed theory, the nonrenormalization theorem~\cite{Seiberg_1994} implies that Eq.~\eqref{eq:N=1 VEV configuration, prelims} holds at all $\mu$. Even when $\mu$ is very large, there exists some (possibly large) energy scale at which the condensate value should have a measurable impact on observables. 

When $m_{3/2}\gg\Lambda_0$, we can integrate out $\phi$ and $\lambda$. Assuming that AMSB does not violate the decoupling theorem, an observer performing experiments at scales well below $m_{3/2}$ should be unable to distinguish the anomaly mediated theory from $\mathcal{N}=1$ SYM. To be a little bit more precise, let us abstractly consider the sets of correlators up to some UV cutoff $\Lambda_{UV}$ in the two theories as set-valued functions of the dimensional parameters, that is, $\mathcal{C}_{\text{AMSB}}(m_{3/2},\Lambda_{UV}, \Lambda_0)$ and $\mathcal{C}_{\text{SW}}(\mu, \Lambda_{UV}, \Lambda_0)$. Let us also consider the set of all correlators in pure $\mathcal{N}=1$ SYM up to the same cutoff, $\mathcal{C}_{\text{SYM}}(\Lambda_{\text{SYM}}, \Lambda_{UV})$. We assume that in the limits $\frac{\mu}{\Lambda_{UV}}$ and $\frac{m_{3/2}}{\Lambda_{UV}}\to\infty$, we have \begin{align}
\mathcal{C}_{\text{SW}} &\to  \mathcal{C}_{\text{SYM}}(\Lambda^{\text{SW}}_{\text{SYM}} )\nonumber \\
\mathcal{C}_{\text{AMSB}}&\to \mathcal{C}_{\text{SYM}}(\Lambda^{\text{AMSB}}_\text{SYM}),
\end{align}
where $\Lambda^{\text{SW}}_{\text{SYM}}/\Lambda^{\text{AMSB}}_{\text{SYM}} $ is equal to some constant which may or may not be 1.
In other words, we assume that the observables of the anomaly mediated theory tend to those of the SW-deformed theory, up to an overall rescaling of $\Lambda_{\text{SYM}}$, which is the only dimensional parameter in $\mathcal{N}=1$ SYM.
When $m_{3/2},\mu\ll\Lambda_0$, this statement  clearly holds in the somewhat trivial sense that both theories have a finite mass gap, so there are only two states in the $\frac{\Lambda_{UV}}{m_{3/2}}\to 0$ limit. But as $m_{3/2}, \mu \to \infty$, we should expect $\mathcal{C}_{SYM}$ to closely approximate $\mathcal{C}_{AMSB},\mathcal{C}_{SW}$ at nonzero $\Lambda_{UV}$, as long as $\frac{\Lambda_{UV}}{m_{3/2}}\sim \frac{\Lambda_{UV}}{\mu}\sim 0$.

 Does the monopole condensate sit at a scale that is accessible without probing the decoupled sector? The cutoff of the decoupled theory is $\sim \mu$, while the scale of the monopole condensate is $\sim \mu\Lambda_0$. Keeping $\Lambda_{\text{SYM}}$ fixed and sending $\mu\to\infty$, $\Lambda_0^2$ must go to 0 to compensate. To achieve this balance, we can write \begin{align}
\Lambda_0^2 \propto \frac{1}{\mu}
\end{align} along this flow, and so the gaugino condensate has a finite scale even in the strict $\mu\to\infty$ limit. By contrast, the monopole condensate scales as $\mu\Lambda_0\propto \sqrt{\mu}$. At finite but very large $\mu$, the monopole condensate sits at a scale far below the UV cutoff and should be measurable. However, in the strict $\mu\to\infty$ limit, the value of the condensate diverges. The AMSB condensates scale in the same manner, and so by our definition of decoupling, we cannot rule out the possibility that both Eq.~\eqref{eq: gaugino condensate for AMSB} and Eq.~\eqref{eq: monopole VEV} are simultaneously true at all $m_{3/2}$.  This would present a curious situation where the anomaly mediated theory appears supersymmetric and SW-like in the extreme IR ($\frac{\Lambda_{UV}}{m_{3/2}}\to 0$) and the extreme UV ($\frac{\Lambda_{UV}}{m_{3/2}}\to \infty$) but in between these limits, the theory breaks supersymmetry and has very different dynamics from the SW-deformed theory.

Although we cannot rule out this possibility on the grounds of decoupling as we defined it, it seems rather unlikely that the condensate mismatch persists, for the following physical reason: perturbation theory is valid at scales $\gg\Lambda_{\text{SYM}}$. At these scales, instantons should have practically no effect on the dynamics. But supposing that the monopole condensate $\sim m_{3/2}\Lambda_0$ sets the characteristic scale at which differences with the SW-deformed theory clearly manifest themselves, this means that the instanton effects substiantially alter the dynamics at this characteristic scale. As we have already shown, fixing $\Lambda_{\text{SYM}}$ and taking $
m_{3/2}$ to infinity, the condensate scales as $\sqrt{m_{3/2}}$, which diverges, albeit slowly. It is difficult to believe that at fixed strong coupling scale, instantons can have a dramatic effect on the dynamics at a scale which can be made arbitrarily large. 

For this reason, we believe it is most likely the case that the monopole condensate simply shifts to conform to the value that one would expect in the SW-deformed theory, as $m_{3/2}$ exceeds the cutoff of the monopole EFT. This is not a phase transition per se, because at any nonzero $m_{3/2}$, the theory is gapped and the far IR theory contains only 2 states --- the vacua --- related by a $\mathbb{Z}_2$ symmetry. 

Our argument for this evolution of vacua is somewhat vague. We are simply assuming that the differences in the theories should manifest themselves clearly at scales $\sim m_{3/2}\Lambda_0$. We do not know if this is correct, nor do we know exactly what the dynamical signature of this difference should be. But in Section~\ref{sec: discussion NF nonzero}, we will conduct a similar analysis on the theories with $N_F\neq 0$. The Higgs branches in the $N_F=2, 3$ theories make this line of argument more concrete.

\section{Applying AMSB to the theories with $N_F>0$}\label{sec: SQCD with AMSB}

In this section, we generalize the results of Section~\ref{sec:AMSB main body} to $N_F=1, 2, 3$. Much of our analysis in Section~\ref{sec:AMSB main body} carries over with minimal modification. The largest difference is the appearance of a Higgs branch in the theories with $N_F=2, 3$.  On the Higgs branches, the gauge symmetry is completely broken and the K{\"a}hler geometry is protected from quantum corrections due to the $\mathcal{N}=2$ nonrenormalization theorems~\cite{Argyres:1996eh}. The Higgs branches make contact with the Coulomb branches at singularities where dyons become light. Though the effective descriptions at these singularities are very similar to those in the $N_F=0$ case, they host multiple flavours of massless dyons that continuously evolve into the moduli of the Higgs branch.

In the UV, we will find that the conclusions of Section~\ref{sec:UVAMSB} are unaffected by the inclusion of hypermultiplets. At the perturbative level, AMSB always introduces an $\mathcal{N}=1$-preserving mass term for $\Phi$, and it has no effect on the hypermultiplets themselves. For all $N_F$, the Coulomb branches are lifted and the theories flow to the singularities. The vacuum field configurations are very similar to the $\mathcal{N}=1$ theory while the spectrum and dynamics are different. We will argue that the Higgs branches are unaffected by AMSB far away from the Coulomb branch, while the near singularity Higgs geometry is deformed in the same manner as in the SW-deformed theory.

\subsection{Generalizing the $N_F=0$ calculations to $N_F>0$}

\label{sec: generalizing Coulomb+singularity AMSB}
In the UV, the derivations of Section~\ref{sec:UVAMSB} largely carry over. The anomalous dimensions of the hypermultiplets $Q_i, \widetilde{Q}_i$ are 0, and so they remain massless. Because there is now a Yukawa term Eq.~\eqref{eq: UV yukawa term}, the trilinear AMSB term from Eq.~\eqref{eq:loopAMSB} now appears in the SUSY-broken Lagrangian. The only nonvanishing anomalous dimension is that of $\Phi$ in canonical normalization, and it can be related to the $\beta$ function as usual. This gives us the following:
\begin{align} \label{eq:UV AMSB NF>0}
m_\Phi^2 &  = \frac{\beta(g)^2 |m_{3/2}|^2}{g^2} \nonumber \\
m_\lambda & = -\frac{\beta(g)}{g} m_{3/2} \nonumber\\
A_{\widetilde{Q}\Phi Q} & = \frac{\beta(g)}{g} m_{3/2} \nonumber \\
m_{Q_i}^2& =m_{\widetilde{Q}_i}^2  = 0.
\end{align}

\begin{figure}
    \centering
    \includegraphics{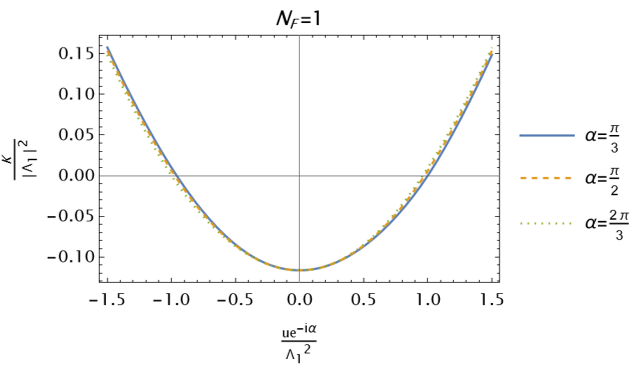}
    \includegraphics{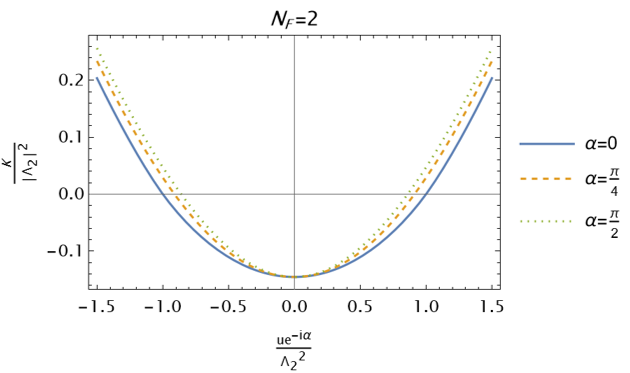}
    \includegraphics{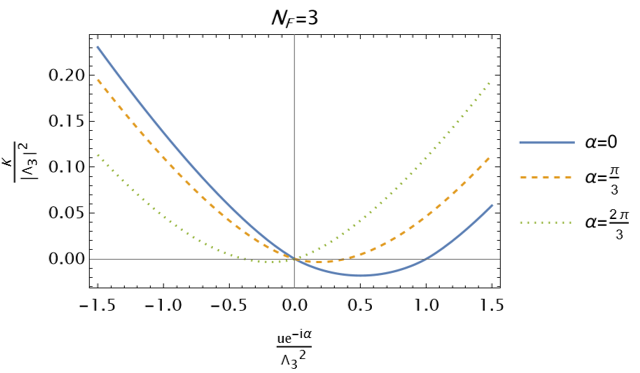}
    \caption{These plots show representative slices of the normalized K{\"a}hler potentials for $N_F=1, 2, 3$. Note the $u$-plane global symmetries listed in Table~\ref{tab:list of singularities vs NF}, which for $N_F=1, 2$ relate the curves shown to curves at larger angles. We use the same normalization used in the $N_F=0$ case, Fig.~\ref{fig:Kahler potential on Coulomb branch for NF=0}. Once again, this means that when $m_{3/2}$ and $\Lambda_{N_F}$ are set to 1, these plots reduce to plots of $K$ versus $u$, and under changes in either parameter, the function has the same overall shape up to scaling and rotation in the complex $u$-plane.}
    \label{fig:Kahler potentials for nonzero NF}
\end{figure}

\begin{figure}
    \centering
    \includegraphics{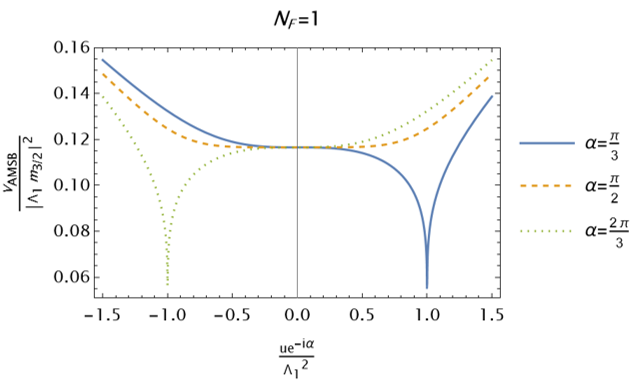}
    \includegraphics{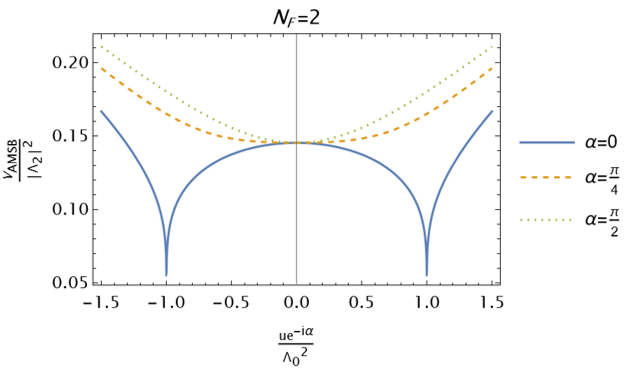}
    \includegraphics{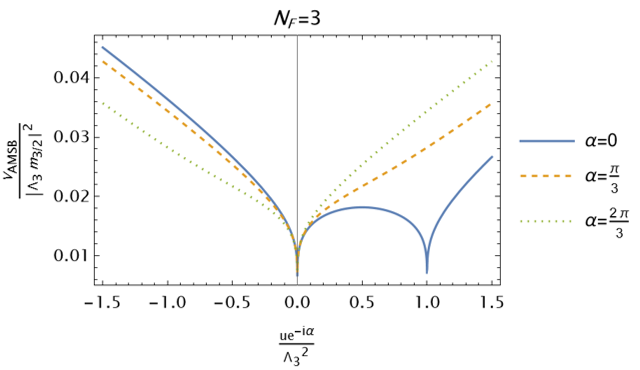}
    \caption{These plots show normalized AMSB potentials for $N_F=1, 2, 3$ along the real axis. We use the same normalization used in the $N_F=0$ case, Fig.~\ref{fig:VAMSB on Coulomb branch for NF=0}. When $\Lambda_0$ and $m_{3/2}$ are set to 1, these plots reduce to plots of $V_{\text{AMSB}}$ versus $u$. Under changes in either parameter, the function has the same overall shape up to scaling and rotation in the complex $u$-plane.}
    \label{fig:Coulomb branch VAMSB for nonzero NF}
\end{figure}

Up to $SU(2)_R$ rotation, these terms can again be concisely packaged in an $\mathcal{N}=1$-preserving mass term of the form of Eq.~\eqref{eq: AMSB perturbative superpotential}. To see that the trilinear term arises, we may write \begin{align}
\int d^2\theta (\sqrt{2}\widetilde{Q}\Phi Q + M\Tr(\Phi^2) ) &\supset \frac{\partial W}{\partial \phi^a}\frac{\partial W^\dagger}{\partial \phi^{\dagger}_a} \nonumber \\& \supset \sqrt{2}M q^\dagger\phi \widetilde{q}^\dagger + h.c.\nonumber \\ &=\sqrt{2}A_{\widetilde{Q}\Phi Q} \,s\{\widetilde{q}\phi q\} ,
\end{align}
where $s$ is the $SU(2)_R$ swap, and $M$ is the mass in Eq.~\eqref{eq: AMSB perturbative superpotential}. The constant $b_0$ in that equation is the coefficient of the leading term in the $\beta$ function. It is altered by the presence of hypermultiplets. The 1-loop $\beta$ function, Eq.~\eqref{eq: 1-loop beta function}, gives the following values for the leading coefficients: 
\begin{align} \label{eq: 1loop beta coefficients}
b_0^{N_F=1} &= \frac{3}{16\pi^2}\nonumber\\
b_0^{N_F=2} &= \frac{1}{8\pi^2} \nonumber\\
b_0^{N_F=3} &= \frac{1}{16\pi^2}. 
\end{align}
Let us now discuss the Coulomb branches of the theories with higher $N_F$. They are very similar to that of the pure Yang Mills theory. The singularities on each moduli space are listed in Table~\ref{tab:list of singularities vs NF}. All of the theories' Coulomb branches are described by K{\"a}hler potentials of the form of Eq.~\eqref{eq: coulomb branch Kahler potential}, with a tree level AMSB potential given by Eq.~\eqref{eq:CoulombAMSB}. The remarks of Appendix~\ref{sec:convexK} still apply: a real convex K{\"a}hler potential necessarily gives rise to an AMSB potential that drives the theory to the singularities. In Figs.~\ref{fig:Kahler potentials for nonzero NF} and~\ref{fig:Coulomb branch VAMSB for nonzero NF}, we plot the K{\"a}hler and AMSB potentials for $N_F=1, 2, 3$. 


Again, the Coulomb moduli roll to the singularities.

The near singularity EFTs all have potentials of the form of Eq.~\eqref{eq: general form for IR AMSB potential}. The K{\"a}hler potentials for each singularity are different, but the arguments of Section~\ref{sec:singularityEFT} are agnostic to the particular form of the K{\"a}hler potential and generalize straightforwardly. For the singularities with lone light dyon hypermultiplets, the derivation in~\ref{sec:singularityEFT} can be carried over without any modification. The cases with multiple light dyons require slightly more care. We will momentarily address this, but first, we state the general result: \begin{align} \label{eq: VEVs, NF nonzero}
a_D&=0\nonumber \\
m_n \widetilde{m}^n& = -m_{3/2}^\dagger\frac{\partial K}{\sqrt{2}}\nonumber\\
m^n m^\dagger_n&=\widetilde{m}^n \widetilde{m}_n^\dagger,
\end{align}
where $m_n, \widetilde{m}^n$ refer to the dyons that become light at a given singularity. The index $n$ indexes the flavours of dyon hypermultiplets where there are multiple. The field $a_D$ refers not strictly to the magnetic dual of the operator $a$ as defined in the UV, but rather to the particular linear combination of $a$ and $a_D$ that set the masses of these dyons. 

For the theories with multiple massless dyons, the $D$-term constraint $|m_n|^2=|\widetilde{m}_n|^2$ can be satisfied by a whole family of vacua. In the absence of AMSB, the VEVs are, up to symmetry transformations, \begin{align} \label{eq: higgs branch moduli space for dyons}
\vec{m} & = (B, 0, 0, ...) \nonumber\\
\vec{\widetilde{m}} & = (0, B, 0, ...).
\end{align}
These moduli transform under the global symmetry group on the Higgs branch, and in fact they are Higgs branch moduli, albeit expressed in variables appropriate to the small $B$ region of the branch. When AMSB is turned on, the constraints in Eq.~\eqref{eq: VEVs, NF nonzero} are satisfied by any of the following \begin{align} \label{eq: deformed higgs vev}
\vec{m} & = (C, 0,...) \nonumber \\
\vec{\widetilde{m}} & = (-\frac{D}{C}, B, ...),
\end{align}
with the third line of Eq.~\eqref{eq: VEVs, NF nonzero} implying that $\left|\frac{D}{C}\right|^2 + |B|^2 =  |C|^2$. Here $D$ is the VEV $m_n\widetilde{m}^n = -D = -m_{3/2}\frac{\partial K}{\sqrt{2}}$. This solution applies also to the SW-deformed theory, except with $-D=-\frac{\mu}{\sqrt{2}}\frac{d u}{da_D}$. As $m_{3/2}\to 0$, or equivalently as $C\to \infty$, the moduli space Eq.~\eqref{eq: deformed higgs vev} reduces to Eq.~\eqref{eq: higgs branch moduli space for dyons}. 

We now briefly outline why the arguments in Section~\ref{sec:singularityEFT} remain robust when multiple light dyons are present. Taking Eq.~\eqref{eq: general form for IR AMSB potential} and replacing $m, \widetilde{m}\to m_n, \widetilde{m}_n$, all but the $\mathcal{N}=2$-preserving terms depend only on the combination $m_n\widetilde{m}^n$. Our arguments for the positivity of the potential thus carry over directly, and it follows that the  potential is globally minimized at $a_D=0$. In that limit, the equations of motion are dominated by the $D$-term $\frac{g_{eff}^2}{2}(|m_n|^2-|\widetilde{m}_n|^2)^2$ and the term responsible for the monopole condensate, $\frac{g_{eff}^2}{2}|\sqrt{2}m_n\widetilde{m}^n +m_{3/2}^\dagger \partial K|^2$. They dominate because of the prefactors $\propto g_{eff}^2$. The term $|a_D|^2(|m_n|^2+|\widetilde{m}_n|^2)$ is not necessarily minimized for arbitrary field configurations satisfying Eq.~\eqref{eq: deformed higgs vev}, unlike in the $N_F=0$ case where the $D$-term constraint also minimizes this term. However, this term smoothly approaches 0 with $a_D$, and so it becomes negligible as we adiabatically bring $a_D\to 0$. Thus, Eq.~\eqref{eq: VEVs, NF nonzero} and~\eqref{eq: deformed higgs vev} correctly characterize the vacua to all orders in the expansion of the local K{\"a}hler potential.

We defer the presentation of the actual condensate values to Section~\ref{sec: discussion NF nonzero}. They can be found in Table~\ref{tab: expected monopole condensates based on perturbation theory}. In the meantime, we proceed to analyze the Higgs branches in Section~\ref{sec:Higgs branch}.

\subsection{AMSB on the Higgs branches} \label{sec:Higgs branch}
We saw in Section~\ref{sec: generalizing Coulomb+singularity AMSB} that the Coulomb branches in theories with $N_F>0$ behave very similar to those of the $N_F=0$ case. But the Higgs branches appearing in the theories with $N_F=2, 3$ have no analog in the $N_F=0$ theory. We briefly reviewed the structure of these branches at the end of Section~\ref{sec: prelims on NF neq 0}. Let us now discuss how they are affected by AMSB. 

For the potential to vanish, the following constraints on the field configurations must be satisfied: \begin{align}
Q_{ia}^\dagger T_{k}^{ab} Q^i_{b} & =\widetilde{Q}_{ia}^\dagger T_{k}^{ab}\widetilde{Q}^i_{b} \nonumber\\
\widetilde{Q}_{ia} T_k^{ab}Q^{i}_{b}&=0,
\end{align}
where $T_k$ is the $k$th generator of the gauge group, $i$ is a flavour index, and $a, b$ are colour indices. In~\cite{Seiberg_1994_II}, the authors work out the space of gauge-equivalent field configurations satisfying this constraint. We will not need the details of this description. However, we will need the fact that there is a single degree of freedom which is a singlet under the global symmetry group. This degree of freedom can be thought of as a radial coordinate on the moduli space~\cite{Antoniadis_1997} which sets the distance from the origin. In analogy to Eq.~\eqref{eq: higgs branch moduli space for dyons}, we refer to this coordinate as $B$.

The modulus $B$ sets the mass scale of the $\mathcal{N}=2$ vector multiplet, which is integrated out in the low-energy description. At the origin of the Higgs branch, there is a singularity that classically signals the vanishing of the $\mathcal{N}=2$ vector multiplet's mass. Quantum mechanically, the origin of the Higgs branch remains singular.\footnote{In the $N_F=2$ theory, there are two disjoint Higgs branches related by a $\mathbb{Z}_2$ transform which make contact at the origin in the classical theory. Quantum mechanically, neither branch is modified but the singularity splits into two. These are the same singularities appearing on the Coulomb branch.} This is the point at which the Higgs branch makes contact with the Coulomb branch. There, the effective description is given by the near singularity theory we have already discussed in the context of the Coulomb branch. Indeed, in the unbroken $\mathcal{N}=2$ theory, the dyon moduli space Eq.~\eqref{eq: higgs branch moduli space for dyons} \textit{is} simply the Higgs branch, albeit in variables appropriate to the near singularity region. Deeper on the Higgs branch, for $B\gg \Lambda_{N_F}$, it is appropriate to instead work with gauge invariant composites build out of the fundamental quark fields $Q_i, \widetilde{Q}_i$. 

There are good reasons to suspect that the AMSB formulae actually vanish when applied to the pure Higgs branch effective description. Crucially, this does not mean that the Higgs branch is completely unaffected by anomaly mediation. The Higgs modulus sets the mass of the $\mathcal{N}=2$ vector multiplet. When it is comparable to (or smaller than) $m_{3/2}$, the pure Higgs description is inappropriate. But it means that to good approximation, the moduli space at large $\frac{B}{m_{3/2}}$ survives and is unaltered by AMSB, regardless of how either scale compares to $\Lambda_{N_F}$. This is in contrast to the Coulomb branch, where the geometry of the branch itself produces a potential, Eq.~\eqref{eq:CoulombAMSB}, which monotonically increases with $|u|$ outside of the strong coupling region and demonstrably drives the moduli towards the singularities, even without knowlegde of what occurs when the light dyons are reintroduced to the theory. Although this sounds strange, it comports with the behaviour of the anomaly mediated UV theory --- the Higgs branch is composed of squark moduli, and the $\mathcal{N}=2$ vector multiplet has a mass set by these moduli. In the UV theory, the $\mathcal{N}=2$ vector multiplet feels anomaly mediation but the hypermultiplets do not. The vanishing of AMSB on the Higgs branch is an IR manifestation of this fact, and it suggests that any differences between the anomaly mediated and SW-deformed Higgs branches can be traced back to the coupling of AMSB to instantons.

We have sketched a proof that the AMSB terms vanish on the Higgs branch. The details are technical and inessential to the main thread of the paper, so we relegate our argument to Appendix~\ref{sec: higgs branch involved derivation}. More important are the effects of AMSB when the Higgs moduli are \textit{not} large compared to $m_{3/2}$. We have already implicitly analyzed the effects of AMSB on the near singularity region $B\ll \Lambda_{N_F}$ when $m_{3/2}\ll \Lambda_{N_F}$. In this regime, the unbroken $\mathcal{N}=2$ Higgs branch is described by the dyonic moduli Eq.~\eqref{eq: higgs branch moduli space for dyons}. Turning on AMSB deforms this moduli space and produces Eq.~\eqref{eq: deformed higgs vev}, which reduces to Eq.~\eqref{eq: higgs branch moduli space for dyons} as $\frac{C}{ m_{3/2}}\to\infty$.\footnote{With the deformation turned on, it is $|C|$ rather than $|B|$ which parameterizes the distance from the singularity on the Higgs branch. To see this, note that the constraint $\left|\frac{D}{C}\right|^2$ below Eq.~\eqref{eq: deformed higgs vev} has the form of the Pythagorean theorem, with $|C|$ playing the role of the hypotenuse.} Since AMSB vanishes on the Higgs branch, this statement holds even when $C\gg \Lambda_{N_F}$, even though the dyonic description of the theory is inapplicable in this regime. 

When $m_{3/2}\gg C\gg \Lambda_{N_F}$, the perturbative AMSB superpotential Eq.~\eqref{eq: AMSB perturbative superpotential} is an excellent approximation. Instantons break the supersymmetry, but instantons are related to the low-energy behaviour of the $\mathcal{N}=2$ vector multiplet whose mass is set by $C$. As long as $C\gg \Lambda_{N_F}$, these corrections are highly suppressed and we can treat AMSB like an SW deformation. The original paper~\cite{Seiberg_1994_II} addressed the large $\mu$ limit of the SW-deformed Higgs branch by introducing the most general possible superpotential consistent with holomorphy and spurious symmetries in each case. Replacing $\mu$ with $b_0^{N_F}m_{3/2}$, these superpotentials can simply be carried over to the anomaly mediated theories when $m_{3/2}\gg C\gg\Lambda_{N_F}$. The superpotentials in question are listed in Appendix~\ref{sec: large SW-deformation Higgs branch superpotentials}.

In fact, it is very likely that this superpotential still gives the correct behaviour even for $C\ll \Lambda_{N_F}$, as long as $m_{3/2}\gg \Lambda_{N_F}$. This point is crucial to our conclusions, and we will argue for it in Section \ref{sec: discussion NF nonzero}, to which we now proceed.

\section{Comparing AMSB to the SW deformation for $N_F>0$} \label{sec: discussion NF nonzero}

Paralleling the discussion of Section~\ref{sec: Nf=0 theory comparison}, here we compare and contrast the $N_F\neq 0$ theories with AMSB to their SW-deformed counterparts. 
We begin by reviewing the behaviour of the $\mathcal{N}=1$ SQCDs recovered in the decoupling limit of the SW deformation. The anomaly mediated theory should have the same decoupling limit, since at very large $m_{3/2}$, the $\mathcal{N}=1$ preserving superpotential Eq.~\eqref{eq: AMSB perturbative superpotential} is a good approximation of the anomaly mediated terms. 

$\mathcal{N}=1$ SQCD with 1 flavour develops a dynamical superpotential known as the Affleck-Dine-Seiberg superpotential~\cite{Affleck:1984xz}. In the absence of bare mass terms for squarks, it destabilizes the vacuum. The lack of a stable moduli space is consistent with the fact that the $\mathcal{N}=2$ theory with $N_F=1$ has no Higgs branch. Since the decoupling limit is unstable, this theory is not particularly well suited to our analysis, and we will largely ignore it.

$\mathcal{N}=1$ SQCD with 2 flavours has a moduli space, but it has only one branch, in contrast to the $\mathcal{N}=2$ theory that has two branches that make contact with the two Coulomb branch singularities.  The two branches merge as the SUSY-breaking parameter is turned up, leaving us with the moduli space of the $\mathcal{N}=1$ theory. The origin of this moduli space is deformed by quantum corrections. This deformation erases the would be singularity and leads to the spontaneous breaking of chiral symmetry everywhere on the moduli space. From the perspective of the SW-deformed $\mathcal{N}=2$ theory, this phenomenon can be explained as a consequence of dyon condensation~\cite{Seiberg_1994_II}. The condensing dyons are charged under the global flavour group $SO(2N_F)$, and they are associated with fermion condensates transforming in the vector representation. Since a condensate with global charges has formed, chiral symmetry is spontaneously broken.  

Lastly, the $N_F=3$ SQCD is s-confining, with no chiral symmetry breaking at the origin. From the perspective of the SW-deformed theory, this is because the singularity with a lone dyon hypermultiplet merges with the Higgs branch as $\mu\to\infty$. The dyon at that singularity is a singlet under $SO(2N_F)$. This is how the $N_F=3$ theory evades chiral symmetry breaking at the origin.

In~\cite{Seiberg_1994_II}, the moduli spaces of $
\mathcal{N}=1$ SQCD with$N_F=2, 3$ are recovered in the large-$\mu$ limit as follows.  For $\mu\gg \Lambda_{N_F}$, the superpotentials given in Appendix~\ref{sec: large SW-deformation Higgs branch superpotentials} can be constructed on the basis of spurious symmetry invariance and holomorphy. The resulting equations of motion reproduce the known results of the $\mathcal{N}=1$ theories in the $\mu\to\infty$ limit. In the small $\mu$ limit, the dyon moduli satisfy Eq.~\eqref{eq:N=1 VEV configuration, prelims}, leading to a moduli space constrained by Eq.~\eqref{eq: deformed higgs vev} in the near singularity region. In either description, the symmetry breaking pattern is the same, and it is believed that the interpolation between the two descriptions is smooth and continuous. 

Let us now compare the two theories, starting with the small $m_{3/2}$ and small $\mu$ limits. As in the $N_F=0$ case, the scalar potential induced by AMSB breaks SUSY completely. Fermionic terms induced by the superpotential $\mu U$ are absent, and though the scalar potential itself resembles that of the $\mathcal{N}=1$ theory at leading order in $\frac{a_D}{\Lambda_{N_F}}$, it is not the same even at the level of relevant operators. But the vacua that we have found are qualitatively similar to those in the $\mathcal{N}=1$ theory, and they are exact as long as $m_{3/2}$ is kept small compared to the cutoff of the near singularity EFT. 

The final column of Table~\ref{tab: expected monopole condensates based on perturbation theory} contains the AMSB condensates at all singularities (after the $SU(2)_R$ rotation). The third column lists the condensates in the perturbatively equivalent SW theory, with superpotential Eq.~\eqref{eq: AMSB perturbative superpotential}. We can obtain these values by plugging $\mu=b_0^{N_F} m_{3/2}$ into Eq.~\eqref{eq:N=1 VEV configuration, prelims}.
Dividing the fourth column of Table~\ref{tab: expected monopole condensates based on perturbation theory} by the third column, we find that the magnitude of the actual calculated condensate is always $1/2$ of the condensate we would expect in the equivalent SW-deformed theory.\footnote{While finalizing our manuscript, we found what we believe to be a minor error in~\cite{csáki2025phasetransitionsunusualvalues}, from which we pulled the expansions for the K{\"a}hler potentials. Their Eq.~(3.28) for $\partial K$ at the second singularity disagrees with the value one obtains by deriving the result from their local prepotential, Eq.~(3.19) by a factor of 2. We believe the former expression is correct, as the latter would make this singularity the only one for which the ratio of condensate magnitudes is \textit{not} $\frac{1}{2}$.}  The phases of the two columns also disagree.

\begin{table}
    \centering
    \begin{tabular}{|c|c|c|c|}
    \hline
 $N_F$ & Massless $(n_m, n_e)$ & Expected $\braket{m\widetilde{m}}$ (subbing $\mu\to b_0^{N_F}m_{3/2}$)  & Actual condensate (after $SU(2)_R$ swap) \\
      \hline
  0 & (0, 1) &      $-m_{3/2}\frac{i\Lambda_0}{2\sqrt{2}\pi^2}$  & $m_{3/2}\frac{i\Lambda_0}{4\pi^2\sqrt{2}}$ \\
     0 & (1, 1) &       $-m_{3/2}\frac{\Lambda_0}{2\sqrt{2}\pi^2}$ & $m_{3/2}\frac{\Lambda_0}{4\pi^2\sqrt{2}}$ \\
 1 & (1, -1)&       $\exp(5i\pi/3)\frac{3m_{3/2}\Lambda_1}{8\pi^2}$ & $\exp(2i\pi/3)\frac{3m_{3/2}\Lambda_1}{16\pi^2}$ \\
   1 & (1, 1) &     $\frac{3m_{3/2}\Lambda_1}{8\pi^2}$ & $\frac{3m_{3/2}\Lambda_1}{16\pi^2}$\\
   1 & (1, 0)  &      $\exp(i\pi /3)\frac{3m_{3/2}\Lambda_1}{8\pi^2}$ & $\exp(4i\pi/3)\frac{3m_{3/2}\Lambda_1}{16\pi^2}$\\
 $2$ & (1, 0)   &      $-\frac{im_{3/2}}{2\sqrt{2}\pi^2}\Lambda_2$ &  $i\frac{m_{3/2}\Lambda_2}{4\pi^2\sqrt{2}}$ \\
2  & (1, 1) &       $-\frac{m_{3/2}}{2\sqrt{2}\pi^2}\Lambda_2$ & $\frac{m_{3/2}\Lambda_2}{4\pi^2\sqrt{2}}$\\
 $3$ & (1, 0)&      $-\frac{m_{3/2}\Lambda_3}{4\pi^2}$ & $\frac{m_{3/2}\Lambda_3}{8\pi^2}$\\
  3 &(2, -1)  &      $\frac{-im_{3/2}\Lambda_3}{8\pi^2}$ & $\frac{-m_{3/2}\Lambda_3}{16\pi^2}$ \\
  \hline
        
    \end{tabular}
    \caption{Table of ``expected" vs calculated monopole condensates. The expected condensates are the values obtained by taking the SW-like Eq.~\eqref{eq:N=1 VEV configuration, prelims} and substituting $\mu$ for the coefficient of $\Tr(\Phi^2)$ in the perturbative AMSB superpotential, Eq.~\eqref{eq: AMSB perturbative superpotential}. The calculated condensates are the conjugates of those we would get by simply evaluating Eq.~\eqref{eq: monopole vacuum summary}. The conjugation is to account for the $SU(2)_R$ swap.}
    \label{tab: expected monopole condensates based on perturbation theory}
\end{table}

As in Section~\ref{sec: Nf=0 theory comparison}, we can perform a $U(1)$ spurion analysis on the $N_F>0$ theories. The analysis carries over without modification. For all $N_F$ there is a mismatch in the appropriate spurionic $U(1)_J$-charge assignment for $m_{3/2}$, depending on whether one is trying to restore spurious $J$-invariance in the perturbative UV theory or in the near singularity dyonic EFT. To give the UV theory a spurious $U(1)_J$, we need the charge of $m_{3/2}$ to flip under $SU(2)_R$ swap transformations. However, the resulting charge assignment does not preserve the spurious $U(1)_J$ symmetry of Eq.~\eqref{eq: monopole vacuum summary}. 

We have again seen that the vacuum structures of the anomaly mediated and SW-deformed theories are qualitatively similar, but not identical. We will assume once more that in the large $m_{3/2}$ and large $\mu$ limits, the theories become indistinguishable. Do the AMSB vacua necessarily need to evolve into the SW-deformed vacua as $m_{3/2}$ becomes large? In the $N_F=0$ theory, we argued that they probably do, on the grounds that failure to do so would imply that instanton suppressed effects can have a substantial effect on scales which are arbitrarily large relative to the $\mathcal{N}=1$ coupling. In the $N_F=2, 3$ theories, the same line of thinking can be made somewhat more precise, because the SW-deformed theory has a nontrivial moduli space in both the small $\mu$ and large $\mu$ limits, and moreover, these limits are believed to be continuously connected~\cite{Seiberg_1994_II}. The anomaly mediated theory also has a nontrivial moduli space in both limits, and arbitrarily low-energy experiments can probe the space's geometry and search for deviations from SW-deformed behaviour.

The near singularity moduli space in the anomaly mediated theory is described by Eq.~\eqref{eq: deformed higgs vev} at small $m_{3/2}$. Up to a redefinition of $D$, this is identical to the constraint in the small $\mu$ SW-deformed theory.  Phase rotations of the dyon VEV leave this defining equation invariant. The compensating rephasing of the moduli amounts to an symmetry of the deformed moduli space. Of course, the equation is \textit{not} invariant under changes in the magnitude of the VEV $D$. As $m_{3/2}$ approaches the cutoff of the dyonic EFT, the deformation grows. The same is true of the SW-deformed theory, but the scale of the deformation is different due to the difference in magnitude between the columns in Table~\ref{tab: expected monopole condensates based on perturbation theory}.

Let us now consider the region of parameter space where $m_{3/2}\gg C\gg\Lambda_{N_F}$. In this regime, the corrections from instantons that differentiate AMSB from the SW-deformed theory are heavily suppressed by the modulus $C$. Since $C$ is large, one expects that the instanton driven breaking of $\mathcal{N}=1$ SUSY is highly suppressed, and we can treat AMSB as though it is a SW deformation with superpotential Eq.~\eqref{eq: AMSB perturbative superpotential}. In~\cite{Seiberg_1994_II}, they capture the effects of a large $\mu$ SW deformation on the Higgs branch via the superpotentials in Appendix~\ref{sec: large SW-deformation Higgs branch superpotentials}. The superpotentials are not $U(1)_J$-invariant, though by construction they are spuriously invariant. It is therefore meaningful to talk about the orbit of deformed moduli spaces under $U(1)_J$ rotations. 

In the SW-deformed theory, the large and small $\mu$ limits are continuously connected. While we cannot explicitly describe the deformed Higgs branch $\mathcal{M}_{\text{SW}}$ at all $\mu$, we can loosely treat $\mathcal{M}_{\text{SW}}$ as a continuous function $\mathcal{M}_{\text{SW}}(\mu)$ that associates a K{\"a}hler manifold to each value of $\mu$. We are not necessarily guaranteed that the analogous function of AMSB-deformed Higgs branch manifolds $\mathcal{M}_{\text{AMSB}}(m_{3/2})$ is continuous in the region $m_{3/2}\sim \Lambda_{N_F}$, but we at least know that such a function exists on either side of the transition region. 
In this language, the assumption of decoupling suggests that in the $m_{3/2}\to\infty$ limit, we have \begin{align}
\lim_{m_{3/2}\to\infty} (\mathcal{M}_{\text{AMSB}}(m_{3/2})) = \lim_{m_{3/2}\to\infty} (\mathcal{M}_{\text{SW}}(Fb_0m_{3/2})),
\end{align}
where $F$ is a constant which may or may not be equal to $1$.

When $m_{3/2}, \mu\ll \Lambda_{N_F}$, the moduli spaces of both theories satisfy a constraint of the form of Eq.~\eqref{eq: deformed higgs vev}, and both tend to the $\mathcal{N}=2$ geometry far away from the origin where $B\sim C \gg m_{3/2}$. As we have discussed, the deformation parameter $D$ in the theory with AMSB is related by a factor of $2e^{i\alpha}$ to the deformation parameter in the SW-deformed theory with $\mu=b_0^{N_F} m_{3/2}$, where $\alpha$ is a phase accounting for both the phase differences between the columns of Table~\ref{tab: expected monopole condensates based on perturbation theory} and any phase differences induced by the $U(1)_J$ spurion mismatch. We can write \begin{align}\label{eq: small-m32 higgs branch abstract manifold comparison}
\mathcal{M}_{\text{AMSB}}(\mathcal{P}(m_{3/2})) = \mathcal{M}_{\text{SW}}(b_0^{N_F}m_{3/2})
\end{align}
where $\mathcal{P}:m_{3/2}\to 2e^{i\alpha}m_{3/2}$. Note that the phase $\alpha$ is not the same from singularity to singularity. 

When $m_{3/2}\gg \Lambda_{N_F}$, the SW deformation superpotential is applicable at least as long as $C\gg \Lambda_{N_F}$. Denoting by $\mathcal{M}^+$ the region of the moduli space with $C\gg \Lambda_{N_F}$, we have that \begin{align} \label{eq: large-B, large SUSY-breaking higgs branch abstract manifold equation}
\mathcal{M}^+_{\text{AMSB}}(m_{3/2}) = \mathcal{M}^+_{\text{SW}}(b^{N_F}_0m_{3/2}).
\end{align}
The question is whether or not we can extend the domain of validity of this equation to \begin{align} \label{eq: Full moduli space equivalence at large-m_{3/2}}
\mathcal{M}_{\text{AMSB}}(m_{3/2}) = \mathcal{M}_{\text{SW}}(b^{N_F}_0m_{3/2}),
\end{align}
as long as $m_{3/2}\gg \Lambda_{N_F}$. This  would be consistent with the coincidence of the perturbative UV descriptions of the two theories and it would give the expected decoupling limit. It would indicate that the small $m_{3/2}$ AMSB moduli space must change as $m_{3/2}$ crosses the confinement scale, in the following sense:

From Eq.~\eqref{eq: small-m32 higgs branch abstract manifold comparison}, the small $m_{3/2}$ AMSB moduli spaces are equivalent to the small $\mu$ SW-like moduli spaces by an identification $\mu = \mathcal{P}^{-1}(b_0m_{3/2})$. Denoting by $\mathcal{M}^-$ a finite region of the moduli space which encloses the origin and is truncated at some maximum $C_{\text{max}}$, we have \begin{align}
\mathcal{M}^-_{\text{SW}} \label{eq: small-B, large m32 hypothetical higgs branch relation}
(b_0^{N_F}m_{3/2})=\mathcal{M}^-_{\text{AMSB}}(\mathcal{P}m_{3/2})
\end{align}
But if Eq.~\eqref{eq: Full moduli space equivalence at large-m_{3/2}} holds at large $m_{3/2}$, this relationship must break down as $m_{3/2}$ crosses the confinement scale.  

If it does not, we would simply have that Eq.~\eqref{eq: small-B, large m32 hypothetical higgs branch relation} and Eq.~\eqref{eq: large-B, large SUSY-breaking higgs branch abstract manifold equation} hold simultaneously. But in the transition region between these two limits, the geometry of $\mathcal{M}_{\text{AMSB}}$ will not be related to any SW-deformed Higgs branch. At least in principle, this warping of the geometry on the moduli space will be measurable.

If we are strict about how we define decoupling and only insist that the two deformations are indistinguishable in the limit of 0 energy (relative to $m_{3/2}$), it is possible that this warping is allowed. The reason is that a perturbative calculation about some vacuum on the Higgs branch will involve a Taylor expansion of the coefficients appearing in the sigma model Lagrangian describing massless fluctuations on the moduli space. For example, a scalar kinetic term can be schematically expanded as \begin{align}
h_{i\bar{j}}\partial_\mu\phi^i \partial^\mu\phi^{\dagger \bar{j}} =( h_{i\bar{j}}|_p + \partial_kh_{i\bar{j}}|_p\delta\phi^k+h.c. + ...)\partial_\mu\phi^i \partial^\mu\phi^{\dagger \bar{j}},
\end{align}
where $p$ is the vacuum about which we are expanding and $\delta\phi^k$ is the perturbation about this vacuum. The leading order dimensionless coefficient is a constant, and the higher order terms are all nonrenormalizable. The dimensionful parameters available are $\Lambda_{N_F}$, the modulus of the background value $|C|$, and the AMSB parameter $m_{3/2}$. At fixed $m_{3/2}$, the only one of these parameters which varies from vacuum to vacuum is $|C|$, and we should expect that the radius of convergence of this expansion depends on the position of this vacuum on the moduli space. A typical operator of dimension $4+n$ will depend on inverse powers of $|C|$. On the grounds of the sigma model terms alone, any differences in the two theories will vanish in the strict IR limit for any finite $C$, and in this sense, the two theories must be in the same universality class. However, this is not a very useful way to characterize the theory, for by this definition, all pure sigma models with the same field content are in the same universality class, up to singular points on the metric. Any metric looks flat if you only expand it to 0th order, and higher order operators are irrelevant. This is also ignoring the SW-like superpotentials, which are renormalizable. If they become modified for $C\ll m_{3/2}$, it is possible that the difference between the theories will appear at the renormalizable level. Moreover, it is typically the case in supersymmetric theories that in decoupling limits, the moduli spaces of the decoupled theories are recovered. All of these considerations suggest that the moduli space should not remain warped as $m_{3/2}\to\infty$.

One could argue that the transition region interpolating between Eq.~\eqref{eq: small-B, large m32 hypothetical higgs branch relation} and Eq.~\eqref{eq: large-B, large SUSY-breaking higgs branch abstract manifold equation} survives at any finite $m_{3/2}$, but moves radially outward on the moduli space so that in the decoupling limit, no finite $|C|$ is able to resolve it. We believe this is highly unlikely, for reasons paralleling our discussion in Section~\ref{sec: Nf=0 theory comparison}. In the $N_F=0$ theory, we assumed that the scale of the monopole condensate should define the scale at which differences between the two theories become clear, in some admittedly vague sense. The condensate varies as $\sqrt{m_{3/2}}$ as we bring $m_{3/2}\to\infty$, and so in the strict limit, it decouples. However, this requires that instantons, which are suppressed by a factor $\sim \exp(-\frac{1}{g^2})$ above $\Lambda_{\text{SYM}}$, have an effect on the dynamics at arbitrarily large scales. In the context of the Higgs branches, the notion that the interpolation region survives but asymptotically approaches $|C|\sim\infty$ is analogous. In this context it would be even more bizarre, because $|C|$ sets the mass of the gauge boson via the Higgs mechanism. One would expect this to further suppress the influence of instantons on the dynamics. For these reasons, we believe, as in the $N_F=0$ case, that Eq.~\eqref{eq: Full moduli space equivalence at large-m_{3/2}} must hold to good approximation relatively quickly after $m_{3/2}$ crosses the strong coupling scale, and the difference between AMSB and the SW deformation must rapidly fall off.

Notice also that the relative phase between the two $N_F=3$ singularities is different in the 3rd and 4th column of Table~\ref{tab: expected monopole condensates based on perturbation theory}. If Eq.~\eqref{eq: small-B, large m32 hypothetical higgs branch relation} holds at large $m_{3/2}$, this phase difference persists even in the decoupling limit. But in that limit, the isolated singularity merges with the continuum to form the $\mathcal{N}=1$ moduli space. It is difficult to see how this can be reconciled with decoupling, unless Eq.~\eqref{eq: small-B, large m32 hypothetical higgs branch relation} breaks down.

Heuristically, there are also dynamical reasons to believe that the transition to an SW-like geometry occurs relatively quickly as $m_{3/2}$ traverses the incalculable region of the theory where $m_{3/2}\sim \Lambda_{N_F}$. In a somewhat imprecise sense, the dynamics of the vector differ most dramatically from the SW-deformed theory exactly when the scale of the dynamics and the scale of $m_{3/2}$ lie in this incalculable region. In the dyonic EFT, we see this in the infinite series of nonrenormalizable operators that differentiate the theory from those of the SW-deformed theory. Ignoring the rephasing and rescaling of the condensate value, the SW-deformed scalar potential Eq.~\eqref{eq: N=1 preserving scalar potential} and the AMSB scalar potential Eq.~\eqref{eq: general form for IR AMSB potential} look similar when $m_{3/2}$ is small and only small fluctuations in $a_D$ are considered, but these terms become more important when we increase $m_{3/2}$, or when we consider larger perturbations in $a_D$. As we push the low-energy EFT to its limits, its dynamics look less and less like the SW-deformed theory. On the other hand, in the perturbative UV theory, the SUSY-breaking mass splitting scales with an exponential instanton factor. This mass splitting becomes more important as we flow to $\Lambda_{N_F}$ from above. When $m_{3/2}\gg \Lambda_{N_F}$, this effect is offset by the dynamical suppression of $\Phi$ and the two theories have similar dynamics. But as $m_{3/2}$ is brought down past the strong coupling scale, this dynamical suppression fails to take place. The perturbative dynamics differ more and more from the SW-deformed dynamics as we flow towards the confinement scale. While these observations do not tell us anything definitive about the vacua of the theory, they reinforce the picture that the anomaly mediated theory differs most violently from any SW-like approximation when $m_{3/2}$ is in the incalculable region.

To summarize, we believe that the AMSB vacua undergo a transition of sorts as $m_{3/2}$ passes through the incalculable region. On either side of this region, the AMSB and SW-deformed theories are approximately equivalent, up to an identification of $\mu$ with $b_0^{N_F}m_{3/2}$ for large $m_{3/2}$ and $\mathcal{P}^{-1}(b_0^{N_F}m_{3/2})$ for small $m_{3/2}$. But in the intermediate regime, the physics of the two deformations are very different, and this is where the transition occurs. 

The transition is not a phase transition in the usual sense, because for both large and small $m_{3/2}$, the massless field content and the global symmetry breaking patterns are the same. Although the large $m_{3/2}$ theory appears to preserve $\mathcal{N}=1$ supersymmetry and the small $m_{3/2}$ theory appears to break it, this is only approximately true at any finite nonzero $m_{3/2}$, and the operators which break SUSY all involve fields which are massive when $m_{3/2}$ is nonzero. The universality class of the theory is independent of $m_{3/2}$, as long as it is finite and nonzero. Nevertheless, the theory does undergo a major change in quantitative behaviour under the flow from one limit to the other. 

This conclusion is interesting, but it also underlines the difficulty of extrapolating from small to large $m_{3/2}$. Although there was no phase transition in the proper sense, our results tell us that the behaviour of AMSB is sensitive to nonperturbative violation of conformal symmetry. In the dyonic EFTs, our solutions are exact and there is no sign whatsoever that the vacua will undergo change. In the perturbative UV theories, we were able to see that SUSY is always broken completely, but only because $\mathcal{N}=2$ SUSY affords us an unusual level of analytic control over the running of the theory. Had we not included the instanton corrections to the $\beta$ function, it would have appeared that the theory is exactly identical to the SW-deformed theory at any $m_{3/2}>\Lambda_{N_F}$. Then, upon crossing that threshold, the theory abruptly changes and breaks global supersymmetry completely. In theories that are less controllable, it is not difficult to imagine that all manner of subtle phenomena are completely invisible to calculations that are restricted to weakly coupled limits.

\section{Summary and conclusion} \label{sec:conclusion}
In this work, we have applied anomaly mediation to $\mathcal{N}=2$ SQCD with gauge group $SU(2)$ and $N_F=0, 1, 2, 3$. When $m_{3/2}\gg\Lambda_{N_F}$, we can use the perturbative description of the theory to find that the corrections from AMSB contribute an SW-like $\mathcal{N}=1$ preserving mass term for the adjoint chiral multiplet $\Phi$, up to a crucial $SU(2)_R$ rotation that exchanges the adjoint fermions. The SW deformation is very well understood, and it played an important role in the seminal works \cite{Seiberg_1994, Seiberg_1994_II}.

However, the anomaly mediated theory is not exactly equivalent to the SW deformation. Instanton corrections to the $\beta$ function induce a mass splitting among the components of $\Phi$, further breaking SUSY down to $\mathcal{N}=0$. For $m_{3/2}\ll \Lambda_{N_F}$, we can apply anomaly mediation to the effective low-energy $\mathcal{N}=2$ theory. The anomaly mediated theory in this regime is qualitatively similar to the SW-deformed theory, but there are various differences. In both theories, the Coulomb branch is lifted and the modulus $u$ rolls to the singularities where magnetically charged dyons become massless. These dyons condense and induce the confinement of electric charge. But in the anomaly mediated theory, the effective Lagrangian breaks SUSY completely and the mass spectrum resulting from the dual Higgs mechanism is different, in part due to these differences in the Lagrangian. Matching the SW-deformation parameter onto the perturbative SW-like contribution from AMSB, we find that the magnitudes and phases of the condensates themselves also differ between the theories. 

For physical reasons related to the decoupling of heavy states and the UV suppression of instantons, we believe that the AMSB condensates must evolve into SW-like condensates as $m_{3/2}$ flows to the UV.  In this limit there is an remnant supersymmetry, because when $\Phi$ decouples we recover $\mathcal{N}=1$ SQCD.

The theory does not exhibit a phase transition per se, because the massless spectrum and the global symmetries are the same at any finite $m_{3/2}$. Nevertheless, our study outlines the subtleties innate to the programme of using AMSB as a tool to study of nonsupersymmetric gauge theories. In the small $m_{3/2}$ theory, our solutions for the vacua are exact and the low-energy theory gives no indication that they will change as $m_{3/2}$ is increased. In the UV, we were only able to identify complete SUSY breaking because the finiteness properties of $\mathcal{N}=2$ gauge theories allows instanton corrections to the $\beta$ function to be sensibly defined~\cite{Seiberg:1988ur}. Were we to neglect the instanton corrections, we would erroneously conclude that SUSY is abruptly restored as $m_{3/2}$ crosses the confinement scale. This demonstrates that anomaly mediation couples to conformal violations induced by nonperturbative effects, and that these effects can break symmetries which are preserved to all orders in perturbation theory. 

Our argument for the shift in vacua under the flow of $m_{3/2}$ is somewhat indirect in nature, and it is not clear how one might go about explicitly describing the physics of the transition. In the IR EFT at small $m_{3/2}$, the vacua are stable at all orders in both perturbation theory and the EFT expansion. In the UV, even with knowledge of the instanton induced mass splitting, the adjoint multiplet can be integrated out, and, if we neglect irrelevant operators, the theory simply reduces to $\mathcal{N}=1$ SQCD. For this reason, among others discussed in Section~\ref{sec: discussion NF nonzero}, it appears that the transition occurs as $m_{3/2}$ flows through the regime $\sim\Lambda_{N_F}$. When $m_{3/2}$ exceeds the UV cutoff of the singularity EFT, corrections from adjacent regions must be included. The degrees of freedom that are weakly coupled and pointlike in these adjacent regions are solitons of nonzero spatial extent in the duality frame of the singularity about which we are expanding. To describe what happens in this regime, one must contend with the interactions between these mutually nonlocal degrees of freedom. This nonlocality in softly broken $\mathcal{N}=2$ theories was pointed out in~\cite{_LVAREZ_GAUM__1996} for generic soft breaking, and considerations from holography and phenomenological models of hadronic dynamics like the Lund model~\cite{Andersson:2002ap} or unquenched quark models~\cite{Santopinto_2015} lead us to speculate that such nonlocal interactions may be a feature of asymptotically free gauge theories more generally. This is an innately difficult regime, and it appears to play a crucial role in determining the fate of the anomaly mediated vacua under flows in $m_{3/2}$. It is difficult to imagine that direct calculations in this regime will be possible any time soon. Nonetheless, we are tentatively hopeful that robust results may be possible via a more systematic understanding of the global symmetry breaking patterns induced by AMSB.

After more than 30 years of remarkable results, it is clear that supersymmetric theories and their softly broken cousins still have plenty to teach us about strongly coupled QFT, and anomaly mediated theories are no exception in this regard. Of course there are limitations to this approach. Without a way to derive $m_{3/2}$ independent results, it is risky to extrapolate small $m_{3/2}$ results to large $m_{3/2}$. It has been demonstrated that in some theories, a phase transition occurs~\cite{Bai_2022, Csaki:2022cyg, delima2024sconfiningsusyqcdanomalymediation}, and our work here has shown that anomaly mediation can produce subtle effects which are invisible in perturbation theory yet have implications for the properties of the vacua and the global symmetries. Nevertheless, the immense analytic power of $\mathcal{N}=2$ SUSY afforded us a modest glimpse into these effects, and there is clearly much more to learn.

\section*{Acknowledgments}
We thank Jacques Distler for helpful conversations, and Nancy Orkish for aid in copy-editing. This work was supported in part by
the Natural Sciences and Engineering Research Council of Canada (NSERC). The work of DS was performed in part
at Aspen Center for Physics, which is supported by National Science Foundation grant
PHY-2210452.

\appendix
\section{Canonical normalization and loop level AMSB} \label{sec:IR rescaling}

Here we discuss loop level AMSB and canonical normalization in the near singularity dyon EFT studied in Section~\ref{sec:singularityEFT}.  We will examine the $N_F=0$ monopole theory, but the calculations here generalize straightforwardly. In the UV, canonical normalization of the fields was crucial. One might imagine the same could be true of the IR. This turns out not to be the case. Nevertheless, the procedure raises a few subtleties worthy of commentary. Moreover, the calculation allows us to verify that the vacua found in Section~\ref{sec:singularityEFT} are stable under loop AMSB contributions. 

As in the UV, the goal is to turn terms like $\frac{1}{g_{eff}^2} |\partial_\mu a_D|^2$ into terms like $|\partial_\mu a_D|^2$. This is a subtler procedure in the IR, because $g_{eff}$ is an explicit function of $|a_D|$. Moreover, $g_{eff}\to 0$ at the origin, and its derivative $\frac{\partial g_{eff}}{\partial |a_D|}\to\infty$ in that limit. Nevertheless, we argue that the Lagrangian behaves in a sensible way as we take the $a_D\to0$ limit. 

The fields cannot be simultaneously canonically normalized on the whole of the moduli space. For supersymmetric sigma models, SUSY-preserving field redefinitions are holomorphic coordinate transformations of the K{\"a}hler manifold. Canonically normalizing the kinetic term $h_{i\bar{j}}\varphi^i(\varphi^\dagger)^{\bar{j}}$ is equivalent to setting $h_{i\bar{j}}\to \delta_{i\bar{j}}$ at a point on the target space. Such a coordinate system is  analogous\footnote{Riemann normal coordinates are not holomorphic in general, and so do not manifestly preserve SUSY. Refs.~\cite{Higashijima_2001, Higashijima_2002} define and discuss K{\"a}hler normal coordinates, an analog of Riemann normal coordinates that manifestly preserve the manifold's complex structure.} to Riemann normal coordinates. Just as is it impossible to make a general spacetime metric globally flat, it is also impossible to make the kinetic term of a generic sigma model globally canonical. This is not a cause for concern, it is simply a consequence of the fact that field strengths vary over the moduli space. 

We proceed with the following field redefinitions: \begin{align}
    a_D & \to \widetilde{g}(\widetilde{a}_D) \widetilde{a}_D\\
    A^\mu &\to \widetilde{g}(\widetilde{a}_D) A^\mu \\
    \psi_{a_D}&\to \widetilde{g}(\widetilde{a}_D)\widetilde{\psi}_{a_D} \\
    \lambda& \to \widetilde{g}(\widetilde{a}_D) \widetilde{\lambda}
\end{align}
where $\lambda$ is the $U(1)$ gaugino and $\psi_{a_D}$ is the fermionic superpartner of $a_D$. The coupling constant $\widetilde{g}$ has the same value as the coupling constant $g_{eff}$ at any $a_D=\widetilde{g}(\widetilde{a}_D)\widetilde{a}_D$. We define \begin{align} \label{eq: effective dyon coupling rescaling for appendix}
\widetilde{g}(\widetilde{a}_D)\equiv g_{eff}(\widetilde{g}\widetilde{a}_D) =g_{eff}(a_D).
\end{align}
We distinguish between the two couplings because while the values coincide, the \textit{form} of $\widetilde{g}$ as a function of the rescaled $\widetilde{a}_D$ is not guaranteed to be identical to the form of $g_{eff}$ as a function of $a_D$. Let us determine the former at leading order. Plugging Eq.~\eqref{eq: leading order magnetic coupling} into Eq.~\eqref{eq: effective dyon coupling rescaling for appendix}, we obtain  
\begin{align}
\widetilde{g}(\widetilde{a}_D)^2=g_{eff}(\widetilde{g}\widetilde{a}_D)^2 \sim \frac{1}{\log|\frac{{\Lambda_0}}{\widetilde{g}(\widetilde{a}_D)\widetilde{a}_D}|^2-2}.
\end{align}
where we have neglected the factor of $32\pi^2$ in the numerator, as our analysis here is largely qualitative and the conclusions are unaffected by its omission. We can write this equation in the form \begin{align}\label{eq: equation satisfied by gtilde which lets us use the lambert function trick}
\widetilde{g}^{-2}e^{-\widetilde{g}^{-2}}=e^2\left|\frac{\widetilde{a}_D}{\Lambda_0}\right|^2,
\end{align}
The product logarithm function $W_k$\footnote{Also known as the Lambert $W$ function.} is defined by the relationship \begin{align}
W_k(z)e^{W_k(z)}=z
\end{align}
where $k$ indexes the branches. This defining relationship is of the form of Eq.~\eqref{eq: equation satisfied by gtilde which lets us use the lambert function trick}, with $W_k(z)=-\widetilde{g}^{-2}$ and $z=-e\left|\frac{\widetilde{a}_D}{\Lambda_0}\right|^2$.
The branch $W_{-1}(z)$, defined in the interval $[-\frac{1}{e}, 0)$, has an asymptotic expansion as $z\to 0$ given by \begin{align}
W_{-1}(x)\sim \log(-x)+...
\end{align}
So, as $\widetilde{a}_D\to 0$, our coupling asymptotes to \begin{align} \label{eq:rescaled coupling}
\widetilde{g}(\widetilde{a}_D)^{-2} \sim -\log\left(e^2\left|\frac{\widetilde{a}_D}{\Lambda_0}\right|^2\right) 
\end{align}
and we have \begin{align}
\widetilde{g}(\widetilde{a}_D)^2 = \frac{1}{\log\left(\left|\frac{\Lambda_0}{\widetilde{a}_D}\right|^2  \right)-2}
\end{align}
We see that the function $\widetilde{g}(\widetilde{a}_D)$ is indeed identical to $g_{eff}(a_D)$ at leading order.

Now, consider the kinetic term for $a_D$, schematically of the form \begin{align}
\frac{1}{g_{eff}(a_D)^2} |\partial_\mu a_D|^2 
\end{align}
Canonically normalizing and using our leading log approximation for $\widetilde{g}$, we find
 \begin{align} \label{eq:rescaled kinetic term}
\frac{1}{g_{eff}(a_D)^2} |\partial_\mu a_D|^2 & = \frac{1}{\widetilde{g}_{eff}(\widetilde{a}_D)^2} |\partial_\mu \left(\widetilde{g}_{eff}(\widetilde{a}_D) \widetilde{a}_D\right)|^2,  \nonumber \\
& =|\partial_\mu \widetilde{a}_D|^2 + \frac{1}{\widetilde{g}^2}((\widetilde{g}(\partial_\mu\widetilde{a}_D)\widetilde{a}_D^\dagger (\partial^\mu \widetilde{g})+h.c.)+(\partial_\mu \widetilde{g})^2 |\widetilde{a}_D|^2.
\end{align}
The first term is our canonical kinetic term. Up to the overall constant we neglected, the spacetime derivative of $\widetilde{g}$ is \begin{align}
\partial_\mu \widetilde{g}=\partial \widetilde{g} \partial_\mu \widetilde{a}_D + h.c. \\
\sim \frac{\widetilde{g}^3}{\widetilde{a}_D}\partial_\mu \widetilde{a}_D +h.c.,
\end{align}
and so the overall expression takes the form \begin{align}\label{eq: rescaled kinetic term for ad in IR EFT canonical}
\frac{1}{\widetilde{g}(\widetilde{a}_D)^2} |\partial_\mu \widetilde{a}_D|^2 \sim |\partial_\mu \widetilde{a}_D|^2 \left(1+2\widetilde{g}^2 + 2\widetilde{g}^4\right)
    + \left(\widetilde{g}^2+\widetilde{g}^4\right)
    \left(\frac{\widetilde{a}_D^\dagger}{\widetilde{a}_D}(\partial_\mu \widetilde{a}_D)^2 + h.c. \right)
\end{align}
where we use $\sim$'' instead of ``$=$'' because we have not kept track of the constant factors arising from the $32\pi^2$ in the numerator of the coupling constant. The leading term in $\widetilde{g}$ is a canonical kinetic term. The subleading terms indicate that the field strength runs with the value of $g$, and they contribute nonrenormalizable interactions which are ill-defined in the $\widetilde{a}_D\to 0$ limit. The other terms proceed similarly. We will always end up with canonically normalized terms at the origin, but we will also generically find $\frac{1}{\widetilde{a}_D}$ divergences in the nonrenormalizable interaction terms. 

How do we interpret these divergences? If we want to perform, say, a scattering calculation, we can Taylor expand the action about some fixed background value of $|\widetilde{a}_D|$. Perturbations about this background value are the $\widetilde{a}_D$ ``particles'' that enter our calculation, and for sufficiently small perturbations of $\widetilde{a}_D$, the action takes the form of a convergent power series in the perturbation. The tree level matrix elements can be read off from the expansion coefficients. 

Let us sketch some of the salient features of this expansion. Any term involving $\widetilde{a}_D$ will also involve $\widetilde{g}$. The Taylor expansion of the action will thus contain factors that look something like the following: \begin{align}\label{eq: taylor expansion of gtilde}
\partial^n \bar{\partial}^m \widetilde{g}|_{\widetilde{a}_D=\epsilon} \propto \epsilon^{-n}\epsilon^{\dagger-m} (\epsilon-\widetilde{a}_D)^n(\epsilon^\dagger-\widetilde{a}_D^\dagger)^m. 
\end{align}
where $\epsilon$ is the background value of $\widetilde{a}_D$. The coefficients of this expansion diverge as $\epsilon\to 0$. In order to keep the series under control for small but nonzero $\epsilon$, the deviation $\widetilde{a}_D-\epsilon$ must be kept small compared to $\epsilon$. This means that $\widetilde{a}_D$ essentially sets the scale for perturbative calculations. The running of the coefficients is set by the geometry of the moduli space itself. At the origin, the action becomes nonanalytic and the expansion unavoidably breaks down. It is natural that this is the case, for at the origin all modes have been integrated out and the action is an exact effective action. It cannot be trusted to calculate anything except VEVs.

The Taylor expansion in Eq.~\eqref{eq: taylor expansion of gtilde} bounds the field strength of fluctuations, given by $|\epsilon-\widetilde{a}_D|$, but the $\frac{\partial_\mu \widetilde{a}_D}{\widetilde{a}_D}$ divergences in Eq.~\eqref{eq: rescaled kinetic term for ad in IR EFT canonical} tell us that if we good behaviour as we adiabatically take the background value of $\widetilde{a}_D$ to small values, we must keep $\partial_\mu \widetilde{a}_D\ll \widetilde{a}_D$. In this way, the UV cutoff set by the background value of $\widetilde{a}_D$ is also imposed on momenta. 

This UV cutoff also applies to the other fields in the theory. For example, consider the four fermion interaction proportional to the holomorphic Riemann tensor, as in equation~\eqref{eq:supersymmetric sigma Lagrangian}. The tensor diverges at the origin before rescaling. Rescaling does not remove the divergence. Denoting the tensor by $R$ since it has only one component, it is given in holomorphic normalization by 
\begin{align}
R \sim -\frac{g_{eff}^3}{|a_D|^2}.
\end{align}
Under rescaling, the power of the coupling changes but the factor of $\frac{1}{|\widetilde{a}_D|^2}$ remains.
The meaning of the divergence is to be gleaned from the equations of motion. At the origin, the Euler-Lagrange equations give the exact vacuum field configurations. Near the origin, with a nonzero background value of $\widetilde{a}_D$, we can take solutions to the equations as a lowest order approximation to the fermionic field configurations, then calculate perturbations about this background. By dimensional analysis, $n$th order corrections to matrix elements from this vertex will scale as $|p|^{2n} R^{n+1}\sim \frac{|p|^{2n}}{|\widetilde{a}_D|^{2(n+1)}}$. We see that the Riemann tensor imposes a UV cutoff $\propto|\widetilde{a}_D|$ on the fermion interactions.

We now turn to loop level AMSB. At some point near, but not at, the singularity, we can truncate at the level of renormalizable operators. Then the theory is reasonably approximated as $\mathcal{N}=2$ SQED with the scale $\mu$ identified with $|\widetilde{a}_D|$. As already discussed, the logarithmic derivative $\frac{d\widetilde{g}}{d\log(|\widetilde{a}_D|)}$ reproduces the $\widetilde{g}^3$ form of the 1-loop SQED $\beta$ function. At leading order, the coefficients are thus \begin{align}
m_{\lambda_D} 
\propto\widetilde{g}(|\widetilde{a}_D|)^2m_{3/2} \nonumber \\
m_{a_D}^2 \propto \widetilde{g}(|\widetilde{a}_D|)^4m_{3/2}^2 \nonumber \\
A_{\widetilde{a}_Dm\widetilde{m}} \propto \widetilde{g}(|\widetilde{a}_D|)^2m_{3/2}
\end{align}
where $\lambda_D$ is the dual gaugino. Since SQED is IR-free, the sign of the fermion mass is negative, unlike in the UV theory. But $m_{
\widetilde{a}_D}^2$ remains positive. At fixed $\widetilde{g}$, these terms do behave like an $\mathcal{N}=1$-preserving mass for $\widetilde{A}_D$. They push the theory towards the origin, and so they do not disrupt the stability of the $a_D=0$ solution described in Section~\ref{sec:singularityEFT}. Since they are proportional to $\widetilde{g}$, they vanish at the vacuum. We conclude that loop corrections do not alter the conclusions of Section~\ref{sec:singularityEFT}. 

Naively, our rescaling raises an apparent paradox. We saw in Section~\ref{sec:singularityEFT} that a monopole condensate forms and gives rise to a mass gap, like in~\cite{Seiberg_1994}. But when we canonically normalize the fields, something strange occurs. Let us examine the couplings that give rise to the Higgsing for the $N_F=0$ case. When we throw away the subleading terms in $m_{3/2}a_D$, the leading-order AMSB potential is  
\begin{align}
V = 2|a_D|^2(|m|^2+|\widetilde{m}|^2) + \frac{g_{eff}^2}{2} (2|m\widetilde{m}|^2+(|m|^2-|\widetilde{m}|^2)^2) + g_{eff}^2|m\widetilde{m}-D|^2,
\end{align}
with $D$ the $m_{3/2}$-dependent VEV. When we canonically normalize the fields, we have \begin{align}
V = \widetilde{g}^2\left(2|\widetilde{a}_D|^2(|m|^2+|\widetilde{m}|^2) + \frac{1}{2} (2|m\widetilde{m}|^2+(|m|^2-|\widetilde{m}|^2)^2) + |m\widetilde{m}-D|^2 \right).
\end{align}
Similarly, the gauge covariant derivatives $\mathcal{D}_\mu m=(\partial_\mu - i A_\mu)m$ under rescaling become $(\partial_\mu - i\widetilde{g}A_\mu)m$; The point is that all the interaction terms through which fields in this theory can be Higgsed are proportional to factors of $\widetilde{g}$. This is true for the theory with AMSB just as well as with the SW-deformed theory.  When $\widetilde{a}_D=0$, as it is in the vacuum configuration, the mass gap appears to vanish! 

This is not the case. The explanation lies in the relative scaling of $\widetilde{a}$ vs $\widetilde{g}$.  As we have just discussed, perturbative calculations away from the origin can be carried out by expanding about some background field. Yet these Taylor expansions involve derivatives of $\widetilde{g}$, which grow as $\frac{1}{\widetilde{a}_D}$ as $\widetilde{a}_D\to 0$. For the expansion to converge, invariant momenta must be bounded by the background value of $|\widetilde{a}_D|$, but the mass gap is proportional to $m_{3/2}\widetilde{g}$. Up to some finite background value $|\widetilde{a}_D|\sim m_{3/2}\widetilde{g}$, the upper bound on momenta set by the background field is smaller than the gap set by that background field. For this reason, the theory develops an IR cutoff. This is how the mass gap retains a finite size under rescaling. Since the mass gap is finite, any process involving an on-shell particle will occur at or above the scale of this gap, and $\widetilde{g}$ will be nonzero. For this reason, the AMSB loop corrections \textit{do} affect the spectrum of the EFT, even though they do not alter the vacuum state itself.

\section{On the critical points of Eq.~\eqref{eq:CoulombAMSB}} \label{sec:convexK}
In this appendix, we analyze the tree level SUSY-breaking potential from Eq.~\eqref{eq:CoulombAMSB}. Leveraging the structure of the Picard-Fuchs equations Eq.~\eqref{eq:picard fuchs}, we show that $V$ has a local minimum on the punctured $u$-plane if and only if that point is a saddle point for $K$. This means that if $K$ is real convex, the Coulomb branch is lifted and the only candidate vacua are the singularities.

Real convexity is a property that can change under K{\"a}hler transformations, so taken at face value this statement is somewhat vacuous. The key point is that the AMSB formulae given in equation~\eqref{eq:treeB} also fail to be K{\"a}hler invariant (see e.g. Ref~\cite{Pomarol_1999} for some discussion of this matter). For example, given a canonical K{\"a}hler potential $K=Q^\dagger Q$, under a K{\"a}hler transformation $K\to K + f(Q)+f^\dagger(Q^\dagger)$, the $|m_{3/2}|^2$ part of Eq.~\eqref{eq:VAMSB when W=0} will transform as $0 \to |m_{3/2}|^2(\frac{df}{dq}\frac{df\dagger}{dq^\dagger}+\frac{\partial K}{\partial q}\frac{df^\dagger}{dq^\dagger} + \frac{\partial K}{\partial q^\dagger}\frac{df}{dq}-(f(q)+f^\dagger(q^\dagger)))$, which will not vanish in general.  When the theory is classically conformal, these formulae should simply return a constant. Thus, we take the correct gauge to be the one in which the behaviour of the K{\"a}hler potential asymptotes towards its classical form, without any extraneous (anti)holomorphic components.   

If we want to analytically find a local extremum of $V_{\text{AMSB}}$, we want to set $\partial V_{\text{AMSB}} = \bar{\partial}V_{\text{AMSB}} = 0$, where $\partial \equiv \frac{\partial }{\partial u}$.
Taking the holomorphic gradient of $V_{\text{AMSB}}$ from Eq.~\eqref{eq:VAMSB when W=0} and using $h=\partial \bar{\partial}K$, we have \begin{align}
\partial V_{\text{AMSB}} = |m_{3/2}|^2\left\{\frac{\partial^2 K \bar\partial K}{\bar\partial\partial K}-\frac{\partial K \bar\partial K}{(\partial\bar\partial K)^2}\bar\partial \partial^2 K \right\}.
\end{align}
 But from Eqs.~\eqref{eq: generic form of coulomb branch kahler potential} and~\eqref{eq:picard fuchs}, we have that \begin{align} \label{eq: second derivative of K is K/p}
\partial^2K = \frac{i}{8\pi}(a^\dagger\partial^2a_D - a_D^\dagger\partial^2a)=\frac{1}{p_{N_F}(u)}K.
\end{align}
where we have also leveraged the fact that $a$ and $a_D$ are holomorphic so that $\partial \bar{a}=\partial\bar{a}_D=0$.
Similarly, $\partial^2\bar{\partial}K = \frac{1}{p_{N_F}(u)}\bar\partial K$. We can also readily conjugate these formulae. Then at a stationary point, the holomorphic component of the gradient of $V_{\text{AMSB}}$ may be written as \begin{align}
\partial V_{\text{AMSB}} = -\frac{V_{\text{AMSB}} \bar\partial K}{p_{N_F}(u) h}.
\end{align}
 By definition, $h$ and $p_{N_F}$ are nonzero and finite away from the singularities, so for a stationary point we have the condition \begin{align}
V_{\text{AMSB}}\bar\partial K = 0, \\
V_{\text{AMSB}} \partial K = 0.
\end{align}
Hence, we have stationary points only at $V_{\text{AMSB}} = 0$ or $\partial K = \bar\partial K=0$. Away from the singularities, $V_{\text{AMSB}}$ is simply the product of a collection of holomorphic and antiholomorphic functions, namely $a, a_D$, and their derivatives and conjugates. Therefore $V_{\text{AMSB}}$ is real analytic and we use the identity theorem to rule out $V_{\text{AMSB}}=0$ at a crticial point as follows. Since $\partial V_{\text{AMSB}} \propto V_{\text{AMSB}}$ (and similar for $\bar\partial$), we have that $\partial^n \bar{\partial}^m V_{\text{AMSB}}\propto V_{\text{AMSB}}$. So if $V_{\text{AMSB}}$ were 0 at some point, this would imply that all its derivatives are 0 at that point, meaning that this is an accumulation point. By the identity theorem, this can only be true if $V_{\text{AMSB}}$ is uniformly 0 on any simply connected domain. But this is not the case. So, we are left to consider stationary points of the K{\"a}hler potential itself.

Suppose we have found such a stationary point at some coordinate $u_0$. To determine whether it is a minimum, let us perform a second derivative test. Our complex and real coordinates are related by $z = x+iy$ and the real Hessian is \begin{align}
    H_{r}=\begin{bmatrix}
    \partial_x^2 & \partial_x\partial_y \\
    \partial_y\partial_x & \partial_y^2
    \end{bmatrix}.
\end{align}
If the determinant of the real Hessian at a critical point is positive, then the point is an extremum. If it is negative, it is a saddle point.
Given the complex Hessian \begin{align}
H_c = \begin{bmatrix}
\partial_z\partial_{\bar{z}} & \partial_z^2 \\
\partial_{\bar{z}}^2 & \partial_{\bar{z}}\partial_z
\end{bmatrix},
\end{align}
we can relate it to the real Hessian via the following Hermitian conjugacy: \begin{align}
H_{r} \propto J^\dagger H_c J \\
J = \begin{bmatrix}
1 & i \\
1 & -i
\end{bmatrix}.
\end{align}
We can straightforwardly relate the determinant of $H_{r}$ to that of $H_c$. We have \begin{align}
\det(H_{r}) = \frac{1}{4}\det(J^\dagger)\det(H_c)\det(J) = \frac{\det(H_c)}{4}. 
\end{align}
We see that the sign of $\det(H_c)$ matches that of $\det(H_r)$. If our $H_c$ for $V_{\text{AMSB}}$ has negative determinant, the point is a saddle. If it has positive determinant, the point is a local extremum.

Evaluating the complex Hessian of the scalar potential, then imposing that the first derivatives of $K$ vanish, we have\begin{align}
H_c(V_{\text{AMSB}})=|m_{3/2}|^2\begin{bmatrix}
\frac{|\partial^2K|^2}{h} & \partial^2 K \\
\bar{\partial}^2 K &  \frac{|\partial^2K|^2}{h}
\end{bmatrix}.
\end{align}
Its determinant is \begin{align}
\det(H_c(V_{\text{AMSB}})) = |m_{3/2}|^4|\partial^2 K|^2\left( \frac{|\partial^2K|^2}{h^2}-1\right).
\end{align}
The determinant of the Hessian of $K$ is \begin{align}
\det(H_c(K)) = h^2-|\partial^2K|^2.
\end{align}
If our point is a minimum of $K$, then \begin{align}
h^2>|\partial^2 K|^2,
\end{align}
and hence $\det(H_c(V_{\text{\bcancel{SUSY}}}))<0$ and the point is a saddle of our potential. Conversely, if it is a saddle of $K$, then \begin{align}\label{eq: appendix, bound on h wrt d2k}
h^2<|\partial^2 K|^2 = \left|\frac{K}{p}\right|^2,
\end{align}
where the equality on the right hand side follows from Eq.~\eqref{eq: second derivative of K is K/p}
and so $\det(H_c(V_{\text{\bcancel{SUSY}}}))>0$, giving us a local extremum. This local extremum must be a minimum, as we now show. 

We will consider small changes in $V$ induced by small displacements about such a point and show that this change must be positive in every direction. 
Consider a perturbation of $V$ about the critical point. The lowest order nonvanishing term is \begin{align}\label{eq: displacement from critical point on coulomb branch}
\frac{dV_{\text{AMSB}}}{|m_{3/2}|^2} &\approx dz^2 H_{zz} + d\bar{z}^2 H_{\bar{z}\bar{z}} +2 dz d\bar{z} H_{z\bar{z}} \\
& = dz^2 \frac{K}{p} + d\bar{z}^2 \frac{K}{\bar{p}} +2 |dz|^2 \frac{K^2}{|p|^2 h} \\
&=2\text{Re}\left(\frac{K}{p}dz^2\right)+\frac{2}{h}\left|dz\frac{K}{p}\right|^2
\end{align}
We don't in general know the sign of $\text{Re}(\frac{K}{p}dz^2)$. But we can show that the strictly positive term is always larger. We can write \begin{align}
\frac{2}{h}\left|dz\frac{K}{p}\right|^2=\left|\frac{\partial^2K}{h}\right| \left|2\frac{K}{p}\right||dz|^2.
\end{align} 
and from Eq.~\eqref{eq: appendix, bound on h wrt d2k}, we have that $\left|\frac{\partial^2K}{h}\right|^2>1$. In general, \begin{align}\left|2\frac{K}{p} \right||dz|^2\geq| 2 \text{Re}(dz^2\frac{K}{p})|.\end{align} 
Therefore, 
\begin{align}
\frac{2}{h}\left|dz\frac{K}{p}\right|^2> \left|2\frac{K}{p} \right||dz|^2\geq\left|2\text{Re}(\frac{K}{p}dz^2)\right|,
\end{align}
We see that even when the first term in Eq.~\eqref{eq: displacement from critical point on coulomb branch} is negative, the second term, which is positive definite, is guaranteed to to be larger. It follows that a saddle of $K$ is a local minimum of $V_{\text{AMSB}}$.

We thus conclude that $V_{\text{AMSB}}$ has a minimum on the smooth region of the Coulomb branch if and only if the K{\"a}hler potential has a saddle point, while an extremum of $K$ is a saddle point in $V_{\text{AMSB}}$.

Unfortunately, the strategy of eliminating second derivatives via the Picard-Fuchs equation does not appear to be sufficient to analytically rule out the possibility of a saddle point in the K{\"a}hler potential. A saddle point occurs when \begin{align}\label{eq: saddle point condition on Hess(K)}
h^2<\frac{|K|^2}{|p|^2}.
\end{align} 
The left hand side of this equation depends on first derivatives of $(a_D(u), a_D(u))$, and their conjugates. The first derivatives obey the same Picard-Fuchs equations with the same monodromies, but they define a different section of the $SL(2, \mathbb{Z})$ bundle over the $u$-plane, and this bound cannot be established via the structure of the Picard-Fuchs equations alone. Perhaps there exists a simple way to analytically show that the inequality Eq.~\eqref{eq: saddle point condition on Hess(K)} does not hold anywhere, but we will not pursue it in this work.

\section{AMSB on the Higgs branch for $B\gg m_{3/2}$} \label{sec: higgs branch involved derivation}

In this appendix, we argue that anomaly mediation on the Higgs branch vanishes when we neglect its coupling to the Coulomb branch moduli at the origin. This is in contrast to the Coulomb branch, where even for small $m_{3/2}$, we find a scalar potential that slopes toward the strong-coupling region. The coupling to the Coulomb moduli is the primary subject of Section~\ref{sec:Higgs branch}.  

On general grounds, there is no superpotential on the Higgs branch~\cite{Tachikawa_2015}. The anomaly mediated potential is therefore of the form of Eq.~\eqref{eq:VAMSB when W=0}. To be explicit, we have \begin{align}\label{eq: VAMSB when W=0 for Higgs branch}
V_{\text{AMSB}}=|m_{3/2}|^2( \partial_iK\bar{\partial}_{\bar{j}} K h^{i\bar{j}}-K)
\end{align}
where the partial derivatives are being taken with respect to the Higgs branch moduli. As with the Coulomb branch, one could consider loop corrections to the effective action obtained by Taylor expanding about a particular vacuum, but these loop corrections cannot change the vacua.

We now argue that Eq.~\eqref{eq: VAMSB when W=0 for Higgs branch} must simply evaluate to 0 on the Higgs branch.  There is a nonrenormalization theorem for the the Higgs branch in $\mathcal{N}=2$ theories~\cite{Seiberg_1994_II, Argyres:1996eh}, and it is crucial to the analysis of this section. 
This nonrenormalization theorem states that the Higgs branch Kahler metric recieves no quantum corrections at all. At an intuitive level, we can argue that $V_{\text{AMSB}}=0$ purely on the basis of this absence of quantum corrections. Since we have set the bare hypermultiplet masses to 0, the UV Lagrangian is classically conformal. Normally, quantum corrections break this conformal invariance and introduce a scale $\Lambda_{N_F}$ to the theory. But on the Higgs branch, there are no quantum corrections, nor are there any classically dimensionful parameters. Thus, there is no conformal anomaly for AMSB to couple to. 

We now make a more precise argument showing that Eq.~\eqref{eq: VAMSB when W=0 for Higgs branch} must be 0. The core idea is that the Higgs branch is simply a subspace of the unconstrained field space, that is, the flat $\mathbb{C}^{2N_F}$ defined by the canonical Kahler potential $K=Q^\dagger_i Q_i + \widetilde{Q}^\dagger_i \widetilde{Q}_i$. The Higgs branch AMSB potential Eq.~\eqref{eq: VAMSB when W=0 for Higgs branch} is just the scalar function one gets by the evaluating Eq.~\eqref{eq: VAMSB when W=0 for Higgs branch} on the unconstrained manifold $\mathbb{C}^{2N_F}$, and restricting its domain to the subspace\footnote{There is not a unique such subspace. Since we quotient the vacuum field configurations by the gauge group, any section of the orbit under the gauge group will do.} that comprises the Higgs branch. Since the unconstrained manifold has a canonical K{\"a}hler potential, Eq.~\eqref{eq: VAMSB when W=0 for Higgs branch} evaluates to 0. This is true on both the unconstrained manifold and the Higgs branch.

It is intuitive that Eq.~\eqref{eq: VAMSB when W=0 for Higgs branch} on the Higgs branch is simply the restriction of the same expression evaluated on $\mathbb{C}^{2N_F}$, but it is not immediately obvious that this intuition is correct. The function in question is somewhat geometric in character, depending on the metric and the derivatives of the K{\"a}hler potential as well as $K$ itself. It vanishes on flat K{\"a}hler manifolds but not necessarily on curved K{\"a}hler manifolds. The classical moduli space, despite being a subspace of a flat space, is not guaranteed to be flat in and of itself.
Nonetheless, using some general properties of K{\"a}hler quotients, we can show that indeed $V_{\text{AMSB}}$ on the moduli space is simply inherited from the same expression evaluated on the unconstrained canonical manifold, the ``ambient space" in which the moduli space is embedded. 

Let us first review the manner in which classical moduli spaces are constructed in $\mathcal{N}=1$ supersymmetric theories. We will generically assume there is no superpotential, and we will show that on these classical spaces, Eq.~\eqref{eq: VAMSB when W=0 for Higgs branch} evaluates to 0. Normally, the moduli space experiences quantum corrections and this result is not particularly reflective of the real behaviour of the theory, but since $\mathcal{N}=2$ Higgs branches are protected from quantum corrections, the classical calculation and the quantum calculation are one and the same. The construction of $\mathcal{N}=2$ moduli spaces is slightly more elaborate than $\mathcal{N}=1$ because of the $F$-flatness conditions required by $\mathcal{N}=2$ supersymmetry, but it will be straightforward to adapt the $\mathcal{N}=1$ argument to the $\mathcal{N}=2$ case.

There are a number of equivalent ways to characterize a classical moduli space. For our purposes, the following construction will suffice: starting with the unrestricted manifold $M$ specified by the canonical UV K{\"a}hler potential, one restricts to the submanifold $M_0$ satisfying the $D$-flatness condition, then quotients by the gauge group $G$ to obtain the moduli space, $\mathcal{M}=\frac{M_0}{G}$. The effective K{\"a}hler potential on $\mathcal{M}$ is then just the canonical K{\"a}hler potential on $M$, restricted to an appropriate subspace and expressed in terms of gauge singlets \cite{Antoniadis_1997}.\footnote{This statement does not quite hold in the presence of Fayet-Iliopolis terms, although it generalizes straightforwardly. Since our gauge group is $SU(2)$, no such terms are present.}

A simple example of this construction is $\mathcal{N}=1$ SQED with a single flavour $(q, \widetilde{q})$. The unconstrained UV manifold $M$ is equal to $\mathbb{C}^2$ and is described by a canonical K{\"a}hler potential $K= |q|^2+|\widetilde{q}|^2$. The $D$-term constraint imposes $|\widetilde{q}|=|q|$. The manifold $M_0$ is the submanifold $\mathbb{C}$ of $
M$ on which this constraint is satisfied. To finally get the low-energy moduli space $\mathcal{M}$, we must quotient by the $U(1)$ gauge group. This is immediately achieved if we directly work in terms of gauge invariant holomorphic monomials $\xi = \widetilde{q}q$. The low-energy K{\"a}hler potential $K'$ can then be written \begin{align}
    K' = 2\sqrt{\xi^\dagger \xi},
\end{align}
The manifold associated with this K{\"a}hler potential is $\mathbb{C}/\mathbb{Z}_2$. This $\mathbb{Z}_2$ quotient is a result of the gauge quotient. To see why, we can write $q =ve^{i\varphi} e^{i\beta}$ and $\widetilde{q} = ve^{i\varphi}e^{-i\beta}$. We can think of this as a parameterization of the $D$-flat manifold $M_0=\mathbb{C}$. Performing a gauge transformation simply changes the angle $\beta$. It leaves $\xi$ invariant, but it acts on $q$ and $\widetilde{q}$ as \begin{align}
q & \to q e^{i\alpha} \nonumber \\
\widetilde{q} & \to \widetilde{q}e^{-i\alpha}.
\end{align}
If we take $\alpha = \pi$, we take $q, \widetilde{q} \to -q, -\widetilde{q}$. This coincides with  a shift of the coordinate $\beta$ by $\pi$, and so coordinates on $M_0$ related by a negative sign are to be identified when passing to $\mathcal{M}$. This is automatically built into the description in terms of $\xi$, since $\xi = v^2 e^{2i\varphi}$.

One can readily check that plugging $K'$ into Eq.~\eqref{eq: VAMSB when W=0 for Higgs branch} yields 0. Of course, in this example, the low-energy manifold is locally flat and one may wonder if the vanishing of the AMSB potential is simply a consequence of this flatness. In Seiberg-Witten theories, the $N_F=2$ Higgs branch is a flat cone  $\frac{\mathbb{C}^2}{\mathbb{Z}_2}$ \cite{Seiberg_1994_II}, but the $N_F=3$ case is significantly more complicated. We now quote some general results from~\cite{Antoniadis_1997} that we can use to demonstrate that this result is not special to theories with flat moduli spaces, and in fact $V_{\text{AMSB}}=0$ on the classical moduli space of \textit{any} classically conformal theory.

Using primes to denote the K{\"a}hler potential, the metric, and the AMSB potential on the low-energy moduli space $\mathcal{M}$, we have \begin{align} \label{eq: differential part of AMSB potential}
V_{\text{AMSB}}'\propto\partial_i K' \bar{\partial}_{\bar{j}} K'h'^{i\bar{j}}-K'
\end{align}
As we briefly stated earlier, $K'$ is a restriction of $K$, where $K$ is the canonical UV K{\"a}hler potential. Our task is then to argue that the other term  $\partial_i K' \bar{\partial}_{\bar{j}} K'h'^{i\bar{j}}$ is also a restriction of its UV analog on $M$ to $\mathcal{M}$. 

Take an atlas of the moduli space. The $\partial_i K', \bar{\partial}_{\bar{j}}K'$ define vector fields on the charts making up this atlas. Denoting them more abstractly by $X', \bar{X}'$, we can expand this expression as \begin{align}
h'(X', \bar{X}') & = g'(X', \bar{X}') + i\omega'(X', \bar{X})' \nonumber \\&
=g'(X', \bar{X}') 
\end{align}
where $g'$ is the K{\"a}hler manifold's Riemannian metric and $\omega'$ its symplectic form. The second line follows from the fact that the first term in Eq.~\eqref{eq: differential part of AMSB potential} is real. From Ref~\cite{Antoniadis_1997}, the scalar function $g'(X', \bar{X}')$ pulls back to $M_0$ straightforwardly: $X'$ and $\bar{X}'$ pull back to vectors $X_0$ and $\bar{X}_0$ that are orthogonal to the killing vectors associated with the action of $G$. Then \begin{align}
    g_0(X_0, \bar{X}_0)=g'(X', \bar{X}'),
\end{align}
where $g_0$ is the restriction of the metric on $M$ to $M_0$. On $M_0$, the K{\"a}hler potential and the K{\"a}hler metric are just the restrictions to $M_0$ of their analogs on $M$. The authors of~\cite{Antoniadis_1997} note that the pullback of $g'$ to $M_0$ is \textit{not} simply the restriction of $g_0$ to $\mathcal{M}$, for the pullback of $g'$ is degenerate along the action of $G$. However, this is not a problem. In an explicit coordinate system, one might imagine that the restriction of $g_0$ to $\mathcal{M}$ would contain extra terms absent in $g'$ associated with variations along the action of $G$. But we know that the metric $g_0$ acting on the (anti)holomorphic gradients of $K_0$ simply returns $K_0$, since both are simply the restrictions of $g$ and $K$ on the unconstrained manifold. Since $K_0$ is invariant under the action of the symmetry group, these additional terms must simply evaluate to 0. It follows that the value of the AMSB potential at a point on the unconstrained moduli space is the value of the AMSB potential on the image of that point under the K{\"a}hler quotient. In plainer language, if $V_{\text{AMSB}}=0$ on the unconstrained canonical space $M$, it immediately follows that $V'_{\text{AMSB}}=0$ on the moduli space. 

We have thus demonstrated that Eq.~\eqref{eq:VAMSB when W=0} evaluates to 0 on the classical moduli space of an $\mathcal{N}=1$ theory with canonical UV K{\"a}hler potential and $W=0$. How does the story change for $\mathcal{N}=2$ theories? They are more constrained than $\mathcal{N}=1$ theories, because $\mathcal{N}=2$ SUSY forces us to include a superpotential giving rise to a particular set of $F$-flatness conditions alongside the $D$-flatness condition. But the $F$-flatness conditions closely resemble the $D$-flatness conditions, and the restriction to $F$-flat subspaces is closely analogous to the restriction to $D$-flat subspaces which we already discussed. See~\cite{Antoniadis_1997} for a detailed discussion. The manifold $M_0$ is defined as the intersection of the $D$ and $F$-flat subspaces of $M$. But \cite{Antoniadis_1997}, the $F$-flat subspace $M_h$ is simply a complex submanifold of $M$ which inherits $M$'s K{\"a}hler structure, similar to how in the $\mathcal{N}=1$ case, the $D$-flat submanifold $M_0$ inherits the metric and K{\"a}hler potential from the canonical $M$. The potential $V_{\text{AMSB}}$ is 0 on $M_h$ just as on $M$, and from there, we can apply the argument we have already laid out. 

Thus, for both $N_F=2$ and $N_F=3$, the AMSB potential on the Higgs branch evaluates to 0. More generally, this result suggests that Eq.~\eqref{eq:VAMSB when W=0} will evaluate to 0 on any $\mathcal{N}=2$ Higgs branch (since hypermultiplet masses preclude massive hypermultiplets from having a nonzero VEV, effectively reducing the canonical UV manifold to the subspace spanned only by the massless hypermultiplets). From the perspective of dimensional analysis, this is not necessarily surprising. The canonical K{\"a}hler potential is conformal, and since the low-energy K{\"a}hler potential is just the same potential restricted to some subspace and quotiented by a group action, no dimensionful constant enters the description. 

We reiterate that this does not mean that the Higgs branch is completely unaffected by AMSB. Rather, it says that the effects are communicated to the Higgs moduli entirely through their coupling to the $\mathcal{N}=2$ vector multiplet at the origin.

\section{Hypermultiplet masses and AMSB} \label{sec: bare masses for hypermultiplets and AMSB}

Here, we note that when the bare hypermultiplet masses in Eq.~\eqref{eq: N=2 hypermultiplet superpotential} are turned on, the preservation of $\mathcal{N}=1$ SUSY that we saw in the massless theories does not occur. From Eq.~\eqref{eq:treeB} one finds the following SUSY-breaking AMSB  potential at tree level: \begin{align}
V_{\text{tree}} = -m_{3/2}\sum_i M^i \widetilde{q}^i q^i.
\end{align}
This potential behaves very differently depending on the phases of $m_{3/2}$ and $M^i$. If both parameters are real and $m_{3/2}>M_i>0$, the squarks become tachyonic. It is possible that this could lead to a runaway potential. 

We do not investigate the massive theories in any more detail, but it is interesting that mass terms for the hypermultiplets cause AMSB to break SUSY down to $\mathcal{N}=0$ at tree level. Classically, the conserved $U(1)_R$ current sits in a SUSY multiplet with the conserved conformal current. Instantons break the $U(1)_R$ symmetry~\cite{Argyres:1996eh}. The $U(1)_R$ anomaly is $1$-loop exact~\cite{Seiberg:1988ur}, and its anomaly sits in a multiplet with the 1-loop $\beta$ function, but instantons destroy this simple relationship~\cite{Seiberg:1988ur,Arkani_Hamed_1998}. Since mass terms explicitly destroy both symmetries, this suggests that the preservation of $\mathcal{N}=1$ SUSY in perturbation theory is closely tied to the supersymmetric relationship between $U(1)_R$ and conformal anomalies. 
These observations are hardly concrete, but they hint that it may be fruitful to explicitly examine AMSB through the lens of its coupling to the supercurrent anomaly multiplet. 

\section{Derivation of AMSB potential} \label{sec:AMSB derivation}
In this appendix, we derive the standard anomaly mediated formulae~\cite{Giudice_1998, Randall_1999, Pomarol_1999, Jung_2009} that we use in Section~\ref{sec:AMSB prelims} via the conformal compensator formalism.

\subsection{Tree level}
At tree level, we multiply the K{\"a}hler and superpotentials by whatever powers of the conformal compensator $\chi = 1+\theta^2F_{\chi}$ and its conjugate are needed to spuriously restore super-Weyl invariance. This gives us \begin{align}
\int d^4\theta K& \to \int d^4\theta \chi^\dagger \chi K, \nonumber \\
\int d^2\theta W& \to \int d^2\theta \chi^3W.
\end{align}
For a generic K{\"a}hler potential as a function of some chiral superfields $Q_i, Q^\dagger_i$, we can work out the bosonic component by Taylor expanding about a point $p$ in field space: \begin{align}
K(Q^i, Q^{j\dagger})& = K|_{p}+\frac{\partial K}{\partial Q^i}|_p Q^i + c.c. + \frac{\partial ^2K}{\partial Q^i\partial Q^{j\dagger }}|_p Q^i Q^{j \dagger} + ... \nonumber \\&
= K|_{p}+\frac{\partial K}{\partial Q^i}|_p (q^i+\theta^2 F_q^i) + c.c. + \frac{\partial ^2K}{\partial Q^i\partial Q^{j\dagger }}|_p (q^i+\theta^2 F_q^i)(q^{j\dagger}+\bar\theta ^2F_q^{j\dagger }) + ... 
\end{align}
Fixing the complex part of $K$ to the expansion point $p$ but allowing the Grassmannian coordinate to vary, we have \begin{align}
K(Q^i, Q^{j\dagger}) = K|_{p}+\frac{\partial K}{\partial Q^i}|_p (\theta^2 F_q^i) + c.c. + \frac{\partial ^2K}{\partial Q^i\partial Q^{j\dagger }}|_p (\theta^2 F_q^i)(\bar\theta ^2F_q^{j\dagger }).
\end{align}
There are no other nonzero terms in this expansion. Now with the conformal compensator plugged in, we have \begin{align}
\int d^4\theta \chi^\dagger \chi K = \int d^4\theta  \left(K|_{p}+\frac{\partial K}{\partial Q^i}|_p (\theta^2 F_q^i) + c.c. + \frac{\partial ^2K}{\partial Q^i\partial Q^{j\dagger }}|_p (\theta^2 F_q^i)(\bar\theta ^2F_q^{j\dagger })\right)\nonumber\\\times(1+\theta^2F_\chi + c.c. + \theta^2\bar\theta^2 |F_\chi|^2).
\end{align}
This gives rise to new SUSY-breaking terms \begin{align}
\int d^4\theta \chi^\dagger \chi K  = ...+\frac{\partial K}{\partial q^i}F_q^i F_\chi^\dagger + c.c. + K(q^i, q^{j\dagger})|F_\chi|^2.
\end{align}
The superpotential is comparatively trivial. Expanding it in a similar fashion, we have \begin{align}
W(Q^i) = W(q^i)|_p +\frac{\partial W}{\partial q^i}|_p \theta^2 F_q^i 
\end{align}
Introducing the conformal compensator, we get \begin{align}
\int d^2\theta \chi^3 W = ...+3W(q^i) F_\chi.
\end{align}
In the absence of supersymmetry breaking, the auxiliary fields are given by \begin{align}
F_q^{i\dagger} = -\partial_k W h^{kj^*},
\end{align}
where we have introduced the usual index notation for the K{\"a}hler manifold, with $h^{kj^*}$ the inverse K{\"a}hler metric. When SUSY breaking is turned on, the auxiliaries instead become \begin{align}
F_q^{j\dagger} = -(\partial_k W +F_{\chi}\partial_kK) h^{kj^*}. 
\end{align}
Plugging these auxiliaries back into the Lagrangian, we recover Eq.~\eqref{eq:treeB}.
\subsection{Loop level}

\subsubsection{Scalar masses}
Consider a canonical K{\"a}hler potential \begin{align}
K =\int d^4\theta \mathcal{Z}(\mu)Q^\dagger Q.
\end{align}

To spuriously preserve super-Weyl invariance, we promote $\mathcal{Z}$ to a superfield 
$\mathcal{Z}(\mu)\to \mathcal{Z}\left(\mu/\sqrt{\chi^\dagger \chi}\right)$. 
Then our K{\"a}hler potential picks up additional $D$-terms, which we may find by Taylor expanding $\mathcal{Z}$ in $\chi$. \begin{align}
& \int d^4\theta \,\mathcal{Z}\left(\frac{\mu}{\sqrt{\chi^\dagger\chi}}\right) Q^\dagger Q \nonumber \\&= \int d^4\theta(\mathcal{Z}|_{\chi=1}+\frac{\partial\mathcal{Z}}{\partial \chi}|_{\chi=1}\theta^2 F_{\chi} + c.c. + \frac{\partial^2 \mathcal{Z}}{\partial\chi^\dagger \partial\chi}|_{\chi=\chi^\dagger =1}\theta^2\bar\theta^2|F_\chi|^2) Q^\dagger Q.
\end{align}
To evaluate the partial derivatives, we can write 
\begin{align}
\frac{\partial \mathcal{Z}\left(\frac{\mu}{\sqrt{\chi^\dagger\chi}}\right)}{\partial \chi} &= 
\frac{\partial \mathcal{Z}}{\partial \mu}
\frac{\partial \mu}{\partial \left(\frac{\mu}{\sqrt{\chi^\dagger\chi}}\right)}
\frac{\partial \left(\frac{\mu}{\sqrt{\chi^\dagger\chi}}\right)}{\partial\chi}\nonumber \\&
=-\frac{1}{2\chi}\frac{\partial \mathcal{Z}}{\partial \log\mu}.
\end{align}
Then the K{\"a}hler potential is \begin{align}
\int d^4\theta \left(\mathcal{Z}|_{\chi=1}-\frac{1}{2\chi}\frac{\partial \mathcal{Z}}{\partial \log\mu}|_{\chi=1}\theta^2F_{\chi}+c.c. + \frac{1}{4|\chi|^2}\frac{\partial^2\mathcal{Z}}{\partial \log\mu ^2}|_{\chi=1}\theta^2\bar{\theta}^2 |F_\chi|^2\right) Q^\dagger Q.
\end{align} 
We may normalize our fields by dividing this K{\"a}hler potential by $\mathcal{Z}|_{\chi=1}$. Doing this and evaluating the Grassmann integral, we get \begin{align}
\int d^4\theta \frac{K}{\mathcal{Z}|_{\chi=1}} = ...-\frac{1}{2}(\gamma q F_{\chi} F_q^\dagger + c.c.)+\frac{1}{4\mathcal{Z}|_{\chi=1}}\frac{\partial^2\mathcal{Z}}{\partial \log\mu ^2}|_{\chi=1}|F_\chi|^2 |q|^2,
\end{align}
where the ... contains the usual supersymmetric Lagrangian. 
Solving for the $F_q$, one finds that they are given by \begin{align}
F_q = \frac{1}{2}\gamma q F_{\chi}-\frac{\partial W^\dagger}{\partial q^\dagger}.
\end{align}
Plugging in and simplifying, the remaining SUSY-breaking term is \begin{align}
V_{\text{scalar mass}}&= -\left(\frac{1}{4\mathcal{Z}|_{\chi=1}}\frac{\partial^2\mathcal{Z}}{\partial \log\mu ^2}|_{\chi=1} -\frac{1}{4} \gamma^2\right)|F_\chi|^2 |q|^2\\&
=-\frac{1}{4}\dot\gamma |F_\chi|^2|q|^2,
\end{align}
giving rise to the scalar mass in Eq.~\eqref{eq:softmass}.

\subsubsection{Trilinear terms}
Now, suppose the superpotential has a Yukawa term. The bare superpotential, given by \begin{align}
W = y^0_{ijk}Q^iQ^jQ^k,
\end{align} is subject to the nonrenormalization theorem. When we renormalize, we instead have \begin{align}
W = y_{ijk}(\mathcal{Z}^i\mathcal{Z}^j\mathcal{Z}^k)^{1/2} Q^i Q^j Q^k.
\end{align}
We are being sloppy with our indices in the above equation, since $\mathcal{Z}^i$ and $Q^i$ share an index. Imbuing the $\mathcal{Z}^i$ with a nonzero Grassmannian component, we have \begin{align}&
y_{ijk}(\mathcal{Z}^i\mathcal{Z}^j\mathcal{Z}^k)^{1/2} Q^i Q^j Q^k \nonumber \\&= y_{ijk}\left((\mathcal{Z}^i\mathcal{Z}^j\mathcal{Z}^k)^{1/2}|_{\chi=1}+\frac{d\sqrt{\mathcal{Z}_i\mathcal{Z}_j\mathcal{Z}_k}}{d\chi}|_{\chi=1}\theta^2F_\chi +...\right)^{1/2} Q^i Q^j Q^k.
\end{align}
The partial derivative is \begin{align}
\frac{d\sqrt{\mathcal{Z}_i\mathcal{Z}_j\mathcal{Z}_k}}{d\chi}& = -\frac{\mu}{2\chi^\dagger}(\mathcal{Z}_i\mathcal{Z}_j\mathcal{Z}_k)^{-1/2}\left(\frac{d\mathcal{Z}_i}{d\mu}\mathcal{Z}_j\mathcal{Z}_k+\mathcal{Z}_i\frac{d\mathcal{Z}_j}{d\mu}\mathcal{Z}_k+\mathcal{Z}_i\mathcal{Z}_j\frac{d\mathcal{Z}_k}{d\mu}\right) \nonumber \\ &= \frac{1}{2\chi^\dagger}\sqrt{\mathcal{Z}_i\mathcal{Z}_j\mathcal{Z}_k}(\gamma_i+\gamma_j+\gamma_k).
\end{align}
Performing the Grassmannian integral, one recovers the trilinear SUSY-breaking term \begin{align}
V_{\text{trilinear}} = -\frac{\sqrt{\mathcal{Z}_i\mathcal{Z}_j\mathcal{Z}_k}}{2}y^{ijk}(\gamma_i+\gamma_j+\gamma_k)F_\chi q^iq^jq^k\end{align}
which defines $A_{ijk}$ as given in Eq.~\eqref{eq:loopAMSB}.

\subsubsection{Gaugino mass}
Similarly to the field strength renormalization, the holomorphic gauge coupling $\tau$ can be treated as a spurion, 
$\tau(\mu)\to\tau(\mu/\chi)$. 
As with the field strength renormalization, we can write \begin{align}
\tau\left(\frac{\mu}{\chi}\right) &= \tau|_{\chi=1}+\frac{\partial\tau}{\partial\chi}|_{\chi=1}\theta^2F_\chi \nonumber\\&
= \tau|_{\chi=1}-\frac{1}{\chi}\frac{\partial \tau}{\partial \log\mu}|_{\chi=1}\theta^2F_\chi,
\end{align}
where the second line follows from a few applications of the chain rule. The $\theta$ angle does not run perturbatively, and so \begin{align}\label{eq: running of tau, appendix deriving AMSB equations}
\frac{d\tau}{d \log\mu}=\frac{d}{d\log\mu}\left(\frac{4\pi i}{g^2}\right) = -\frac{8\pi i}{g^3}\beta(g).
\end{align}
The lowest component of the gauge kinetic superfield $\mathcal{W}^\alpha$ is the gaugino component $\lambda^\alpha$. Thus $F_\chi$ contributes a gaugino mass given by 
\begin{align}
\frac{-1}{8\pi}\Im\left[\int d^2\theta \Tr (\tau \mathcal{W}^\alpha\mathcal{W}_\alpha)\right]  = ...+
\left(F_\chi \frac{\beta}{g}\right)\left(\frac{1}{g^2}\right)(\lambda^\alpha\lambda_\alpha )+h.c.,
\end{align}
where $...$ is the usual supersymmetric Lagrangian, along with the other AMSB terms. Passing to canonical normalization and setting $F_\chi=m_{3/2}$, we recover the gaugino mass given in Eq.~\eqref{eq:gauginomass}.

\section{Explicit expressions for Seiberg Witten periods}\label{sec:Hypergeo}
The periods $a(u), a_D(u)$ can all be expressed in terms of hypergeometric functions. We pulled the formulae from~\cite{csáki2025phasetransitionsunusualvalues}. A few comments are warranted. First, the authors of~\cite{csáki2025phasetransitionsunusualvalues} use a different notation, in which the periods are indexed by singularity and $a$ always refers to the local weakly coupled degree of freedom while $a_D$ always reduces to the photon at infinity. Their conventions also introduce a minus sign to the definition of the K{\"a}hler potential in terms of the periods.\footnote{The $N_F=3$ case is exceptional in their formalism, and neither of these caveats apply.} We use their solutions, but we label them in our conventions. 

It should also be noted that the expressions in this section are not unique. Indeed, \cite{csáki2025phasetransitionsunusualvalues} associates different expressions with different singularities. Hypergeometric functions are formally defined in terms of series solutions to the hypergeometric equation, but these series do not converge everywhere on the domain. Rather, different solutions converge in different regions. However, they can all be analytically continued over the entire punctured plane. The analytic continuations are multivalued, and at loci where branches meet, the various local solutions can be stitched together. The branched structure is how the solutions encode electromagnetic duality.  If one wants an effective weakly coupled expansion of the action near a particular singularity, it is necessary to use a branch that admits a convergent series expansion in the neighbourhood of that singularity. This is what motivates the authors of~\cite{csáki2025phasetransitionsunusualvalues} to consider different solutions in different regions.

However, both the K{\"a}hler potential and the AMSB potential on the Coulomb branch are duality invariant. For our purposes in Section~\ref{sec:coulomb branch AMSB}, we are free to arbitrarily pick a branch and stick with it throughout, secure in the knowledge that other branches would produce identical results. In this appendix, we only list the particular choices that we made. 

In all of the following, $z\equiv \frac{u}{\Lambda_{N_F}^2}$ and $\Lambda_{N_F}$ is given in Eq.~\eqref{eq:LambdaNF}. 

$\mathbf{N_F=0}$ \newline \newline
The periods for the $N_F=0$ case are given by
\begin{align}
a_D(z) &= -\Lambda_0\sqrt{\frac{z+1}{2}}  \,  
{_2F_1}\left(-1/2, 1/2,1 ; \frac{2}{z+1}\right), \\
a(z) &= \Lambda_0\left(\frac{-i(z-1)}{2}\right) \,    
{_2F_1}\left(1/2, 1/2,2 ; \frac{1-z}{2}\right).
\end{align}

$\mathbf{N_F=1}$ \newline \newline
In terms of auxiliary functions $b_1(z)$ and $c_1(z)$, the periods can be written
\begin{align}
a(z) &= c_1(z), \\
a_D(z) &= \begin{cases}
-b_1(z)+c_1(z) & -\pi < arg(z)\leq -\frac{2\pi}{3} \\
e^{-\frac{2i
\pi}{3}}b_1(z) & -\frac{2\pi}{3}<arg(z)\leq 0 \\
e^{-\frac{i\pi}{3}}b_1(z) -2c_1(z) & 0<arg(z)\leq\frac{2\pi}{3} \\
b_1(z)-2c_1(z) & \frac{2\pi}{3}<arg(z)\leq \pi.
\end{cases}
\end{align}
The auxiliary functions are \begin{align}
b_1(z) &= \Lambda_1\frac{z^3+1}{3\sqrt{8}}\,
{_2F_1}\left(\frac{5}{6}, \frac{5}{6}, 2;z^3+1\right), \\
c_1(z) &= \Lambda_1\sqrt{\frac{z}{2}}\,
{_2F_1}\left(-\frac{1}{6}, \frac{1}{6},1;-\frac{1}{z^3} \right).
\end{align}

$\mathbf{N_F=2}$ \newline \newline
The periods for $N_F=2$ are closely related to those of $N_F=0$. Here, and only here, we denote the latter by $a_D^{N_F=0}$ and $a^{N_F=0}$. We simply have \begin{align}
a_D^{N_F=2} &= \frac{\Lambda_2}{\Lambda_0}a_D^{N_F=0}, \\
a^{N_F=2} &= \frac{\Lambda_2}{\Lambda_0}a^{N_F=0}.
\end{align}

$\mathbf{N_F=3}$ \newline \newline
In terms of auxiliary functions $b_3(z)$ and $c_3(z)$, the periods are \begin{align}
a_D(z) &= c_3(z), \\
a(z) &= -b_3(z) + \frac{1}{2}c_3(z).
\end{align}
where the auxiliary functions are \begin{align}
b_3(z) &= \Lambda_3 i\frac{z-1}{2^{5/2}}\,
{_2F_1}\left(\frac{1}{2}, \frac{1}{2}, 2;1-z\right)\\
c_3(z) &= \Lambda_3\sqrt{\frac{z}{2}}{2^{5/2}}\,
{_2F_1}\left(-\frac{1}{2}, \frac{1}{2}, 1;\frac{1}{z}\right)
\end{align}

\section{Large $\mu$ superpotentials on the Higgs branch}\label{sec: large SW-deformation Higgs branch superpotentials}
In~\cite{Seiberg_1994_II}, the authors introduce superpotentials on the Higgs branches to describe the effects of the SW deformation when $\mu$ is large. These superpotentials are written in terms of an antisymmetric tensor $V^{rs}\equiv P^{ra}P_{sa}$, which can be used to parameterize the Higgs branches. Here $r$ is a flavour index, while $a$ is a colour index. $P^r$ is an $SO(2N_F)$ vector formed from the quarks $Q, \widetilde{Q}$, which can be put in an $SO(2N_F)$ vector because of the isomorphism between fundamental and antifundamental representations of the gauge group $SU(2)$ \cite{Seiberg_1994_II}.

For $N_F=2$, the superpotential is given by
\begin{align}
W=X(\text{Pf}(V)-\mu^2\Lambda^2_2)+\frac{1}{\mu}V^2 
\end{align}
where $X$ is a Lagrange multiplier. For $N_F=3$, it is given by 
\begin{align}
W=-\frac{1}{\mu^2\Lambda_3}\text{Pf}(V)+\frac{1}{\mu}V^2
\end{align}
Solving the equations of motion and taking the large $\mu$ limit recovers known results in  $\mathcal{N}=1$ SQCD.

\section{Corrections to the dual prepotential}\label{sec: higher corrections to the dual prepotential}
From~\cite{Lerche_1997}, the dual prepotential has the general form\begin{align}
\mathcal{F}_D(a_D)=\frac{a_D^2}{4\pi i}\log(\frac{a_D}{\Lambda_0})-\frac{\Lambda_0^2}{2\pi i}\sum_{l=0}c^D_l\left(\frac{ia_D}{\Lambda_0}\right)^l.
\end{align}
Plugging this expansion into Eq.~\eqref{eq: handy singularity EFT formulae}, it is straightforward to show that the dual coupling constant has the form \begin{align}\label{eq: general form of expansion for effective magnetic coupling}
\text{Im}(\tau_D)\,\propto\,\left(\frac{1}{g_{eff}}\right)^2\,\propto&\, \text{Im}\left[-i\left(\log(\frac{a_D}{\Lambda_0})+3/2-\Lambda_0^2\sum_{l=2}c_l^D\left(\frac{i}{\Lambda_0}\right)^la_D^{l-2}l(l-1)\right)\right] \nonumber \\
\, g_{eff}^2&\, \propto \left[\log\left(\left|\frac{\Lambda_0}{a_D}\right|^2\right)-3-\sum_{l=2}l(l-1)\left(c_l^D\left(\frac{ia_D}{\Lambda_0}\right)^{l-2} -h.c.\right)\right]^{-1}.
\end{align}
The leading logarithmic term is responsible for the SQED-like running of the coupling, Eq.~\eqref{eq: running of effective magnetic coupling}. We see that the $-2$ in the denominator of Eq.~\eqref{eq: leading order magnetic coupling} is altered. This has a minimal effect on the dynamics when $|\Lambda_0/a_D|\gg 1$ because the expression is dominated by the logarithm, but as we approach the Landau pole, this change has a rather dramatic effect. In particular the pole's location is exponentially sensitive to this change. The $l=2$ term in Eq.~\eqref{eq: general form of expansion for effective magnetic coupling} further changes this constant. All other terms contribute to nonrenormalizable interactions in the Lagrangian. Take, for example, the bosonic kinetic term $\sim \frac{1}{g_{eff}^2} |\partial^\mu a_D|^2$. The polynomial part of Eq.~\eqref{eq: general form of expansion for effective magnetic coupling} contributes terms of the form $a_D^l|\partial_\mu a_D|^2$. Calculating the $\beta$ function with these higher order terms added, one finds \begin{align}
\frac{\partial g}{\partial \log(a_D)} \sim g^3(1+P(a_D, a_D^\dagger)),
\end{align}
where $P(a_D, a_D^\dagger)$ is a polynomial in $a_D$ and its conjugate. This indicates that nonrenormalizable operators contribute to the running of other nonrenormalizable operators, and all $\beta$ functions are proportional to the leading order SQED-like $\beta$ function.

\bibliographystyle{apsrev-title}
\bibliography{./biblio}

\end{document}